\pdfoutput=1
\documentclass{article}

\usepackage{amsmath,amssymb,amsthm,mathtools} 
\usepackage[title]{appendix}
\usepackage[affil-it]{authblk}
\usepackage{color}    
\usepackage{epsfig}   
\usepackage{graphics,graphicx} 
\usepackage[margin=2cm]{geometry} 
\usepackage[acronym,toc,shortcuts]{glossaries}
\usepackage{ifplatform}
\usepackage{import}
\usepackage{mathptmx} 
\usepackage{natbib}
\usepackage[section]{placeins} 
\usepackage{siunitx}  
\usepackage{subcaption}
\usepackage{substr}   
\usepackage{times}    
\usepackage{units,xspace}
\usepackage{verbatim} 
\usepackage{xifthen}  
\usepackage{xstring}  

\usepackage{relsize} 
\DeclareMathSizes{7}{7}{5}{5}

\newcommand{\rulesep}{\unskip\ \vrule\ } 

\newcommand{\processnext}[1]{%
  \ifx\listfinish#1\empty\else\listact{#1}\expandafter\processnext\fi}

\newcommand{\verylargespace}{\ensuremath{\qquad}}

\ifx\isEmbedded\undefined

\else

\fi
\newcommand{\app}[1]{Appendix~#1}

\newcommand{\eq}[1]{Eq.~(#1)}
\newcommand{\eqs}[2]{Eqs.~(#1)~and~(#2)}

\newcommand{\fig}[1]{Fig.~#1}

\newcommand{\figpanel}[1]{\textbf{(#1)}}

\renewcommand{\sec}[1]{Section~#1}

\makeatletter

\newcommand{\Rmnum}[1]{\expandafter\@slowromancap\romannumeral #1@}
\makeatother

\DeclareMathAlphabet{\mathpzc}{OT1}{pzc}{m}{it} 
\DeclareMathAlphabet{\mathcal}{OMS}{cmsy}{m}{n} 

\DeclarePairedDelimiter{\floor}{\lfloor}{\rfloor}
\DeclarePairedDelimiter{\ceil}{\lceil}{\rceil}
\DeclarePairedDelimiter\abs{\lvert}{\rvert}%
\DeclarePairedDelimiter\norm{\lVert}{\rVert}%
\makeatletter
\let\oldfloor\floor
\def\floor{\@ifstar{\oldfloor}{\oldfloor*}}

\makeatletter
\let\oldceil\ceil
\def\ceil{\@ifstar{\oldceil}{\oldceil*}}

\makeatletter
\let\oldabs\abs
\def\abs{\@ifstar{\oldabs}{\oldabs*}}

\let\oldnorm\norm
\def\norm{\@ifstar{\oldnorm}{\oldnorm*}}
\makeatother

\newcommand*\xoverline[2][0.75]{%
  \sbox{\myboxA}{$\math#2$}%
    \setbox\myboxB\null
    \ht\myboxB=\ht\myboxA%
	\dp\myboxB=\dp\myboxA%
	\wd\myboxB=#1\wd\myboxA
	\sbox\myboxB{$\math\overline{\copy\myboxB}$}
	\setlength\mylenA{\the\wd\myboxA}
	\addtolength\mylenA{-\the\wd\myboxB}%
	\ifdim\wd\myboxB<\wd\myboxA%
	  \rlap{\hskip 0.5\mylenA\usebox\myboxB}{\usebox\myboxA}%
    \else
	  \hskip -0.5\mylenA\rlap{\usebox\myboxA}{\hskip
	  		  0.5\mylenA\usebox\myboxB}%
    \fi
}

\newcommand{\avg}[1]{\ensuremath{\mean{#1}}}
\newcommand{\conv}{\ensuremath{\star}}
\newcommand{\conj}[1]{\ensuremath{\left(#1\right)^*}}

\newcommand{\erf}[1]{\mathrm{erf}\ensuremath{\left(#1\right)}}
\newcommand{\erfc}[1]{\mathrm{erfc}\ensuremath{\left(#1\right)}}
\newcommand{\erfi}[1]{\mathrm{erfi}\ensuremath{\left(#1\right)}}
\renewcommand{\exp}[1]{\mathrm{exp}\ensuremath{\left(#1\right)}}

\newcommand{\ii}{\mathrm{i}}
\newcommand{\mean}[1]{\ensuremath{\overline{#1}}}

\newcommand{\vect}[1]{{\bf #1}}
\newcommand{\icol}[2]{
   \left( #1, #2 \right)
}

\newcommand{\todo}[1]{\color{blue}TO DO: #1\color{black}\newline}
\newcommand{\units}[2]{\ensuremath{\left[#1\right]=#2}}

\newcommand{\del}{\ensuremath{\delay{}{}{}{}}}
\newcommand{\delpsp}{\ensuremath{\delta}^{\layer{B}\layer{C}}_{\del,\psp}}
\newcommand{\delsym}{\ensuremath{\Delta}}
\newcommand{\delay}[4] 
{ \ifthenelse{\isempty{#1}}%
   {\ensuremath{\delsym}}
   {\ifthenelse{\isempty{#3}}%
	  {\ensuremath{\delsym_{#1#2}}}
      {\ensuremath{\delsym_{#1#2}^{#3#4}}}
   }
}
\newcommand{\DelClayer}[2]{\ensuremath{\kappa^{BC}_{\del_{#1#2}}}} 
\newcommand{\Dellayer}[2]{\ensuremath{\kappa^{#1#2}_{\del}}} 
 
\newcommand{\delaymean}{\ensuremath{D}}

\newcommand{\distCOV}{\ensuremath{r'}}
\newcommand{\distancesym}{\ensuremath{d}}
\newcommand{\distance}[4] 
{ \ifthenelse{\isempty{#3}}
	{\ensuremath{\distancesym_{#1#2}}\xspace}
    {\ensuremath{\distancesym_{#1#2}^{#4}}\xspace}
}
\newcommand{\distfunc}{\ensuremath{g}}
\newcommand{\distlayer}[2]{\ensuremath{\distancesym^{#1#2}}}
\newcommand{\distlayerSqu}[2]{
{ \ifthenelse{\isempty{#1}}
	{\ensuremath{\distancesym^2}}
    {\ensuremath{(\distancesym^{#1#2})^2}}
}}	
\newcommand{\expect}[1]{\ensuremath{\mathrm{E}\Bigl[#1\Bigr]}}

\newcommand{\poscont}{\ensuremath{\vect{x}}} 

\newcommand{\posri}[2]{\ensuremath{r_{#1#2}}} 

\newcommand{\posth}{\ensuremath{\theta}}
\newcommand{\posthi}[2]{\ensuremath{\theta_{#1#2}}} 

\newcommand{\postwocont}{\ensuremath{\vect{x}'}} 
\newcommand{\posintr}{\ensuremath{\tilde{r}}}

\newcommand{\postwoth}{\ensuremath{\tilde{\posth}}}

\newcommand{\posx}[2]{\ensuremath{x_{#1#2}}}
\newcommand{\posy}[2]{\ensuremath{y_{#1#2}}}
\newcommand{\posxcont}{\ensuremath{x}} 
\newcommand{\posycont}{\ensuremath{y}} 
\newcommand{\posxtwocont}{\ensuremath{\tilde{x}}} 
\newcommand{\velocity}{\ensuremath{v}}

\newcommand{\atDist}[1]{\ensuremath{{#1}\left(\distance{}{}{}{}\right)}} 
 
\newcommand{\atFreq}[1]{\ensuremath{#1\left(\freq\right)}} 

\newcommand{\atFreqAndRadius}[3]{\ensuremath{#1^{#2#3}\left( \freq \right)}} 

\newcommand{\timesym}{\ensuremath{t}} 
\newcommand{\atTime}[1]{\ensuremath{#1\left( \timesym \right)}} 
\newcommand{\atDelayTime}[2]{\ensuremath{#1\left( \timesym - #2\right)}} 
\newcommand{\atCorrTime}[2]{ \ensuremath{#1\left( \timesym + #2\right)}} 
\newcommand{\filt}{\ensuremath{\psp}}
\newcommand{\fourier}[1]{\ensuremath{{\displaystyle\mathcal{F}\left\{#1\right\}}}}
\newcommand{\freq}{\ensuremath{k}}
\newcommand{\freqT}{\ensuremath{k'}}
\newcommand{\lag}{\ensuremath{p}}
\newcommand{\psp}{\ensuremath{\epsilon}}
\newcommand{\PSPlayer}{\ensuremath{\kappa^{\psp}}} 
\newcommand{\PSP}{\ensuremath{E}}
\newcommand{\Tmax}{\ensuremath{T}}
\newcommand{\xcorr}[3]{\ensuremath{\gamma_{#1,#2}\left(#3\right)}}

\newcommand{\scorr}[2]{\ensuremath{\mathrm{corr}\left(#1, \, #2\right)}}
\newcommand{\scov}[2]{\ensuremath{\mathrm{cov}\left(#1, \, #2\right)}}
\newcommand{\cov}[2]{\ensuremath{\mathrm{cov}\left(#1, \, #2\right)}}
\newcommand{\var}[1]{\ensuremath{\mathrm{var}\left(#1\right)}}

\newcommand{\corr}[2]{\ensuremath{\rho\left(#1, \, #2\right)}}

\newcommand{\taul}[2]{\ensuremath{\tau_l}} 
\newcommand{\taur}[2]{\ensuremath{\tau_r}} 
\newcommand{\taupsp}{\ensuremath{\tau_{\epsilon}}}

\newcommand{\ensemble}[1]{\ensuremath{\left\langle #1 \right\rangle}}
\newcommand{\etalayer}[1]{\ensuremath{\eta_{#1}}}

\newcommand{\cosbr}[2]  
{ \ifthenelse{\isempty{#2}}%
	{\ensuremath{\cos\left(#1\right)}}
	{\ensuremath{\cos^{#2}\left(#1\right)}}
}

\newcommand{\eigvalue}[2]{
{ \ifthenelse{\isempty{#1}}
	{\ensuremath{\lambda}}
    {\ifthenelse{\isempty{#2}}
		{\ensuremath{\lambda_{#1}}}
		{\ensuremath{\lambda_{#1,#2}}}}
}}

\newcommand{\fo}[1]
{ \ifthenelse{\isempty{#1}}%
	{\ensuremath{f_{0}^2}}
	{\ifthenelse{\equal{\detokenize{#1}}{\detokenize{-}}}
		{\ensuremath{f_{0}^{-2}}}
        {\ensuremath{f_{0,#1}^2}}
	}
}

\newcommand{\gridspace}[1]{\ensuremath{\mu^{#1}}}

\newcommand{\laguerre}[3]
{ \ifthenelse{\isempty{#3}}
	{\mathrm{L}\ensuremath{_{#1}^{#2}}}
	{\mathrm{L}\ensuremath{_{#1}^{#2}\left(#3\right)}} 
}
\newcommand{\inputlayer}{\ensuremath{\layer{A}}}

\newcommand{\kone}[2]{\ensuremath{k_1^{#1#2}}}
\newcommand{\ktwo}[2]{\ensuremath{k_2^{#1#2}}}
\newcommand{\layerrate}[1]{\ensuremath{\lambda^{#1}}}
\newcommand{\layerrateSq}[1]{\ensuremath{\left(\lambda^{#1}\right)^2}}
\newcommand{\layervar}[1]{\ensuremath{(\tilde{\sigma^{#1})^2}}}

\newcommand{\layer}[1]{\textit{#1}} 
\newcommand{\Lcov}[3]{\ensuremath{Q_{#2#3}^{#1}}} 
\newcommand{\lcov}[2] 
{ \ifthenelse{\isempty{#1}}
	{\ensuremath{Q}}
	{\ensuremath{Q\left({#1,#2}\right)}}
}
\newcommand{\LRATE}[2]{\ensuremath{F_{#1}^{#2}}}  
\newcommand{\Lrate}[3]  
{ \ifthenelse{\isempty{#3}}
	{ \ensuremath{f_{#1}^{#2}}\xspace}
	{ \ifthenelse{\equal{ \detokenize{#3} }{ \detokenize{\filt} } } 
	    { \ensuremath{ ( f_{#1}^{#2} \conv \filt ) } }
		{ \ensuremath{ ( f_{#1}^{#2(\delsym)} \conv \filt ) } }
	}
}

\newcommand{\Lwgtsym}{\ensuremath{w}}
\newcommand{\Lweight}[4]{\ensuremath{{\Lwgtsym}^{#1#2}_{#3#4}}} 
\newcommand{\Lwchange}[4]{\ensuremath{\dot{\Lwgtsym}^{#1#2}_{#3#4}}} 
\newcommand{\LwgtMin}[2]{\ensuremath{\Lwgtsym_{\textrm{min}}^{#1#2}}} 
\newcommand{\LwgtMax}[2]{\ensuremath{\Lwgtsym_{\textrm{max}}^{#1#2}}} 
\newcommand{\Lwgtcont}[2]{ 
{\ifthenelse{{\isempty{#2}}}
	{\ensuremath{\Lwgtsym\left({#1}\right)}}
	{\ensuremath{\Lwgtsym\left({#1,#2}\right)}}
}}
\newcommand{\Lwderiv}[2]{
{\ifthenelse{{\isempty{#2}}}
	{\ensuremath{\dot{\mathlarger{v}}\left(#1\right)}}
	{\ensuremath{\dot{\mathlarger{v}}\left(#1,#2\right)}}
}}
\newcommand{\FP}[1]{\ensuremath{#1_*}}
\newcommand{\meanCov}[3]{\ensuremath{\avg{\Lcov{#1#2}{}{#3}}}} 
\newcommand{\meanLweight}[2]{\ensuremath{\avg{\Lwgtsym}^{#1#2}}} 
\newcommand{\meanLwchange}[2]{\ensuremath{\dot{\avg{\Lwgtsym}}^{#1#2}}} 
\newcommand{\meanrate}[1]{\ensuremath{\mean{\Lrate{}{#1}{}}}}

\newcommand{\N}[2]{\ensuremath{N^{#1#2}}} 

\newcommand{\Nprepost}[4]{\ensuremath
{\ifthenelse{{\isempty{#3}}}
	{\ensuremath{N_{#1#2}}}
	{\ensuremath{N_{#1#2}^{#3#4}}}
}} 
\newcommand{\Nshared}[2] 
{\ifthenelse{{\isempty{#2}}}
	{\ensuremath{N^{#1#1}}}
	{\ensuremath{N^{#1#2}}}
}

\newcommand{\preN}{\ensuremath{m}}
\newcommand{\preNtwo}{\ensuremath{n}}
\newcommand{\postN}{\ensuremath{i}}
\newcommand{\postNtwo}{\ensuremath{j}}
\newcommand{\postpostN}{\ensuremath{p}}

\newcommand{\radsym}{\ensuremath{\sigma}}
\newcommand{\radius}[2]{\ensuremath{\radsym^{#1#2}}}
\newcommand{\radiusvar}[2]{\ensuremath{(\radsym^{#1#2})^2}}

\newcommand{\radiusquad}[2]{\ensuremath{(\radsym^{#1#2})^4}}

\newcommand{\Ra}[1]
{ \ifthenelse{\isempty{#1}}%
	{\ensuremath{R_a}}
    {\ensuremath{R_a^{#1}}}
}
\newcommand{\Rb}[2]{}
\newcommand{\RbSq}[2]{}
\newcommand{\RF}[1]{\ensuremath{r_{\textrm{on}}^{#1}}}
\newcommand{\RFvar}[1]{\ensuremath{\left(\RF{#1}\right)^2}}

\newcommand{\SumT}[1]{\ensuremath{{\sum\limits_{t=0}^{#1-1}}}}
\newcommand{\Sumt}{\ensuremath{\sum\limits_{t}}}
\newcommand{\Sum}[3]{\ensuremath{\sum\limits_{#1=#2}^{#3}}}

\newcommand{\Sumk}{\ensuremath{\sum\limits_{\freq}}}
\DeclareMathAlphabet{\mathpzc}{OT1}{pzc}{m}{it} 

\newcommand{\tab}{\hspace{5mm}}

\newcommand{\dist}[2]{\ensuremath{\distfont{#1}\left(#2\right)}}
\newcommand{\distfont}[1]
{ \ifthenelse{\isempty{#1}}%
	{\mathscr{#1}}
	{\ifthenelse{\equal{#1}{t}}
		{\ensuremath{\mathlarger{\mathlarger{\mathpzc{#1}}}}}
		{\ifthenelse{\equal{#1}{Lapl}}
			{\ensuremath{\mathlarger{\mathlarger{\mathpzc{#1}}}}}
			{\ifthenelse{\equal{#1}{Poisson}}
			   {\ensuremath{\mathlarger{\mathpzc{Poisson}}}}
			   {\ensuremath{\mathpzc{#1}}}
			}
		}
	}
}

\newcommand{\probdist}[3]
{ \ifthenelse{\isempty{#3}}%
	{p_{#1}\left(#2\right)}
    {p_{#1}\left(#2\,;\,#3\right)}
}
\newcommand{\fnin}[1]
{ \ifthenelse{\isempty{#1}}%
	{}
	{\ensuremath{_{#1}}}
}
\newcommand{\sfnin}[1]
{ \ifthenelse{\isempty{#1}}%
	{}
	{\ensuremath{\left({#1}\right)}}
}
\newcommand{\ssym}[1]
{ \ifthenelse{\isempty{#1}}%
	{}
	{\ensuremath{\mathrm{#1}}}
}
\newcommand{\psym}[1]
{ \ifthenelse{\isempty{#1}}%
	{}
	{\ensuremath{\hat{#1}}}
}

\bibpunct{(}{)}{;}{a}{,}{,} 

\makeglossaries

\newacronym{if}{IF}{integrate-and-fire}
\newacronym{lgn}{LGN}{lateral geniculate nucleus}
\newacronym{psp}{PSP}{post-synaptic potential}
\newacronym{rf}{RF}{receptive field}
\newacronym{stdp}{STDP}{spike-timing-dependent plasticity}
\newacronym{vc}{VC}{visual cortex}
\newacronym{v1}{V1}{primary visual cortex}
\newacronym{wgn}{WGN}{white Gaussian noise}

\makeatletter

\newcommand{\isEmbedded}{true} 

\patchcmd{\NAT@test}{\else \NAT@nm}{\else \NAT@nmfmt{\NAT@nm}}{}{}

\DeclareRobustCommand\citepos
  {\begingroup
   \let\NAT@nmfmt\NAT@posfmt
   \NAT@swafalse\let\NAT@ctype\z@\NAT@partrue
   \@ifstar{\NAT@fulltrue\NAT@citetp}{\NAT@fullfalse\NAT@citetp}}

\let\NAT@orig@nmfmt\NAT@nmfmt
\def\NAT@posfmt#1{\NAT@orig@nmfmt{#1's}}

\begin{document}


\title{Impact of axonal delay on structure development in a multi-layered network}
\author[a]{Catherine E. Davey}
\author[a]{David B. Grayden}
\author[a]{Anthony N. Burkitt}
\affil[a]{Department of Biomedical Engineering, The University of Melbourne, VIC 3010, Australia}
\maketitle


{\bf Keywords:} neural network, rate-based neural plasticity, axonal propagation delay, spatial opponent cells

\section*{Abstract}

The mechanisms underlying how random 
activity in the visual pathway could give rise through neural plasticity to many 
of the features observed experimentally in the early stages of visual processing 
was provided by Linkser in a seminal, three-paper series. Owing to the complexity 
of multi-layer models, by Linkser in a licit assumption in Linsker's and subsequent 
papers has been that propagation delay is homogeneous and, consequently, plays little 
functional role in neural behaviour. In this paper, we relax this assumption to 
examine the impact of distance-dependent axonal propagation delay on neural learning. 
We show that propagation delay induces low-pass filtering by dispersing the 
arrival times of spikes from presynaptic neurons, providing a natural correlation 
cancellation mechanism for distal connections. The cut-off frequency decreases 
as the radial propagation delay within a layer increases relative to propagation 
delay between the layers, introducing an upper limit on temporal resolution. 
Given that the \gls{psp} also acts as a low-pass filter, we show that the effective 
time constant of each should enable the processing of similar scales of temporal 
information. This result has implications for the visual system, in which receptive 
field size and, thus, radial propagation delay, increases with eccentricity.
Furthermore, the network response is frequency dependent since higher frequencies 
require increased input amplitude to compensate for attenuation. This concords 
with frequency-dependent contrast sensitivity in the visual system, which changes 
with eccentricity and receptive field size. We further show that the proportion of
inhibition relative to excitation is larger where radial propagation delay is long
relative to inter-laminar propagation delay. We show that the addition of 
propagation delay reduces the range in the size of the spatial opponent cell's 
on-center size, providing stability to variations in homeostatic parameters 
across a feature map. 

   \glsresetall 

\section{Introduction}\label{sec:intro}
   
Synaptic plasticity describes the process by which changes to synaptic weights 
between neurons occur in response to their activities. The development of mathematical 
models of plasticity has been instrumental in furthering our understanding of 
the formation of receptive fields of simple cells in \gls{v1}. 
\citet{Lin86a,Lin86b,Lin86c} showed how cells sensitive to simple features, such 
as orientation, can develop across several neural layers in the absence of 
structured input, demonstrating how spatial structure in cortical connections is 
sufficient to drive a self-organization learning process. There has subsequently 
been much progress in our understanding of the plasticity mechanisms that 
drive learning of simple features. An important component of this 
work has been the development of a rigorous mathematical framework from which to 
interrogate properties of plasticity and neural network development 
\citep{MacMil90,WimGerHem98}.

Development of a rigorous mathematical framework to express Hebbian learning on 
a network scale has been crucial in advancing our understanding of the formation 
of simple cells and feature maps in the early stages of learning \citep{Mill90}. 
Owing to the complexity of models resulting from multiple layers of structured 
input, there has been limited work that has addressed the impact of axonal delay on the 
development of simple features, with none incorporating the impact of delay 
both within and between layers. \citet{WimWenMilvanH97b} showed the development 
of direction selective cells in response to a combination of lagged and non-lagged 
input via empirical simulation, although the impact of propagation delay was 
assumed to be negligible. \citet{LeiKemVan01} provided a mathematical description 
of the evolution of a temporal feature map, highlighting the importance of temporal 
parameters in the neuron model and learning function but not incorporating 
spatially dependent delay. \citet{LeiVan01} showed the evolution of temporal 
feature maps for sound localization, employing a one-dimensional spatial structure. 

An implicit assumption of most of these papers has been that spike propagation 
is instantaneous or, equivalently, that the interlaminar distance is sufficiently 
greater than radial distances between neurons to permit the assumption that 
propagation time is approximately equal for all spikes received by a neuron. This 
assumption enables correlation between neural outputs to be evaluated without spatially 
induced delay. However, experimental work demonstrates that propagation delay is 
important. For example, \cite{SaaEck00} showed that spike conduction velocity 
impacts the spatial range of synchronization and the subsequent receptive field 
sizes of cells. \citet{KrePerMonAloAerFreMas16} noted the importance of controlling 
the delay of excitation and inhibition in broadening the range of statistics that 
sensory cortical neurons can process. The importance of propagation delay has 
also been highlighted in the context of synaptic plasticity.  \citet{EguNeyStr14} 
discuss the role of propagation delay in developing clusters of cells via \gls{stdp}, 
while \citet{AslValTas17} explore the impact of propagation delay on emerging 
network structure with \gls{stdp} as a result of spike arrival order being disrupted. 
Furthermore, the advantageous arrangement of conduction delay via propagation 
length or axon diameter to cause coincident spike arrival at postsynaptic neurons 
has been long noted, such as in giant squid axons \citep{PumYou38} and the 
barn owl \citep{CarKon88}. 

While the importance of spike timing for synaptic plasticity has been well motivated, 
the modelling of propagation delay to understand its impact on the evolution of 
receptive field structure is still in its infancy. In this paper, we examine the 
impact of three-dimensional propagation delay on the emergence of \gls{v1} simple 
cells using Linsker's three layer network model. We analytically derive the
impact of spike propagation delay on the neural activity of postsynaptic cells
and determine an expression for covariance between neurons that incorporates 
radial propagation delay determined from synaptic connectivity radius, spike 
propagation velocity, cell density, inter-laminar propagation delay, and an 
exponentially decaying \gls{psp}. We show that propagation delay effectively
acts as a low-pass filter, filtering out high frequencies of the inputs by 
spreading out the arrival time of presynaptic signals to the cell. We further
show that the ratio of propagation delay between neural layers to the radial
delay within a layer is the most important factor in determining the extent of
attenuation of the high frequencies. In deriving the impact of propagation delay
on covariance between presynaptic inputs and, hence, on neural plasticity, we
show that, while delay spreads out the arrival times of spikes and, hence,
decimates covariance between presynaptic neurons, the \gls{psp} is crucial in
restoring covariance, as it temporally spreads the impact of a single spike. 
Given that a \gls{psp} also acts as a low-pass filter, we propose that the 
cut-off frequencies for each will be compatible. This may partly explain the 
different synaptic time constants found in peripheral and foveal ganglion cells, 
which can differ by up to a factor of 2 \citep{ZhaHatOkaWan16}. It may also 
explain why contrast sensitivity is frequency dependent in the periphery, and 
why the cut-off frequency increases with eccentricity \citep{VirRovLauNan82}. 
It also provides an insight into why spatial and temporal frequency processing 
in the visual cortex are not independent \citep{ZhaHatOkaWan16,VenLewUnsLun17}.
To simplify application of propagation delay in neural models, we derive a
mathematically simple approximation and demonstrate its accuracy across a range
of parameter configurations. 

Finally, we show that the size of the on-centre compared to the off-surround of 
a resultant \gls{v1} cell is a function of the fixed point mean synaptic weight, 
which can be determined from the homeostatic constants and covariance between 
presynaptic inputs. For smaller fixed point mean weights, there will be relatively 
fewer excitatory inputs and more inhibitory inputs. This is in agreement with the 
synaptic arrangement of ganglion cells, which have significantly more inhibitory 
inputs in the periphery, where the cells are more spread across the laminar and 
the impact of radial propagation delay will be greatest \citep{SinHooBauOkaWonRie17}.  
We show that propagation delay limits the range of receptive field sizes that 
can emerge from learning. 

   \glsresetall 

\section{Methods}

\subsection{Specification of the network}\label{sec:ntwkSpec}
   
\begin{figure}[ht!]
\centering
	\includegraphics[width=0.6\textwidth]{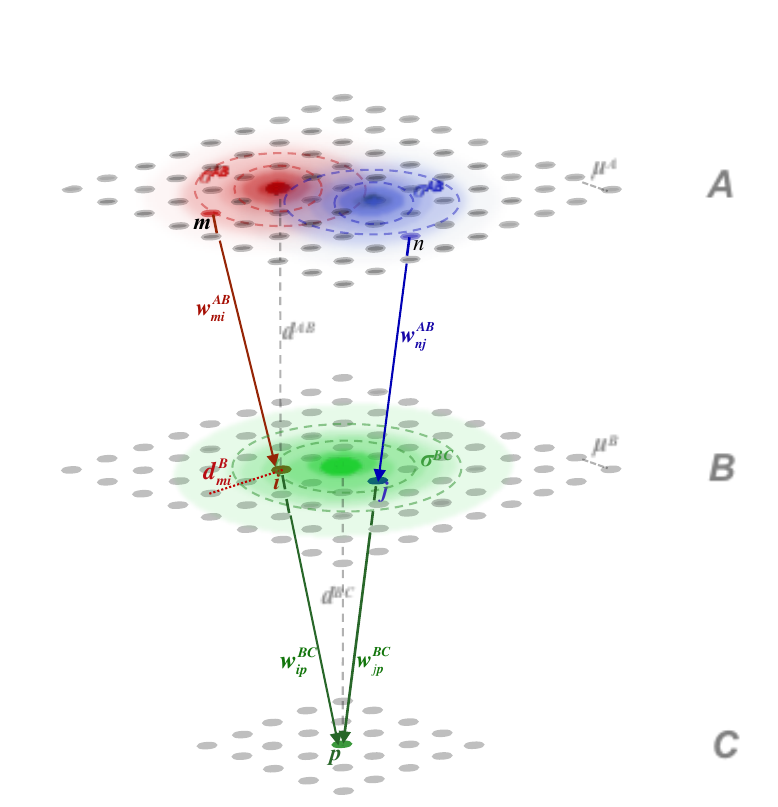}
	\caption{Diagram of a three layer feed-forward network, with neurons $\preN$ 
		and $\preNtwo$ in layer $\layer{\inputlayer}$ connecting to neurons 
		$\postN$ and $\postNtwo$, respectively, in layer $\layer{B}$; both neurons 
		$\postN$ and $\postNtwo$ in layer $\layer{B}$ connect to neuron $\postpostN$ 
		in layer $\layer{C}$. Each postsynaptic neuron has been color coded to 
		aid interpretation, such that neuron $\postN$ has been colored red, 
		neuron $\postNtwo$ is blue, and neuron $\postpostN$ is green. 
		The synaptic connection density of a postsynaptic neuron determines the 
		probability of a neuron in the presynaptic layer connecting to it and 
		is depicted by the density of that neuron's color in the presynaptic 
		layer. Connection density is modelled as Gaussian, with parameters 
		constant across a layer, so that $\sigma^{{\inputlayer}{\layer{B}}}$ 
		denotes the radius, or standard deviation, of connections between layers 
		$\layer{\inputlayer}$ and $\layer{B}$, while $\sigma^{\layer{B}\layer{C}}$
		denotes the radius of connections between layers $\layer{B}$ and $\layer{C}$. 
        The first and second standard deviations of the Gaussian connection probability 
		for a postsynaptic neuron are represented by colored, dashed concentric circles 
		in the presynaptic layer of the neuron. Interlaminar distance between layers 
		$\layer{\inputlayer}$ and $\layer{B}$ is denoted $d^{\inputlayer\layer{B}}$, 
		while distance between layers $\layer{B}$ and $\layer{C}$ is denoted by 
		$d^{\layer{B}\layer{C}}$. The radial distance between two neurons within 
		a lamina, for example between neuron $\preN$ from layer $\inputlayer$ 
		and $\postN$ from layer $\layer{B}$, is denoted $d_{\preN\postN}^{\layer{B}}$. 
		Distance metrics are given in grid spaces, $\gridspace{\inputlayer}$ and 
		$\gridspace{\layer{B}}$, respectively, which are in themselves measured 
		in \si{\micro\meter}, to separate out the effect of cell density. 
		The strength of a synaptic connection between two neurons, for example between 
		neurons $\postNtwo$ and $\postN$, is given by 
		$w_{{\postNtwo}{\postN}}^{{\layer{B}}{\layer{C}}}$. Not shown in the diagram 
		are $\N{\inputlayer}{\layer{B}}$ and $\N{\layer{B}}{\layer{C}}$, which 
		represent the expected number of synaptic connections to each postsynaptic 
		neuron in layers $\layer{B}$ and $\layer{C}$, respectively.
		\label{fig:network}}
\end{figure}

We employ the topographic network proposed by \citet{Lin86a}, which comprises 
three layers, $\layer{A}$, $\layer{B}$, and $\layer{C}$, with feedforward 
connections from layer $\layer{A}$ to $\layer{B}$ and from layer $\layer{B}$ 
to $\layer{C}$, as illustrated in \fig{\ref{fig:network}}. We remain faithful 
to Linsker's notion of a simple network with a minimal set of assumptions to 
identify the fundamental principles driving the emergence of cortical structure 
\citep{Lin86a}. The layers are positioned as planes parallel to one another with 
neurons equispaced in a square grid within each lamina. Connections between layers 
are learned sequentially, such that synapses from $\layer{A}$ to $\layer{B}$ are 
learned to maturity before connections from $\layer{B}$ to $\layer{C}$ are learned. 
To aid interpretation of distance metrics, we measure distances across a lamina 
in terms of number of grid spaces. Connections between layers have a spatial 
distribution such that nearby neurons in the presynaptic layer are more densely 
connected to a postsynaptic neuron in the overlying layer. 

As the radial distance from the postsynaptic neuron increases, connection density 
decreases as a Gaussian function of the distance. Define the two-dimensional radial 
distance of presynaptic neuron, $\preN$, from postsynaptic neuron, $\postN$, by a 
vector, $\icol{\posx{\preN}{\postN}}{\posy{\preN}{\postN}}$. The density of connections 
from layer $\inputlayer$ to the subsequent layer \layer{B} is parameterized by 
the radius, $\radiusvar{\inputlayer}{\layer{B}}$, which is the distance-dependent 
variance of the Gaussian distributed connection probability, measured in grid 
spaces, $\gridspace{\inputlayer}$, to ensure scaling with cell density 
in the lamina. Consequently, the probability of presynaptic neuron, $\preN$, 
in layer $\layer{\inputlayer}$ generating a synaptic connection to postsynaptic 
neuron, $\postN$, in layer \layer{B} is given by 
\begin{align}\label{eq:2DGauss}
	\probdist{N}{ \icol{ \posx{\preN}{\postN} }{ \posy{\preN}{\postN}} }
				{ \frac{1}{2} \radiusvar{\layer{A}}{\layer{B}} }
  &= 
	\frac{1}{\pi\radiusvar{\inputlayer}{\layer{B}}} 
	\exp{- \frac{ \posx{\preN}{\postN}^2 
				+ \posy{\preN}{\postN}^2 }
				{ \radiusvar{\inputlayer}{\layer{B}}} }.
\end{align}
Note that this definition differs from the standard definition by a factor of 
$\sqrt{2}$; it has been specifically chosen for later convenience. 

Similarly, the distribution of presynaptic connections from layer \layer{B} to 
layer \layer{C} is Gaussian, parameterized by $\radiusvar{\layer{B}}{\layer{C}}$. 
Without loss of generality, we can assume that a layer \layer{C} cell is in the 
center of the lamina, and write the probability of a \layer{B} cell making a 
connection to the layer $\layer{C}$ cell using polar coordinates, 
\begin{align}\label{eq:2DGauss_polar}
	\probdist{N}{ \icol{\posri{\postN}{\postpostN} }{ \posthi{\postN}{\postpostN}} }
				{ \radiusvar{\layer{B}}{\layer{C}} }
  &= 
	\frac{1}{ \pi \radiusvar{\layer{B}}{\layer{C}} } 
	\exp{- \frac{ \posri{\postN}{\postpostN}^2 }
				{ \radiusvar{\layer{B}}{\layer{C}}} },
\end{align}
where $\posri{\postN}{\postpostN}$ is the radial distance from the center of the 
laminar to $\postN$ and $\posthi{\postN}{\postpostN}$ is the angle to $\postN$ 
within the two-dimensional lamina.

\ifx\isEmbedded\undefined
	\subsubsection{Shared inputs}
	
To examine network dynamics, it is necessary to ascertain the expected number of 
shared connections between two neurons. The number of shared connections from a 
presynaptic layer to two neurons in the postsynaptic layer, say $\postN$ and 
$\postNtwo$, depends on the radial distance between them since the synaptic 
connection density for each is a Gaussian function of distance 
(see \fig{\ref{fig:network}}). We assume for simplicity and without loss of 
generality that $\postN$ and $\postNtwo$ differ only in their $\posx{}{}$ 
coordinate so that 
$ \distance{\postN}{\postNtwo}{\layer{B}}{\layer{B}} 
= \posx{\preN}{\postN} - \posx{\preN}{\postNtwo}$.

Center the Cartesian coordinates describing a neuron's position in the 
laminar on one of the postsynaptic neurons, say $\postN$, so that the other 
postsynaptic neuron, say $\postNtwo$, lies on the $x$ axis. From 
\eq{\ref{eq:2DGauss}}, neuron, $\preN$, 
in layer $\inputlayer$, has a probability of connecting to neuron $\postN$ 
in layer \layer{B} of 
$\probdist{N}{\posx{\preN}{\postN},\posy{\preN}{\postN}}
             {\vect{0},\Sigma^{\inputlayer}}$ 
and a probability of connecting to neuron $\postNtwo$ in layer $\layer{B}$ of 
$ \probdist{N}{\posx{\preN}{\postNtwo},\posy{\preN}{\postNtwo}}
              {\vect{0},\Sigma^{\inputlayer}}
= \probdist{N}{\posx{\preN}{\postN} - 
	           \distance{\postN}{\postNtwo}{\layer{B}}{\layer{B}},
	           \posy{\preN}{\postN}}
              {\vect{0},\Sigma^{\inputlayer}}$.
The probability of the presynaptic neuron connecting to both postsynaptic neurons 
$\postN$ and $\postNtwo$ is simply the product of the probability of each individual 
connection being made. The expected number of common connections can be determined 
by summing this joint probability over the layer or, in the continuous limit, 
integrating the joint probability over the layer of presynaptic 
neurons. If $\N{\inputlayer}{\layer{B}}$ denotes the number of synaptic connections 
from layer $\inputlayer$ to a layer $\layer{B}$ neuron and 
$\atDist{\Nshared{\layer{B}}{}}$ the number of shared connections between 
two postsynaptic neurons in layer $\layer{B}$ separated by a distance of 
$\distance{}{}{}{}$, then in the continuous limit, 
\begin{align}
    \atDist{\Nshared{\layer{B}}{}}
 &= (\N{\inputlayer}{\layer{B}})^2   
    {\mathlarger{\iint\limits_{\posx{}{}\posy{}{}}}}
    \probdist{N}{\posx{}{},\posy{}{}}
                {\vect{0},\Sigma^{\inputlayer}}
    \probdist{N}{\posx{}{} - \distance{}{}{}{},\posy{}{}}
                {\vect{0},\Sigma^{\inputlayer}}
    \,d\posx{}{} \,d\posy{}{}, 
\end{align}
where the sub- and super-scripts on distance parameters have been dropped to aid 
readability. This can be expanded as  
\begin{align}
   \atDist{\Nshared{\layer{B}}{}}
 &= 
  (\N{\inputlayer}{\layer{B}})^2
  {\mathlarger{\iint\limits_{\posx{}{}\posy{}{}}}}
	 \frac{1}{\left(\pi\radiusvar{\inputlayer}{\layer{B}}\right)^2} 
	 \exp{- \frac{\posx{}{}^2 + \posy{}{}^2}
			           {\radiusvar{\inputlayer}{\layer{B}}} }
	 \exp{- \frac{\left(\posx{}{}-\distance{}{}{}{}\right)^2 + \posy{}{}^2} 
			     {\radiusvar{\inputlayer}{\layer{B}}} }
   \,d\posx{}{} \,d\posy{}{}    \notag \\
 &= 
  \frac{(\N{\inputlayer}{\layer{B}})^2}{\left(\pi\radiusvar{\inputlayer}{\layer{B}}\right)^2} 
  {\mathlarger{\iint\limits_{\posx{}{}\posy{}{}}}}
   \exp{- \frac{ 2 \posx{}{}^2 + 2\posy{}{}^2 
				 + \distance{}{}{}{}^2
				 - 2\posx{}{}\distance{}{}{}{} }
			   {\radiusvar{\inputlayer}{\layer{B}}} }
   \,d\posx{}{} \,d\posy{}{}  \notag   \\
 &= 
  \frac{(\N{\inputlayer}{\layer{B}})^2}{\left(\pi\radiusvar{\inputlayer}{\layer{B}}\right)^2} 
  {\mathlarger{\iint\limits_{\posx{}{}\posy{}{}}}}
	 \exp{-\frac{ 2 \left( \left(\posx{}{}-\frac{\distance{}{}{}{}}{2}\right)^2 
				               + \posy{}{}^2 + \frac{\distance{}{}{}{}^2}{4}
						  \right)}
			           {\radiusvar{\inputlayer}{\layer{B}}} }
   \,d\posx{}{} \,d\posy{}{}.
\end{align}
Introduce $\posx{}{}' = \posx{}{} - \nicefrac{\distance{}{}{}{}}{2}$, so that 
\begin{align}\label{eq:sharedConns}
   \atDist{\Nshared{\layer{B}}{}}
 &= 
   \exp{-\frac{\distance{}{}{}{}^2}{2\radiusvar{\inputlayer}{\layer{B}}}}
   \frac{(\N{\inputlayer}{\layer{B}})^2}{\left(\pi\radiusvar{\inputlayer}{\layer{B}}\right)^2} 
  {\mathlarger{\iint\limits_{\posx{}{}\posy{}{}}}}
   \exp{-\frac{ 2 \left( \posx{}{}'^2 + \posy{}{}^2 \right)}
			  {\radiusvar{\inputlayer}{\layer{B}}} }
   \,d\posx{}{} \,d\posy{}{}                                         \notag \\
 &= 
   \exp{-\frac{\distance{}{}{}{}^2}{2\radiusvar{\inputlayer}{\layer{B}}}}
   \frac{(\N{\inputlayer}{\layer{B}})^2}{\left(\pi\radiusvar{\inputlayer}{\layer{B}}\right)^2} 
   \sqrt{\frac{\pi\radiusvar{\inputlayer}{\layer{B}}}{2}}
   \sqrt{\frac{\pi\radiusvar{\inputlayer}{\layer{B}}}{2}}	                     \notag \\
 &=
   \frac{(\N{\inputlayer}{\layer{B}})^2}{2\pi\radiusvar{\inputlayer}{\layer{B}}}
   \exp{-\frac{\distance{}{}{}{}^2}{2\radiusvar{\inputlayer}{\layer{B}}}},
\end{align}
using the identity
$\int_{-\infty}^{\infty}\exp{-ax^2}=\sqrt{\nicefrac{\pi}{a}}$.

This result demonstrates that the number of shared connections between two neurons 
with Gaussian synaptic connection densities is itself a Gaussian function of the 
radial distance between the neurons with a variance that is half the value of 
the synaptic connection density radius. This means that a postsynaptic neuron 
is expected to have the most common connections with itself, for which 
$\distance{}{}{}{}=0$. Additionally, for small variance or connection radius, 
a postsynaptic neuron will share many connections with proximate neighbors, with the 
number of shared connections falling off quickly with distance. Since the expected 
number of synaptic inputs is constant, a large connection radius implies that the 
neuron will have shared connections with neurons comparatively distal to it, since 
nearby neurons will have comparatively fewer shared connections.

\else
	The expected number of shared presynaptic inputs between two postsynaptic neurons  
	in layer $\layer{B}$ can easily be shown to be (see \app{\ref{app:sharedInputs}}  
	for full derivation)
	\begin{align}\label{eq:sharedInputs}
	   \expect{\atDist{\Nshared{\layer{B}}{}}}
	 &=
	   \frac{ (\N{\inputlayer}{\layer{B}})^2 }{ 2\pi\radiusvar{\inputlayer}{\layer{B}} }
	   \exp{ -\frac{ (\distance{\postN}{\postNtwo}{\layer{B}}{\layer{B}})^2 }
		           {2\radiusvar{\inputlayer}{\layer{B}}}} \, ,
	\end{align}
	where  $\distance{\postN}{\postNtwo}{\layer{B}}{\layer{B}}$ 
	depicts the distance between neurons $\postN$ and $\postNtwo$ such that 
    $  \distance{\postN}{\postNtwo}{\layer{B}}{\layer{B}} 
	 = \sqrt{\posx{\preN}{\postN}^2 + \posy{\preN}{\postN}^2 } $, measured 
	 in grid spaces, $\gridspace{\inputlayer}$. 
\fi

An action potential takes time to propagate along an axon. It is generally accepted 
that, for myelinated axons, transmission time is linearly proportional to the distance 
propagated \citep{AslValTas17}. We introduce a model of transmission delay in 
which delay is a deterministic and linear function of the three-dimensional distance 
between the presynaptic and postsynaptic neurons, including both inter- and 
intra-lamina distances, and is inversely proportional to the speed of propagation. 
Given that axonal propagation delay typically dominates dendritic propagation
delay, we make an assumption that dendritic delay is approximately equal for all
cells and, therefore, that distant-dependent delay can be calculated from axonal 
propagation delay. 

Let the radial distance of a presynaptic neuron, $\preN$, in layer $\layer{A}$ 
to a postsynaptic neuron, $\postN$, in layer $\layer{B}$ be given by the magnitude 
of the $\icol{\posx{\preN}{\postN}}{\posy{\preN}{\postN}}$ vector and designated 
\distance{\preN}{\postN}{\inputlayer}{\layer{B}}. 
For Gaussian \posx{\preN}{\postN} and \posy{\preN}{\postN}, this 
distance is characterized by a Rayleigh distribution, so that if 
\radiusvar{\inputlayer}{\layer{B}} denotes the connection radius of the Gaussian 
connections, then the radial distance between neurons in the presynaptic layer 
$\layer{\inputlayer}$ connecting to postsynaptic neuron $\postN$ has distribution 
$\distance{\preN}{\postN}{\inputlayer}{\layer{B}}  
\sim \dist{Rayl}{0,\frac{\radiusvar{\inputlayer}{\layer{B}}}{2}}$. 
Furthermore, the probability of neuron $\preN$ making a synaptic connection to 
neuron $\postN$ from a radial distance 
$\distance{\preN}{\postN}{\inputlayer}{\layer{B}}$ is 
\begin{align}\label{eq:dist_Rayl}
       \probdist{Rayl}{\distance{\preN}{\postN}{\inputlayer}{\layer{B}}}
                      {\frac{\radiusvar{\inputlayer}{\layer{B}}}{2}} 
   \sim 
       \frac{2\distance{\preN}{\postN}{\inputlayer}{\layer{B}}}
            {\radiusvar{\inputlayer}{\layer{B}}}
       \exp{-\frac{\left(\distance{\preN}{\postN}{\inputlayer}{\layer{B}}
				         \right)^2}
			            {\radiusvar{\inputlayer}{\layer{B}}}}.
\end{align}

Denoting transmission delay between neuron $\preN$ in layer $\layer{\inputlayer}$ 
and neuron $\postN$ in layer $\layer{B}$ by 
$\delay{\preN}{\postN}{\inputlayer}{\layer{B}}$ and assuming an interlaminar  
distance of $\distlayer{\inputlayer}{\layer{B}}$, propagation delay from $\preN$  
to $\postN$ can be expressed as 
\begin{align}\label{eq:dist_3D}
   \delay{\preN}{\postN}{\inputlayer}{\layer{B}} 
 &= 
   \frac{ \left(\distlayerSqu{\inputlayer}{\layer{B}}
		      +(\distance{\preN}{\postN}{\inputlayer}{\layer{B}})^2
	      \right)^{\nicefrac{1}{2}} \gridspace{\inputlayer} }
        { \velocity } \, ,
\end{align}
where $\velocity$ is the spike propagation velocity and $\gridspace{\inputlayer}$ 
is the distance between neurons in the laminar in \si{\micro\meter}. The 
distance metrics, including $\distlayer{\inputlayer}{\layer{B}}$ and 
$\distance{\preN}{\postN}{\inputlayer}{\layer{B}}$, are measured in grid spaces, 
$\gridspace{\inputlayer}$, to separate out the effect of cell density from other 
distance measures such as connectivity radii and inter-laminar distance.

\subsection{Neuron model} \label{sec:neuronModel}
   
To simplify the analysis of a multi-layer neural network with propagation 
delay, we employ a Poisson neuron model. Poisson models  
are simple in that activity is a linear sum of inputs weighted by synaptic 
strength and that spike thresholds and reset are not modeled. Although it is 
a simple model, \citet{KemGerHem99} noted that large networks of 
integrate-and-fire neurons exhibit Poisson firing characteristics. 
The network is driven by spontaneous activity from layer $\layer{\inputlayer}$, 
modelled as a homogeneous Poisson process with rate, $\layerrate{\inputlayer}$.

\subsubsection{Linsker's neuron model} \label{sec:neuronModel_Linsker}
Using a Poisson model, 
the firing rate of the postsynaptic neurons can be described by
\begin{subequations}\label{eq:Linsker_rate}
\begin{align}
  \atTime{ \Lrate{\postN}{\layer{B}}{} }
     &= \Ra{\layer{B}} + \Rb{\layer{\inputlayer}}{\layer{B}}
	    \sum\limits_{\preN} \atTime{\Lweight{\inputlayer}{\layer{B}}{\preN}{\postN}}
		                    \atTime{\Lrate{\preN}{\inputlayer}{}} \, ,
        \label{eq:Linsker_rate_AB}  \\
  \atTime{ \Lrate{\postpostN}{\layer{C}}{} }
     &= \Ra{\layer{C}} + \Rb{\layer{\layer{B}}}{\layer{C}}
	    \sum\limits_{\postNtwo} 
		    \atTime{\Lweight{\layer{B}}{\layer{C}}{\postNtwo}{\postpostN}}
		    \atTime{\Lrate{\postNtwo}{\layer{B}}{}} \, ,
        \label{eq:Linsker_rate_BC}
\end{align}
\end{subequations}
where a neuron, $\preN$, in layer $\layer{\inputlayer}$ provides feedforward 
input to a neuron, $\postN$, in layer $\layer{B}$, and a neuron, $\postNtwo$, 
in layer $\layer{B}$ provides feed-forward input to a neuron, $\postpostN$, in 
layer $\layer{C}$, under an assumption of instantaneous propagation, 
$\Ra{\layer{B}}$ denotes background activity, and 
$\Lweight{\inputlayer}{\layer{B}}{\preN}{\postN}$ depicts synaptic strength 
between neurons $\preN{}$ and $\postN{}$. Note that this differs from 
\citet{Lin86a} by a dimensionless scale factor of $R_b^{{\inputlayer}{\layer{B}}}$ 
applied to the sum that we have absorbed into the weights. The calculation of 
the variance of neural activity in $\layer{B}$ is given in \app{\ref{app:var}}.

It is useful to introduce frequency domain definitions of the temporal signals. 
Spectral signals will, in general, be denoted by the upper-case counterpart to 
the lower-case symbol designating the temporal variable. For example, 
$\LRATE{\preN}{}$ denotes the frequency domain variable of neuronal rate, 
$\Lrate{\preN}{}{}$. From the definition of the Fourier transform 
and under the assumption of stationarity (i.e. slow weight evolution), the 
unfiltered rate can be expressed as a function of spectral variates using the 
inverse Fourier transform, 
\begin{align}\label{eq:rate_FFT}
   \atTime{\Lrate{\preN}{\layer{B}}{}}
 &=
   \frac{1}{\Tmax} \Sumk \atFreq{\LRATE{\preN}{\layer{B}}}
                      \exp{ 2\pi \ii \timesym \freqT},
   \tab
   \atFreq{\LRATE{\preN}{\layer{B}}{}}
  =
   \Sumt \atTime{\Lrate{\preN}{\layer{B}}{}}
		 \exp{-2\pi \ii \timesym \freqT},
\end{align}
where $\Sumk{}$ and $\Sumt$ are shorthand notations for $\Sum{\freq}{0}{\Tmax-1}$ 
and $\SumT{\Tmax}$ and $\freqT = \frac{\freq}{\Tmax}$, respectively, to aid 
readability.

\subsubsection{Network incorporating propagation delay and a general postsynaptic response}
\label{sec:neuronModel_DelPSP}

To model the temporal dynamics of a synapse's membrane potential, we introduce 
the \gls{psp}, $\psp{}$, which, for rate-based neuronal activity, this can be 
interpreted as spikes having a probability distribution over time. For example, 
the spike rates from \eq{\ref{eq:Linsker_rate}} are generalised to incorporate 
an arbitrary \gls{psp} by
\begin{subequations}{\label{eq:Linsker_ratePSP}}
\begin{align}
  \atTime{\Lrate{\postN}{\layer{B}}{\filt}}
     &= \Ra{\layer{B}} + \Rb{\inputlayer}{\layer{B}}
	    \sum\limits_{\preN} \Lweight{\layer{A}}{\layer{B}}{\preN}{\postN}
		    \atTime{ \left(\Lrate{\preN}{\inputlayer}{} \conv \psp \right)} \, ,
        \label{eq:rateB_PSP}  \\
  \atTime{\Lrate{\postpostN}{\layer{C}}{\filt}}
     &= \Ra{\layer{C}} + \Rb{\layer{B}}{\layer{C}}
	    \sum\limits_{\postN} \Lweight{\layer{A}}{\layer{C}}{\postN}{\postpostN}
		    \atTime{ \left(\Lrate{\postN}{\layer{B}}{} \conv \psp \right)} \, ,
        \label{eq:rateC_PSP}
\end{align}
\end{subequations}
where $\conv{}{}$ is the convolution operator and $\Lrate{\postN}{\layer{B}}{\filt}$ 
denotes firing rate that has been filtered by the \gls{psp}. The \gls{psp} is 
typically assumed to integrate to 1; i.e., $\int_0^{\infty} \psp(s)ds = 1$. 

Since convolution in the temporal domain is multiplication in the frequency 
domain, neural output that has been filtered by a \gls{psp} kernel, as defined 
in \eq{\ref{eq:Linsker_ratePSP}}, can be expressed using frequency variables by 
employing the Fourier transform from \eq{\ref{eq:rate_FFT}} to give 
\begin{align}\label{eq:rate_FFT_PSP}
	\atTime{ \Lrate{\postN}{\layer{B}}{\filt} }
 &=
   \frac{1}{\Tmax} \Sumk \atFreq{\LRATE{\postN}{\layer{B}}} \atFreq{\PSP}
                      \exp{2\pi \ii t {\freqT}},
\end{align}
where $\atFreq{\PSP}$ denotes the PSP in the frequency domain at frequency $\freq$.
Note that since the temporal signals are real, the imaginary frequency components 
of the PSP filter are conjugate symmetric around 0.

Relaxing the assumption of homogeneous propagation and denoting the rate for neuron 
$\postN{}$ that is a function of both the synapse response kernel and propagation 
delay by $\Lrate{\postN}{\layer{B}}{\del,\filt}$, the expressions for neuronal 
spiking rates in layers $\layer{B}$ and $\layer{C}$ as a function of input become
\begin{subequations}{\label{eq:Linsker_rateDelayPSP}}
\begin{align}
  \atTime{\Lrate{\postN}{\layer{B}}{\del,\filt}}
     &= \Ra{\layer{B}} + \Rb{\inputlayer}{\layer{B}}
	    \sum\limits_{\preN} \Lweight{\inputlayer}{\layer{B}}{\preN}{\postN}
		    \atDelayTime{\Lrate{\preN}{\inputlayer}{}}
		  			    {\delay{\preN}{\postN}{\inputlayer}{\layer{B}}} 
		    \conv \atTime{\psp} \, , 
        \label{eq:rateB_DelayPSP}  \\
  \atTime{\Lrate{\postpostN}{\layer{C}}{\del,\filt}}
     &= \Ra{\layer{C}} + \Rb{\layer{B}}{\layer{C}}
	    \sum\limits_{\postN} \Lweight{\layer{B}}{\layer{C}}{\postN}{\postpostN}
			\atDelayTime{\Lrate{\postN}{\layer{B}}{\del,\filt}}
						{\delay{\postN}{\postpostN}{\layer{B}}{\layer{C}}} 
			\conv \atTime{\psp} \, .
        \label{eq:rateC_DelayPSP}
\end{align}
\end{subequations}

For simplicity, we suppose that neurons in all layers have the same \gls{psp}. 
However, it is straightforward to incorporate layer-specific \gls{psp} functions. 

By the linearity of the Fourier transform, \eq{\ref{eq:rate_FFT}}, rate can 
be written as a sum of the input spectral variates. For unitary weights from layer 
$\layer{\inputlayer} \rightarrow \layer{B}$ and letting 
$\freqT=\frac{\freq}{\Tmax}$ for readability, the neuronal activity model that 
incorporates propagation delay in \eq{\ref{eq:Linsker_rateDelayPSP}} can be 
written as 
\begin{align}\label{eq:rate_FFT_DelayPSP}
   \Lrate{\postN}{\layer{B}}{\del,\filt} 
 &= 
   \frac{1}{\Tmax} 
   \Sumk \exp{2\pi \ii t\freqT}
         \sum\limits_{\preN} \Rb{\inputlayer}{\layer{B}} 
             \atFreq{\LRATE{\preN}{\inputlayer}} \atFreq{\PSP}
             \exp{-2\pi \ii \delay{\preN}{\postN}{}{} \freqT} \notag \\
\end{align}

\subsection{Neural learning in {L}insker's network}\label{sec:Linsker_plasticity}

  \subsubsection{Covariance of neural activity}
     
\citet{Lin86a} showed that spatial structure in synaptic connections between layers 
creates temporal covariance in neural activity between cells in layer $\layer{B}$ 
and onward. A Gaussian synaptic connection density ensures that layer $\layer{B}$ 
neurons situated nearer to each other in the lamina have more layer $\layer{\inputlayer}$
connections in common and, consequently, the neural activities of these neurons are more 
correlated. Layer $\layer{B}$ neurons are the presynaptic inputs to layer $\layer{C}$ 
neurons, also connected with a Gaussian density, parameterized by a unique 
connectivity radius, $\radiusvar{\layer{B}}{\layer{C}}$. The layer $\layer{B}$ neurons 
located radially closer to the postsynaptic neuron in layer $\layer{C}$ will have 
more nearby neighbors that are also connected to the postsynaptic layer $\layer{C}$ 
neuron and, being located near each other, will have more layer $\layer{\inputlayer}$ 
connections in common and so will be more temporally correlated. Thus, a spatial 
structure in temporal covariance is created. \citet{Lin86a} demonstrated that 
the resulting structure in temporal covariance can generate the emergence of
receptive field structure in deeper layers via Hebbian learning, showing the 
creation of spatial opponent cells in layer $\layer{C}$ and orientation selective 
cells in layers deeper in the hierarchy. 

Since learning is driven by the covariance of presynaptic neural activity 
\citep{KemGerHem99}, we derive an expression for covariance between layer $\layer{B}$ 
neurons, as given by \citet{Lin86a}. Initially, we consider learning in an ideal 
network, such as \citet{Lin86a} employed, in which there is uniform propagation 
delay across all neurons and an implied delta function for the postsynaptic 
response. 

Using the expression for the expected number of shared connections between layer 
$\layer{B}$ neurons in \eq{\ref{eq:sharedInputs}}, the covariance is given by 
(see \app{\ref{app:cov}}) 
\begin{align}\label{appEq:Linsker_cov}
   \scov{\Lrate{\postN}{\layer{B}}{}}{\Lrate{\postNtwo}{\layer{B}}{}}
 &= 
   \frac{(\Rb{\inputlayer}{\layer{B}} \N{\inputlayer}{\layer{B}})^2 \layerrateSq{\inputlayer}}
        {2\pi\radiusvar{\inputlayer}{\layer{B}}}
   \exp{-\frac{\distance{}{}{}{}^2}{2\radiusvar{\inputlayer}{\layer{B}}}} . 
\end{align}
This expression for covariance between layer \layer{B} neurons is pivotal in 
determining the final synaptic structure for a stable network \citep{Lin86a}.

\ifx\isEmbedded\undefined

We express covariance between two stationary, real variates, $x$ and $y$, as a 
function of their frequency domain counterparts, $X$ and $Y$.
\begin{align}\label{eq:xcorr}
   \xcorr{x}{y}{\tau} &= \frac{1}{\Tmax} \Sumt{\Tmax} \atTime{x} \atCorrTime{y}{\tau}.
\end{align}
Similar to the convolution theorem, the cross-correlation theorem states that 
\citep{Kum09} 
\begin{align}\label{eq:xcorr_FFT}
  \xcorr{x}{y}{\tau} = \frac{1}{\Tmax^2} \Sumk \conj{\atFreq{X}} \atFreq{Y} 
                                            \exp{2\pi \ii \tau \frac{\freq}{\Tmax}},
\end{align}
where $\atFreq{X}$ and $\atFreq{Y}$ denote the spectral variates of $x$ 
and $y$, respectively, at frequency $\freq$.

	Since instantaneous covariance is equivalent to the cross-correlation of 
    zero-meaned variates at constant lag, 
\else
	Since instantaneous covariance is equivalent to the cross-correlation of 
    zero-meaned variates at lag $0$ (see \app{\ref{app:covFreq}} for details), 
\fi
we can use \eq{\ref{eq:rate_FFT}} to express covariance between two layer $\layer{B}$ 
neurons as 
\begin{align}\label{eq:cov_FFT}
   \cov{\Lrate{\postN}{\layer{B}}{}}{\Lrate{\postNtwo}{\layer{B}}{}}
 &= 
   \frac{1}{\Tmax^2} \Sumk \conj{\atFreq{\LRATE{\postN}{\layer{B}}}}
                        \atFreq{\LRATE{\postNtwo}{\layer{B}}}
						\exp{2\pi \ii 0 \frac{\freq}{\Tmax}}	  \notag \\
 &= 
   \frac{1}{\Tmax^2} \Sumk \conj{\atFreq{\LRATE{\postN}{\layer{B}}}}
                        \atFreq{\LRATE{\postNtwo}{\layer{B}}}.
\end{align}

\ifx\isEmbedded\undefined
	
To derive the variance in neural activity for a layer $\layer{B}$ neuron, it is 
necessary to consider that a single presynaptic neuron, say neuron $\preN$, in 
layer $\layer{A}$ may have multiple synaptic connections to a neuron $\postN$ 
in layer $\layer{B}$ (see \fig{\ref{fig:network}}). Using the synaptic connection 
density in \eq{\ref{eq:2DGauss}}, we know that the expected number of synaptic 
connections between $\preN$ and $\postN$, separated by a distance of 
$[\posx{\preN}{\postN},\posy{\preN}{\postN}]$, is 
\begin{align}\label{eq:repeated_conns}
  \Nprepost{\preN}{\postN}{\layer{A}}{\layer{B}}
 &= 
  \frac{\N{\inputlayer}{\layer{B}}}{\pi\radiusvar{\inputlayer}{\layer{B}}} 
  \exp{-\frac{\posx{\preN}{\postN}^2 
			 +\posy{\preN}{\postN}^2}
	         {\radiusvar{\inputlayer}{\layer{B}}}}.
\end{align}
From \eq{\ref{eq:Linsker_rate_AB}}, the expected variance of a neuron in layer 
\layer{B} can be written as, 
\begin{align}\label{eq:rate_var}
    \var{\ensemble{\Lrate{\postN}{\layer{B}}{}}}
 &= \var{\ensemble{
	       \Ra{\layer{B}} + 
	       \sum\limits_{\preN}  
	           \Rb{\inputlayer}{\layer{B}} 
			   \Lweight{\layer{B}}{\layer{C}}{\preN}{\postN} 
			   \Lrate{\preN}{\layer{A}}{} } }				  \notag \\
 &=  \RbSq{\inputlayer}{\layer{B}} \N{\inputlayer}{\layer{B}} \layerrate{\layer{A}}
  + 2\RbSq{\inputlayer}{\layer{B}}
     \sum\limits_{\preN} \sum\limits_{\preNtwo < \preN}
		 \ensemble{\cov{\Lrate{\preN}{\layer{A}}{}} {\Lrate{\preNtwo}{\layer{A}}{}}},
\end{align}
where we have assumed that weights between layers \layer{A} and \layer{B} have 
evolved to the upper bound of $1$ (cf. \citet{Lin86a}) and used the fact that
$  \var{ \sum\limits_{i=1}^{n} a_i x_i } 
 = \sum\limits_{i=1}^n a_i^2 \var{x_i} 
 + 2 \sum\limits_j \sum\limits_{i \leq n} a_i a_j \cov{x_i}{x_j} $. 

To determine the contribution of the covariance term in \eq{\ref{eq:rate_var}}, 
we note that each pair of synapses originating from the same presynaptic neuron will 
be fully correlated, while synapses stemming from different presynaptic neurons 
in layer \layer{A} will be uncorrelated. For neuron $\preN{}$ in layer $\layer{A}$ 
with \Nprepost{\preN}{\postN}{\layer{A}}{\layer{B}} synaptic connections to neuron 
\postN{} in layer \layer{B}, there will be 
${\Nprepost{\preN}{\postN}{\layer{A}}{\layer{B}} \choose 2}
= \Nprepost{\preN}{\postN}{\layer{A}}{\layer{B}}
  \left( \Nprepost{\preN}{\postN}{\layer{A}}{\layer{B}} - 1 \right)/2 
\approx  \left(\Nprepost{\preN}{\postN}{\layer{A}}{\layer{B}}\right)^2/2$ 
synapse pairs contributing to the covariance sum. Consequently, 
the total contribution of neuron \preN{} to the covariance term in 
\eq{\ref{eq:rate_var}} will be 
$\frac{(\Nprepost{\preN}{\postN}{\layer{A}}{\layer{B}})^2}{2}
\layerrate{\layer{A}}$.
To obtain an expression for the variance of layer \layer{B} neurons, we evaluate  
this expression using \eq{\ref{eq:repeated_conns}} and substitute it into 
\eq{\ref{eq:rate_var}}. Applying a continuous spatial approximation to the layer, 
we can integrate to get the total expected contribution of all presynaptic neurons 
to the covariance term and evaluate \eq{\ref{eq:rate_var}}, 
\begin{align}\label{eq:rate_varB}
    \var{\ensemble{\Lrate{\postN}{\layer{B}}{}}}
 &= \RbSq{\inputlayer}{\layer{B}} \N{\inputlayer}{\layer{B}} \layerrate{\layer{A}}
  + \RbSq{\inputlayer}{\layer{B}} \layerrate{\layer{A}}
   {\mathlarger{\iint\limits_{\posx{}{}\posy{}{}}}}
         \Nprepost{\preN}{\postN}{}{}^2 d\posx{}{}d\posy{}{}           \notag \\
 &= \RbSq{\inputlayer}{\layer{B}} \N{\inputlayer}{\layer{B}} \layerrate{\layer{A}}
  + \RbSq{\inputlayer}{\layer{B}} \layerrate{\layer{A}}
   {\mathlarger{\iint\limits_{\posx{}{}\posy{}{}}}}
	     \left( \frac{\N{\inputlayer}{\layer{B}}}{\pi\radiusvar{\inputlayer}{\layer{B}}} \right)^2 
	     \exp{- 2\frac{\posx{}{}^2 + \posy{}{}^2}
			          {\radiusvar{\inputlayer}{\layer{B}}} } d\posx{}{}d\posy{}{} \notag \\
 &= \RbSq{\inputlayer}{\layer{B}} \layerrate{\layer{A}}
    \left( \N{\inputlayer}{\layer{B}} 
		 + \frac{ (\N{\inputlayer}{\layer{B}})^2 } {2\pi\radiusvar{\inputlayer}{\layer{B}}} \right). 
\end{align}

\else
\fi

In the following, when considering neural learning of synapses between layers 
$\layer{B}$ and $\layer{C}$, it is assumed that synapses between layers 
$\layer{\inputlayer}$ and $\layer{B}$ have already evolved to maturity such that 
all synapses are excitatory with unity weight (see \citet{Lin86a} for conditions 
required for this to occur). Therefore, we will consider the plasticity of 
$\layer{B}$ to $\layer{C}$ connections in this work.

  \subsubsection{Learning equation}\label{sec:Linsker_learningEqn}
     
Synaptic plasticity occurs adiabatically when compared with neuronal dynamics as 
it is the result of many incremental changes in synapse strength. Furthermore, 
each neuron within a layer has identical spatial connectivity and firing rate 
statistics. Consequently, the system can be considered ergodic. This is significant 
because it enables statistics, such as covariance in neuronal firing rate, to be 
evaluated either by taking an ensemble average or a temporal average over a 
trial or epoch. 

\citet{Lin86a} assumed that a synapse between two neurons was potentiated if the 
product of the pre- and post-synaptic neuronal activity was above average, 
indicating that their activity positively covaried, and vice-versa. 
We can take advantage of the slow change in synaptic weights and the ergodicity 
of the system by taking an ensemble average of weights across the lamina. 
The learning equations for synaptic weights can then be expressed as first-order 
linear differential equations \citep{Lin86a}, 
\begin{subequations}\label{eq:learningRule_Linsker}
\begin{align}
    \Lwchange{\inputlayer}{\layer{B}}{\preN}{\postN} 
  \triangleq
    \etalayer{} \ensemble{\left( \Delta \Lweight{\inputlayer}{\layer{B}}{\preN}{\postN} 
                          \right)}
 &= 
    \etalayer{} \left( 
		\kone{\inputlayer}{\layer{B}} 
	  + \frac{1} {\N{\inputlayer}{\layer{B}}}
		\sum \Lweight{\inputlayer}{\layer{B}}{\preNtwo}{\postN}
			 \left( \lcov{ \Lrate{\preNtwo}{\inputlayer}{} }{ \Lrate{\preN}{\inputlayer}{} }
		   + \ktwo{\inputlayer}{\layer{B}} \right)
	\right) , 
 \verylargespace 
       \LwgtMin{\inputlayer}{\layer{B}} 
  \leq \Lweight{\inputlayer}{\layer{B}}{\preN}{\postN} 
  \leq \LwgtMax{\inputlayer}{\layer{B}}, 
 \label{eq:learningRuleAB_Linsker} \\
    \Lwchange{\layer{B}}{\layer{C}}{\postN}{\postpostN} 
  \triangleq
    \etalayer{} \ensemble{\left( \Delta \Lweight{\layer{B}}{\layer{C}}{\postN}{\postpostN} 
                          \right)}
 &= 
    \etalayer{} \left( 
		\kone{\layer{B}}{\layer{C}} 
	  + \frac{1}{\N{\layer{B}}{\layer{C}}}
		\sum \Lweight{\layer{B}}{\layer{C}}{\postNtwo}{\postpostN}
			 \left( \lcov{ \Lrate{\postNtwo}{\layer{B}}{} }{ \Lrate{\postN}{\layer{B}}{} }
		          + \ktwo{\layer{B}}{\layer{C}} 
		     \right)
	\right) , 
 \verylargespace 
       \LwgtMin{\layer{B}}{\layer{C}} 
  \leq \Lweight{\layer{B}}{\layer{C}}{\postN}{\postpostN} 
  \leq \LwgtMax{\layer{B}}{\layer{C}}, 
 \label{eq:learningRuleBC_Linsker}
\end{align}
\end{subequations}
where $\etalayer{} \ll 1$ is the learning rate, chosen to be sufficiently small 
such that the weights are quasi-constant on the timescale of neuronal dynamics, 
$\Lweight{\inputlayer}{\layer{B}}{\preNtwo}{\postN}$ depicts the weight of the 
synapse connecting presynaptic neuron $\preNtwo$ in layer $\inputlayer$ to 
postsynaptic neuron $\postN$ in layer $\layer{B}$, and, similarly, 
$\Lweight{\layer{B}}{\layer{C}}{\postNtwo}{\postpostN}$ depicts the weight of 
the synapse connecting presynaptic neuron $\postNtwo$ in layer $\layer{B}$ 
to postsynaptic neuron $\postpostN$ in layer $\layer{C}$. The parameters 
$\kone{\inputlayer}{\layer{B}}$, $\ktwo{\inputlayer}{\layer{B}}$, and 
$\kone{\layer{B}}{\layer{C}}$, $\ktwo{\layer{B}}{\layer{C}}$ are learning constants 
that are homogeneous across all synapses connecting two neural populations. 
$\lcov{\Lrate{\preNtwo}{\inputlayer}{} }{\Lrate{\preN}{\inputlayer}{} }$ and 
$\lcov{\Lrate{\postNtwo}{\layer{B}}{} }{\Lrate{\postN}{\layer{B}}{} }$ denote the 
expected covariance in neural activity between neurons $\preNtwo$ and $\preN$ 
in layer $\inputlayer$ and neurons $\postN$ and $\postNtwo$ in layer $\layer{B}$, 
respectively. The definition for each follows the same structure that, for layer 
$\layer{B}$ covariance as an example, is 
$  \lcov{ \Lrate{\postNtwo}{\layer{B}}{} }{ \Lrate{\postN}{\layer{B}}{} } 
 = \fo{-} 
   \ensemble{\Lrate{\postN}{\layer{B}}{}    - \meanrate{\layer{B}}}
   \ensemble{\Lrate{\postNtwo}{\layer{B}}{} - \meanrate{\layer{B}}}$, where  
\ensemble{} denotes an ensemble average, $\Lrate{\postN}{\layer{B}}{}$ is the rate 
of activity of neuron $\postN$ in layer $\layer{\layer{B}}$, $\meanrate{\layer{B}}$ is 
the temporal average of layer $\layer{\layer{B}}$ spiking rates in an ergodic system, 
and $\fo{}$ is a scaling factor to normalize the covariance matrix, $\Lcov{}{}{}$. 
From the expected value of covariance in \eq{\ref{appEq:Linsker_cov}}, it can be 
seen that to normalize the covariance requires that 
\begin{align}\label{eq:cov_temp2ensemble}
   \lcov{\Lrate{\postNtwo}{\layer{B}}{} }{\Lrate{\postN}{\layer{B}}{} }   
 &= 
   \fo{-} 
   \scov{\Lrate{\postN}{\layer{B}}{}}{\Lrate{\postNtwo}{\layer{B}}{}}  \notag \\
 &= 
   \exp{-\frac{\distance{}{}{}{}^2}{2\radiusvar{\inputlayer}{\layer{B}}}} 
   \in \left[0, 1\right] ,
\end{align}
such that 
\begin{align}\label{eq:f0_normCov}
   \fo{} 
 &= 
   \frac{2\pi\radiusvar{\inputlayer}{\layer{B}}}
        {(\Rb{\inputlayer}{\layer{B}} \N{\inputlayer}{\layer{B}})^2 
		  \layerrateSq{\inputlayer}}. 
\end{align}

Inspection of the weight 
\eqs{\ref{eq:learningRuleAB_Linsker}}{\ref{eq:learningRuleBC_Linsker}} shows that 
$\kone{}{}$ and $\ktwo{}{}$ do not depend on input and, therefore, do not drive
learning, but rather regulate overall activity and determine homeostasis. Without 
the dependence of the learning equation on input covariance, all weights would 
evolve to either the upper or lower bound because no presynaptic neuron would 
be more competitive than another and thus all neurons would change equally. 

The weight bounds, $\LwgtMin{}{} \leq \Lweight{}{}{}{} \leq \LwgtMax{}{}$, 
are determined by the proportion of excitatory to inhibitory neurons \citep{Lin86a}. 
\citet{Lin86a} showed that, for a linear network such as this, having a fraction,  
$\LwgtMax{}{}$, of excitatory synapses with limits of $0$ and $1$, 
and a fraction, $\LwgtMin{}{} = 1 - \LwgtMax{}{}$, 
of inhibitory synapses with limits of $-1$ and $0$, is equivalent to all synapses 
having limits of $\LwgtMin{}{}$ and $\LwgtMax{}{}$. 

In this work, the term `receptive field' refers to an inter-layer 
receptive field; i.e., rather than being defined by stimulus space, it is 
defined by a cortical neuron's input synaptic weight structure. 
The receptive field learned by a neuron is described by the set of synaptic weights
from its presynaptic neurons after the system has converged to equilibrium. For neuron 
$\postN$ in layer $\layer{B}$, the set of presynaptic neurons can be described by 
$\{\preN \mid \preN \text{ has a synaptic connection to }\postN \}$. Denote by 
$\FP{(\Lweight{\inputlayer}{\layer{B}}{\preN}{\postN})}$ the fixed point for 
the synapse weight from neuron $\preN$ in layer $\inputlayer$ to neuron $\postN$ 
in layer $\layer{B}$. The fixed point is assumed to be reached once all synapse 
weights, or all but one, are no longer changing \citep{Lin86a}. The set of fixed 
point weights describe the receptive field learned by neuron $\postN$. If a synapse 
has a weight of $0$ then it is not considered to be part of the receptive field.

As per \citet{Lin86a}, the parameters for the layer $\inputlayer$ to $\layer{B}$ 
connection are chosen such that the weights are unstable and all, or all but one, 
reach the upper bound, $\LwgtMax{\inputlayer}{\layer{B}}$. In this way, correlations 
in the neural activity of layer $\layer{B}$ cells emerge from spatial structure 
in the presynaptic input from layer $\inputlayer$, rather than structure in 
the connection weights.


\section{Results}\label{sec:analytical_results}
   
We derive an analytical expression for neural activity of layer $\layer{B}$
neurons in the presence of propagation delay. To facilitate the use of this
expression in analytical neural networks, we determine a simplified form for 
the impact of propagation delay. An expression for covariance between layer 
$\layer{B}$ neurons is calculated for the case in which an arbitrary \gls{psp} 
is included in the neuron model, \eq{\ref{eq:Linsker_ratePSP}}, and then for 
the case in which a \gls{psp} and three-dimensional propagation delay is 
incorporated into the neuron model, \eq{\ref{eq:Linsker_rateDelayPSP}}. Finally, 
since we assume that the plasticity mechanism is initiated near the site of a 
layer $\layer{C}$ cell body, covariance is calculated for the neural activity 
of two presynaptic layer $\layer{B}$ cells at the time this activity is received 
by the postsynaptic neuron in layer $\layer{C}$.

Using the expression for covariance when employing a generalized neuron model 
that incorporates both propagation delay and arbitrary \gls{psp}, a learning 
equation for synaptic weights is specified. We calculate the expected size of 
a layer $\layer{C}$ neuron's receptive field when \citepos{Lin86a} neuron model 
is employed, \eq{\ref{eq:Linsker_rate}}; i.e., when there is an implicit delta 
function model of a \gls{psp} and inter-lamina delay is implicitly assumed to 
dominate propagation delay sufficiently such that radial propagation delay is 
negligible. 

Using the expressions derived for covariance between layer $\layer{B}$ neurons 
when activity is filtered by an arbitrary \gls{psp}, we calculate receptive 
field size for a layer $\layer{C}$ neuron. Finally, we analytically determine 
the receptive field size of a layer $\layer{C}$ cell in the presence of 
three-dimensional propagation delay.

   \subsection{Neural firing rate with propagation delay}\label{sec:DelPSP_rate} 
     
Our first result is to analytically derive the output spiking rate of neurons 
in layer $\layer{B}$ in the presence of propagation delay. It is known that 
neurons have an upper bound on the temporal resolution of the neural code 
carried by the input spikes. The aim of this section is to explore the 
impact of propagation delay on this upper bound. 

For a layer $\layer{B}$ neuron that receives input spikes delayed according to 
the spatial layout of its presynaptic neurons and convolved with a \gls{psp} 
kernel (\eq{\ref{eq:rate_FFT_DelayPSP}}), the output spiking rate can be 
expressed via frequency variables as
\begin{align}\label{eq:rate_FFT_DelayPSP_layerB}
   \Lrate{\postN}{\layer{B}}{\del,\filt} 
 &= 
   \frac{1}{\Tmax} 
   \Sumk \exp{2\pi \ii t\freqT}
         \sum\limits_{\preN} \Rb{\inputlayer}{\layer{B}} 
             \atFreq{\LRATE{\preN}{\inputlayer}} \atFreq{\PSP}
             \exp{-2\pi \ii \delay{\preN}{\postN}{}{} \freqT} \notag \\
 &=
   \frac{1}{\Tmax} 
   \Sumk \exp{2\pi \ii t\freqT}
         \expect{ \exp{-2\pi \ii \delay{\preN}{\postN}{}{} {\freqT}} }
		 \atFreq{ \PSP }
         \sum\limits_{\preN} \Rb{\inputlayer}{\layer{B}} 
             \atFreq{\LRATE{\preN}{\inputlayer}}			  \notag \\
 &=
   \frac{1}{\Tmax} 
   \Sumk \exp{2\pi \ii t\freqT}
         \expect{ \exp{-2\pi \ii \delay{\preN}{\postN}{}{} {\freqT}} }
		 \atFreq{ \PSP }
		 \atFreq{\LRATE{\postN}{\layer{B}}} .
\end{align}
Note that all layer $\layer{A}$ spectral variates are multiplied by the same 
PSP filter and thus it can come out of the inner sum, leaving a sum of layer 
$\layer{A}$ input spectral variates, which gives the same layer $\layer{B}$ 
spectral variate as for the no \gls{psp} and no delay case. Therefore, it 
remains to determine the impact of the expected value of delay in the frequency 
domain to understand its impact on the output rate of a layer $\layer{B}$ neuron. 

Derivation of the expected value of delay in the frequency domain is given in 
\app{\ref{app:delayDerivation}}. The derivation integrates the Fourier transform 
of propagation delay, 
$\exp{-2\pi \ii \delay{\preN}{\postN}{}{} {\freqT}}$, 
between each presynaptic neuron $\preN$ and the given postsynaptic neuron 
$\postN$ over the presynaptic layer, weighted by the probability of the 
presynaptic neuron having a connection to the postsynaptic neuron, 
\eq{\ref{eq:dist_Rayl}}. Note that delay is determined from the 
three-dimensional distance between the pre- and post-synaptic neurons, comprised 
of the radial distance, $\distance{\preN}{\postN}{\layer{A}}{\layer{B}}$, 
the interlaminar distance, $\distlayer{\layer{A}}{\layer{B}}$, and the speed 
of propagation, $\velocity$, \eq{\ref{eq:dist_3D}} (see \fig{\ref{fig:network}} 
for schematic). Given that neurons that are the same distance from the postsynaptic 
neuron will have the same propagation delay and the same probability of connection, 
the integration is done in concentric rings of increasing distance from the 
postsynaptic neuron. Since the integral is over a complex domain, it is solved 
using contour integration, 

\begin{align}\label{eq:avgDelayAB}
	\atFreqAndRadius{\delaymean}{\inputlayer}{\layer{B}}
 =& \,  
	\expect{\exp{-2\pi \ii \del {\freqT}}}    \notag \\
 =& \,   
	\exp{-2\pi \ii {\freqT} 
		   \frac{\distlayer{\inputlayer}{\layer{B}} \gridspace{} }
				{\velocity} }
  - \pi^{\nicefrac{3}{2}} {\freqT} 
	\frac{\radius{\inputlayer}{\layer{B}} \gridspace{} }{\velocity}
	\exp{\frac{\distlayerSqu{\inputlayer}{\layer{B}}}
			  {\radiusvar{\inputlayer}{\layer{B}}}} 
	\exp{ -\left( \pi {\freqT} 
		   \frac{\radius{\inputlayer}{\layer{B}} \gridspace{} }
				{\velocity} \right)^2}					   \notag \\
  & 
	\Bigg( \ii\erfc{\frac{\distlayer{\inputlayer}{\layer{B}}}
						 {\radius{\inputlayer}{\layer{B}}}}  
		 + \erf{ \frac{\distlayer{\inputlayer}{\layer{B}}}
					  {\radius{\inputlayer}{\layer{B}}}
			   + \ii\pi {\freqT} 
				 \frac{ \radius{\inputlayer}{\layer{B}} \gridspace{} }
					  { \velocity } } 
		 - \erf{ \frac{\distlayer{\inputlayer}{\layer{B}}}
					  {\radius{\inputlayer}{\layer{B}}}} 
	\Bigg).
\end{align}

The expression for the mean value of delay in the frequency domain is essentially 
a function of two parameters,
\begin{align}\label{eq:layerDelay}
    \taul{\inputlayer}{\layer{B}} 
  &= 
    \frac{ \distlayer{\inputlayer}{\layer{B}} \gridspace{} }{ \velocity }
	( \units{\taul{\inputlayer}{\layer{B}}}{s} ), \qquad \qquad
    \taur{\inputlayer}{\layer{B} \gridspace{} } 
   = 
    \frac{ \radius{\inputlayer}{\layer{B}} \gridspace{} }{ \velocity }
	( \units{\taur{\inputlayer}{\layer{B}}}{s} ),
\end{align}
where $\taul{\inputlayer}{\layer{B}}$ depicts the propagation time between layers 
and $\taur{\inputlayer}{\layer{B}}$ represents the radial propagation time within 
the lamina. Consequently, mean delay in frequency can be written  
\begin{align}\label{eq:avgDelayAB_reducedDim}
   \atFreqAndRadius{\delaymean}{\inputlayer}{\layer{B}}
 &=  
   \exp{-2\pi \ii {\freqT} \taul{\inputlayer}{\layer{B}} }
  -
	\pi^{\nicefrac{3}{2}} {\freqT} \taur{\inputlayer}{\layer{B}} 
	\exp{ \left( \frac{\taul{\inputlayer}{\layer{B}}}{\taur{\inputlayer}{\layer{B}}} \right)^2 } 
	\exp{ -\left( \pi {\freqT} \taur{\inputlayer}{\layer{B}} \right)^2}
	\left( \ii\erfc{ \frac{\taul{\inputlayer}{\layer{B}}}{\taur{\inputlayer}{\layer{B}}} }  
		 + \erf{ \frac{\taul{\inputlayer}{\layer{B}}}{\taur{\inputlayer}{\layer{B}}} 
		 + \ii\pi {\freqT} \taur{\inputlayer}{\layer{B}} } 
		 - \erf{ \frac{\taul{\inputlayer}{\layer{B}}}{\taur{\inputlayer}{\layer{B}}} } \right).
\end{align}

\begin{figure}[!htb]
\centering
	\includegraphics[width=0.99\textwidth]{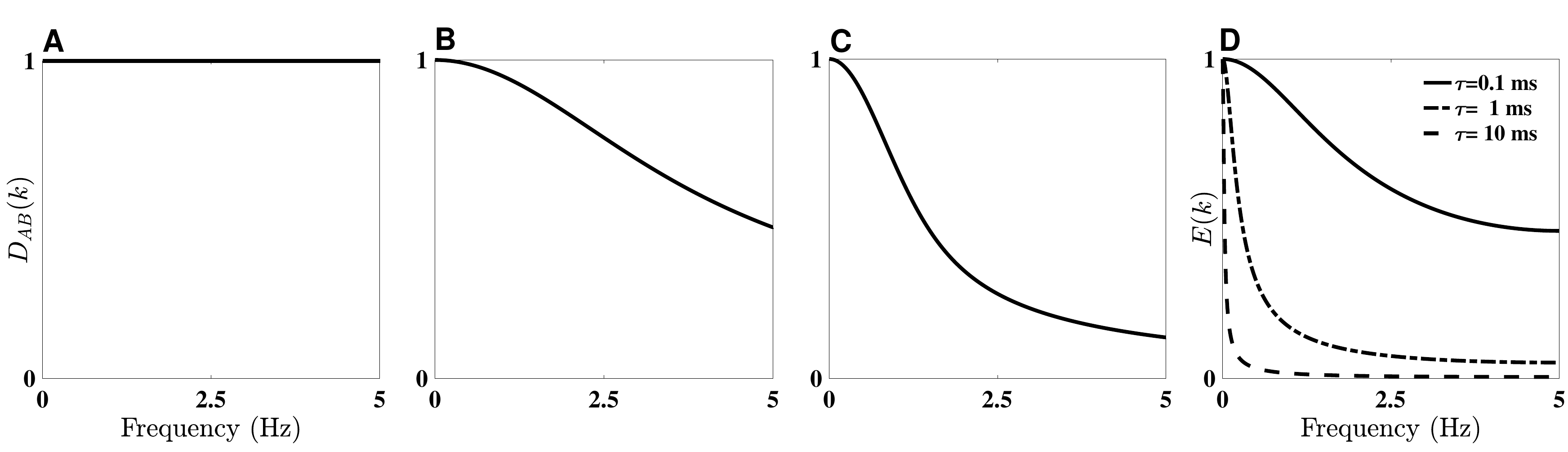}
	\caption{\figpanel{A-C} Magnitude of the expected value of delay 
		\eq{\ref{eq:avgDelayAB_reducedDim}}, in the frequency domain for different 
		radial propagation delays, $\taur{\inputlayer}{\layer{B}}$. The x-axis shows 
		frequencies, $\freqT$, up to the Nyquist frequency, from $0$ to $5$ 
		\si{\kilo\hertz}. The y-axis is 
		$\atFreqAndRadius{\delaymean}{\inputlayer}{\layer{B}} \in [0, 1]$. The 
		inter-layer propagation delay was $\taul{\inputlayer}{\layer{B}} = 500$ 
		\si{\micro\second}, derived from velocity, $\velocity = 2$ \si{\meter\second}, 
		and neural density, $\gridspace{} = 10$ \si{\micro\meter}. The radial propagation 
		delay was 
		\figpanel{A} $\taur{\inputlayer}{\layer{B}} =  50$\si{\micro\second}, 
		\figpanel{B} $\taur{\inputlayer}{\layer{B}} = 250$\si{\micro\second}, and
		\figpanel{C} $\taur{\inputlayer}{\layer{B}} = 500$\si{\micro\second}. 
		\figpanel{D} The magnitude of an example \gls{psp} in the frequency 
		domain, normalized between $0$ and $1$, to enable comparison with frequencies 
		retained in rate-based neural activity after incorporating the impact 
		of propagation delay. In this example, the \gls{psp} is modelled with a 
		single parameter, the decay time, $\tau$, such that 
		$\psp \propto \exp{- \frac{t}{\tau}}$, so that in the frequency domain 
		$\PSP \propto \frac{ \tau }{ 1 + \ii \freqT \tau }$, and normalized to $1$. 
		\label{fig:delCoeffAB_velocity}}
\end{figure}

\fig{\ref{fig:delCoeffAB_velocity}} shows the mean delay as a function of different 
radial propagation velocities, which is determined by the connection radius, the 
neuron density, and spike propagation velocity, as detailed in 
\eq{\ref{eq:layerDelay}}. For comparatively small radial propagation delays 
relative to inter-layer propagation 
delay (\fig{\ref{fig:delCoeffAB_velocity}}A), the delay of all spikes from the 
presynaptic layer to a postsynaptic neuron is approximately equal and the impact 
of delay is negligible. As the radial propagation delay increases relative to the 
inter-layer propagation delay, the propagation time from presynaptic neurons to a 
postsynaptic neuron can no longer be assumed approximately equal and the spread 
in arrival times to the postsynaptic neuron effectively smooths the input to  
layer $\layer{B}$ postsynaptic neurons (\fig{\ref{fig:delCoeffAB_velocity}}B,C). 
Example frequency profiles for the \gls{psp} are shown in 
\fig{\ref{fig:delCoeffAB_velocity}}D, in which the \gls{psp} is modelled as a 
single exponential function with the decay parameter chosen to reflect the effective 
membrane time constant, which is shorter than the membrane time constant due 
to the neuron becoming more leaky as more synaptic channels open 
\citep{BurMefGra03}. For \glspl{psp} with long decay times, its Fourier 
transform contains only low frequencies and, for inter-lamina distances that 
dominate radial distances, the  propagation delay between layers dominates 
the radial propagation delay sufficiently such that the mean delay in frequency, 
$\atFreqAndRadius{\delaymean}{\inputlayer}{\layer{B}} \in [0, 1]$, 
attenuates only the very highest frequencies. Consequently, propagation delay 
does not destroy frequencies that are present in the neuronal signal. 

The most important factor in determining the impact of propagation delay on
neural firing is the ratio of inter-lamina delay to the radial propagation
time. 
Application of the expected value of delay in the frequency domain, 
\eq{\ref{eq:avgDelayAB_reducedDim}}, to the derivation of neural response in the
presence of distance-dependent delay gives 
\begin{align}\label{eq:rate_FFT_meanDelayPSP_layerB}
   \Lrate{\postN}{\layer{B}}{\del,\filt} 
 &=
   \frac{1}{\Tmax} 
   \Sumk \exp{2\pi \ii t\freqT}
         \atFreqAndRadius{\delaymean}{\inputlayer}{\layer{B}}
		 \atFreq{ \PSP }
		 \atFreq{\LRATE{\postN}{\layer{B}}} . 
\end{align}

\begin{figure}[!htb]
\centering
	\includegraphics[width=0.99\textwidth]{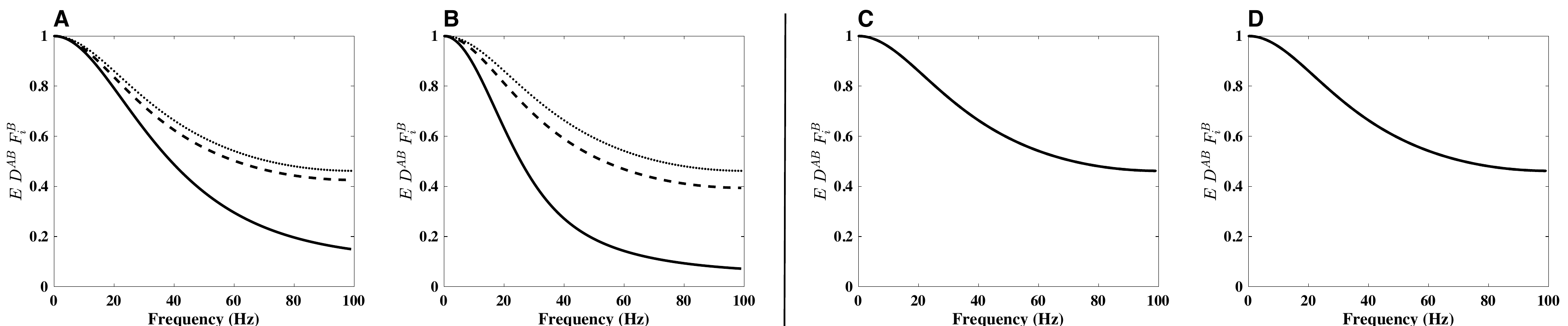}
	\caption{Frequency response of spike rate for any postsynaptic neuron  
		$\postN$ in layer $\layer{B}$, normalised to 1, showing the impact of 
		distance dependent propagation delays, as derived analytically in 
		\eq{\ref{eq:rate_FFT_meanDelayPSP_layerB}}. Given that white noise is 
		input to the neurons, the normalised values of the frequency
		response assuming no or equal delay, 
		$\atFreq{\LRATE{\postNtwo}{\layer{B}}}$, for any postsynaptic neuron 
		$\postNtwo$ will be $1$. Therefore, only the distance-dependent 
		propagation delay, given by 
		$\atFreqAndRadius{\delaymean}{\inputlayer}{\layer{B}}$ for frequency
		$\freq$, and the postsynaptic potential, $\atFreq{\PSP}$, impact the 
		frequency spectra.
		In all figures, the solid line denotes neural response when the time
		constant describing propagation delay between layers, 
		$\taul{\inputlayer}{\layer{B}}$, is equal to the time constant of radial 
		propagation delay, $\taur{\inputlayer}{\layer{B}}$
		(\eq{\ref{eq:layerDelay}}). The dashed line shows the case where the 
		inter-layer delay is ten times larger than the radial propagation delay. 
		The dotted line shows the frequency response of the postsynaptic potential 
		function, $\atFreq{\PSP}$.
		\figpanel{A,B} Frequency responses of neurons assuming a spike velocity of 
		$\velocity = 10\textrm{ms}^{-1}$.
		\figpanel{C,D} Frequency responses of neurons assuming a spike velocity of 
		$\velocity = 100\textrm{ms}^{-1}$.
		\figpanel{A} Frequency responses of neurons for cell density of $7,000$ 
		cells/mm$^2$, an exponential postsynaptic potential, and velocity of 
		$\velocity = 10\textrm{ms}^{-1}$.
		\figpanel{B} Frequency responses of neurons for cell density of $1,000$ 
		cells/mm$^2$, an exponential postsynaptic potential, and velocity of 
		$\velocity = 10\textrm{ms}^{-1}$.
		\figpanel{C} Frequency responses of neurons for cell density of $7,000$ 
		cells/mm$^2$, an exponential postsynaptic potential, and velocity of 
		$\velocity = 100\textrm{ms}^{-1}$. Not all lines are visible as they
		coincide. 
		\figpanel{D} Frequency responses of neurons for cell density of $1,000$ 
		cells/mm$^2$, an exponential postsynaptic potential, and velocity of 
		$\velocity = 100\textrm{ms}^{-1}$. Not all lines are visible as they
		coincide. 
		\label{fig:spikeRateFreq}}
\end{figure}

\fig{\ref{fig:spikeRateFreq}} shows the frequency response for neural spiking 
rate in layer $\layer{B}$, derived in \eq{\ref{eq:rate_FFT_meanDelayPSP_layerB}}, 
in response to white noise input from layer $\layer{A}$ and assuming a \gls{psp} 
modelled as a decaying exponential. The figure shows the mitigating impact of 
increasing propagation velocity on retaining power across all frequencies, as 
expected. Also intuitive is the observation that increasing the distance between 
layers relative to the radial connectivity distance (i.e., increasing the ratio 
of $\taul{}{}$ to $\taur{}{}$) reduces the attenuation of high frequencies, though 
this has a smaller impact than spike propagation velocity. In 
\fig{\ref{fig:spikeRateFreq}}A and B, propagation delay dominates the frequency 
response of the layer $\layer{B}$ neurons, while for \fig{\ref{fig:spikeRateFreq}}C 
and D, the postsynaptic potential dominates the frequency response. That is, 
whichever filters out more of the incoming signal -- propagation delay or the 
postsynaptic potential -- dominates the frequency response of the postsynaptic 
neuron.

   \subsection{Approximating the impact of propagation delay}\label{sec:approxDelay}
	 
\begin{figure}[htb] \centering
  \begin{subfigure}[t]{0.49\textwidth}
      \centering
      \includegraphics[width=0.9\textwidth]{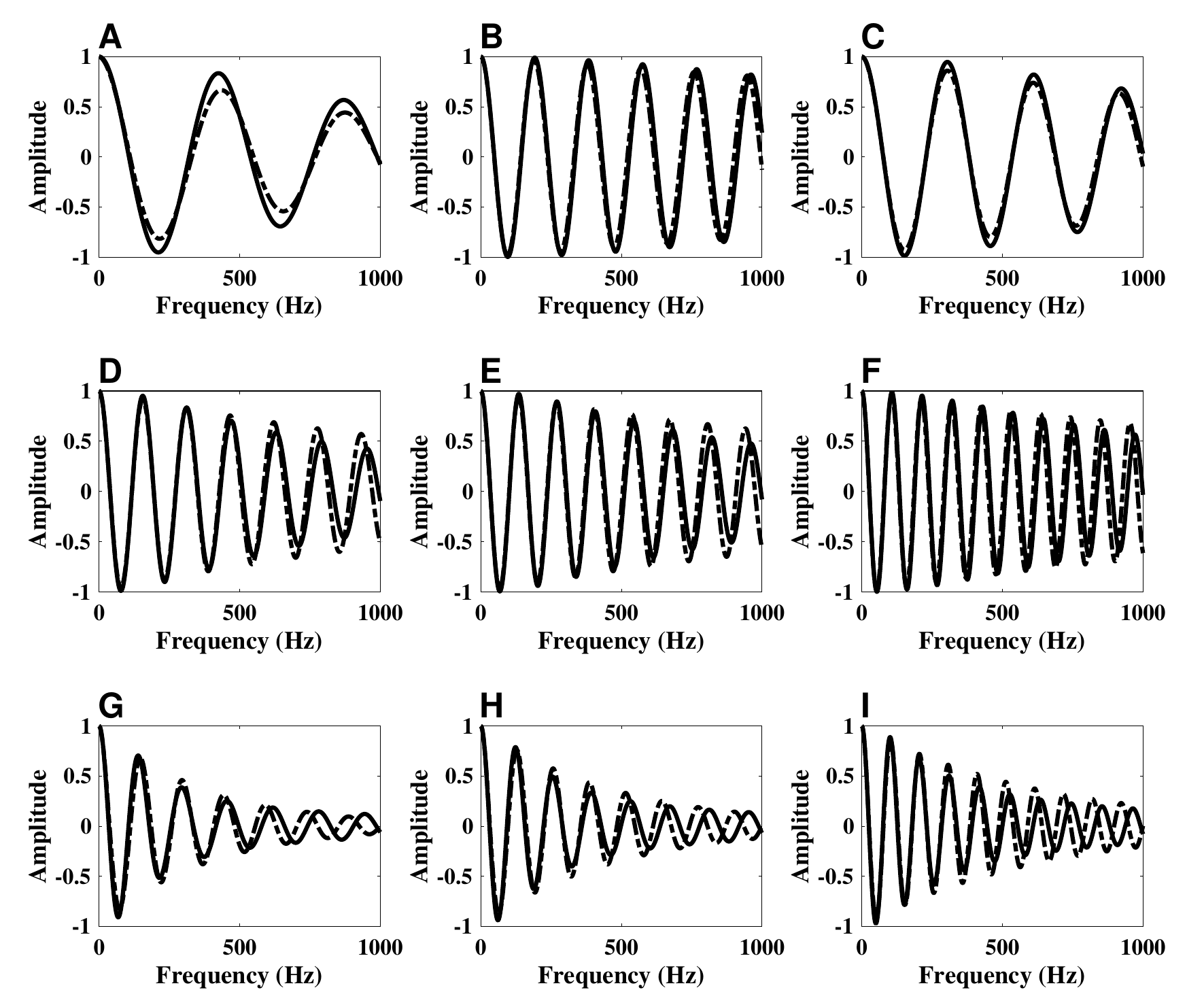}
   \end{subfigure}   
   \rulesep
   \begin{subfigure}[t]{0.49\textwidth}
      \centering
      \includegraphics[width=0.9\textwidth]{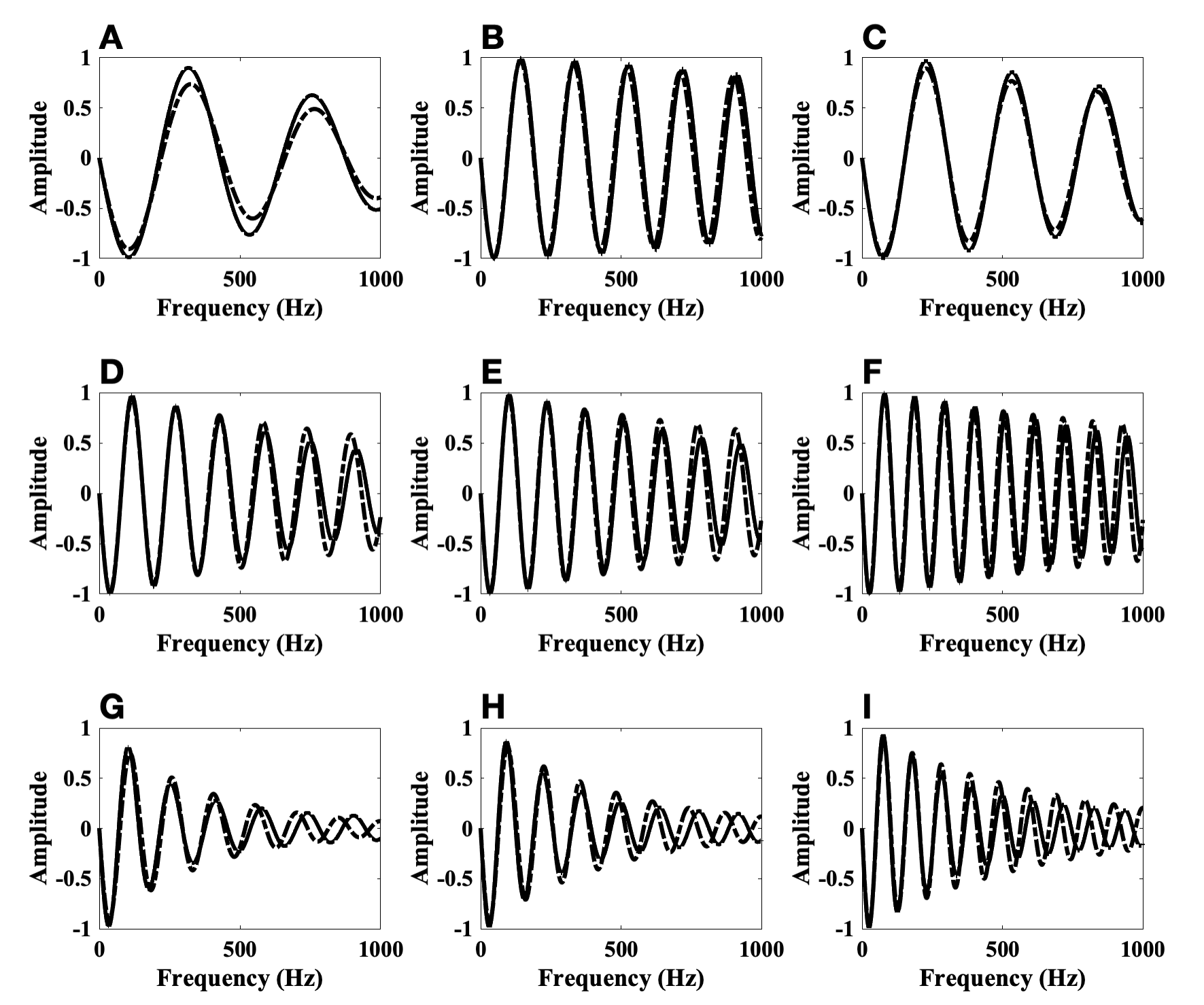}
   \end{subfigure}  
   \caption{Approximation of the complex mean delay function, 
			\eq{\ref{eq:layerDelay}}, describing the impact of distance-dependent 
			propagation delay on the postsynaptic neural rate response by a 
			decaying sinusoid, \eq{\ref{eq:approxDelay}}, for various values of 
			inter-laminar propagation delay, $\taul{}{}$, and radial propagation 
			delay, $\taur{}{}$. Real values are on the left while corresponding 
			imaginary values are on the right. Note that the oscillation frequency 
			of the real and imaginary components is the same. For both panels,
			the radial propagation delay is constant within a row, while the
			inter-lamina propagation delay is constant within a column, such
			that
			\figpanel{A} $\taul{}{}=400$\si{\micro\meter}, $\taur{}{}=100$\si{\micro\meter},
			\figpanel{B} $\taul{}{}=700$\si{\micro\meter}, $\taur{}{}=100$\si{\micro\meter},
			\figpanel{C} $\taul{}{}=900$\si{\micro\meter}, $\taur{}{}=100$\si{\micro\meter},
			\figpanel{D} $\taul{}{}=400$\si{\micro\meter}, $\taur{}{}=200$\si{\micro\meter},
			\figpanel{E} $\taul{}{}=700$\si{\micro\meter}, $\taur{}{}=200$\si{\micro\meter},
			\figpanel{F} $\taul{}{}=900$\si{\micro\meter}, $\taur{}{}=200$\si{\micro\meter},
			\figpanel{G} $\taul{}{}=400$\si{\micro\meter}, $\taur{}{}=400$\si{\micro\meter},
			\figpanel{H} $\taul{}{}=700$\si{\micro\meter}, $\taur{}{}=400$\si{\micro\meter}, and
			\figpanel{I} $\taul{}{}=900$\si{\micro\meter}, $\taur{}{}=400$\si{\micro\meter}.
			}
			\label{fig:meanDelayApprox}
\end{figure}

Examination of \fig{\ref{fig:spikeRateFreq}} suggests that the shape of the 
mean propagation delay in frequency closely matches the \gls{psp} spectrum of 
the one-sided exponential in \fig{\ref{fig:spikeRateFreq}}. To simplify 
incorporation of propagation delay into network models, we developed an 
approximation to the complicated \eq{\ref{eq:layerDelay}} based on 
the Fourier transform of a right-sided exponential decay\footnote{ 
	The Fourier transform of a right-sided exponential is 
	$\mathcal{F}\left[ \exp{-at}u(t) \right] = \frac{1}{a + \ii \omega}$. 
}, to aid in an analytical 
evaluation of the impact of propagation delay. The real and imaginary components 
of the mean delay were modelled as a decaying sinusoid, parameterized by a decay 
term, $\tau$, and a  frequency term, $\omega$, given by 
\begin{align}\label{eq:approxDelay}
	 \mathcal{R} \{ \atFreqAndRadius{\delaymean}{\inputlayer}{\layer{B}} \}
  &= 
	 \mathcal{R}\bigg\{ \frac{1}{ 1 + \ii 2\pi\freq\tau_D } \bigg\}
	 \cos{ \left( \omega \freq \right) }  \notag \\
	 \mathcal{I} \{ \atFreqAndRadius{\delaymean}{\inputlayer}{\layer{B}} \}
  &= 
	-\mathcal{R}\bigg\{ \frac{1}{ 1 + \ii 2\pi\freq\tau_D } \bigg\} 
	 \sin{ \left( \omega \freq \right) }  \notag \\
	 \conj{ \atFreqAndRadius{\delaymean}{\inputlayer}{\layer{B}} }
	        \atFreqAndRadius{\delaymean}{\inputlayer}{\layer{B}}
  &= 
	  \frac{1}{ 1 + \left( 2\pi\freq\tau_D \right)^2 }  \notag \\
	  \textrm{ for } \omega 
  &=  2\pi\taul{}{} + \taur{}{}			\notag \\
	  \textrm{ and } \tau_D 
  &= 
	  \frac{ \taur{}{} }{ \sqrt{2} } 
	  \exp{ - \frac{ \taul{}{} }{ 2 \taur{}{} } } .
\end{align}
The analytically derived impact of delay, \eq{\ref{eq:avgDelayAB_reducedDim}}, 
and its simpler approximation, \eq{\ref{eq:approxDelay}}, are shown in 
\fig{\ref{fig:meanDelayApprox}}, verifying that the approximation is accurate 
across a range of inter-laminar and radial propagation delays. Minor errors 
are introduced at high frequencies when there is significant attenuation 
resulting from either comparatively long radial propagation delays,
$\taur{}{}$, or short delays between layers, $\taul{}{}$. The error arises 
from slight changes in the oscillating frequency of the real and imaginary 
components of the mean delay across the spectrum. 

The time constant that controls decay of the mean delay frequency spectrum, $\tau_D$,
decreases with increasing inter-lamina delay, $\taul{}{}$, which is expected because 
radial delay, $\taur{}{}$, becomes comparatively smaller and, therefore, 
propagation delay between the layers becomes more uniform. This results in 
less low-pass filtering of the input and retention of more high frequency 
components. Conversely, as the presynaptic axons become more spread out 
radially, the time constant that controls decay of the mean delay spectra 
becomes larger, resulting in increased attenuation of high frequency components. 
This mirrors what happens with an exponential \gls{psp}: as the time constant 
of a postsynaptic potential increases in the time domain, the input signal 
becomes increasingly smoothed and so, in the frequency domain, more of the 
high frequencies are attenuated. 

   \subsection{Neural learning with propagation delay}\label{sec:DelPSP_plasticity} 

   \subsubsection{Covariance of neural activity with propagation delay}\label{sec:covDelPSP}
     
In the network Linsker studied, there was equal transmission delay between 
neurons and rate changes were instantaneous, implying that the \gls{psp} 
of cells was implicitly a delta function. In this study, we develop a 
framework for incorporating delay and a non-trivial \gls{psp} function, and 
derive expressions for the covariance between layer $\layer{B}$ neurons and 
neural learning in synapses connecting layers $\layer{B}$ and $\layer{C}$. 
This is a first step towards understanding neural learning with temporal dynamics, 
such as processing moving images. We assess the impact of delay and \gls{psp} 
function on the covariance of neurons and, hence, on the development of structure 
in a Linsker-type network. In order to understand how various parameters, such as 
propagation delay, impact the evolution of spatial opponent cells, it is necessary 
to derive equations for the expected size of a neuron's receptive field and the 
learning time constant. 

Using the expression for covariance between layer $\layer{B}$ neurons, we then 
determine the expected receptive field size of a layer $\layer{C}$ neuron as 
a function of both temporal and spatial parameters. In this way, we determine 
the impact of propagation delay on the emerging structure of the layer 
$\layer{C}$ cells.

\paragraph{Network incorporating a general post-synaptic response}

To determine an expression for covariance between neurons whose activities have been 
filtered by a \gls{psp} kernel, \eq{\ref{eq:Linsker_ratePSP}}, we extend the 
frequency domain expression for covariance between Poisson neurons with an 
implicit delta function model of the \gls{psp}, \eq{\ref{eq:cov_FFT}}, to obtain 
\begin{align}\label{eq:covPSP}
   \cov{\Lrate{\postN}{\layer{B}}{\filt}}{\Lrate{\postNtwo}{\layer{B}}{\filt}}
 &= 
	\frac{1}{\Tmax^2} \sum\limits_{\freq=0}^{\Tmax}
	                    \conj{\LRATE{\postN}{\layer{B}}} 
                              \LRATE{\postNtwo}{\layer{B}} 
						\conj{\atFreq{\PSP}} \atFreq{\PSP}  \notag \\
 &= 
   \frac{1}{\Tmax^2} \expect{ \conj{\LRATE{\postN}{\layer{B}}} 
                           \LRATE{\postNtwo}{\layer{B}} }
				  \Sumk \conj{\atFreq{\PSP}} \atFreq{\PSP}  \notag \\
 &= 
   \cov{\Lrate{\postN}{\layer{B}}{\filt}}{\Lrate{\postNtwo}{\layer{B}}{\filt}}
   \frac{ \Sumk \conj{\atFreq{\PSP}} \atFreq{\PSP} }{\Tmax}	\notag \\
 &=
   \PSPlayer \;
   \cov{\Lrate{\postN}{\layer{B}}{\filt}}{\Lrate{\postNtwo}{\layer{B}}{\filt}}, 
\end{align}
where the second-last line follows from \eq{\ref{eq:rate_FFT}} and we have 
introduced $\PSPlayer$ such that 
\begin{align}\label{eq:covPSPcoeff}
   \PSPlayer
 &= 
   \frac{ \Sumk \conj{\atFreq{\PSP}} \atFreq{\PSP} }{\Tmax} 
   \leq 1.	
\end{align}

For a \gls{psp} modeled as a one-sided exponential exponential decay function
parameterized by $\taupsp$ such that 
$\psp = \taupsp^{-1} \exp{-\frac{\timesym}{\taupsp}}$ and 
$  \conj{\atFreq{\PSP}} \atFreq{\PSP} 
=  \frac{1}{1 + \left( 2\pi {\freqT} \taupsp \right)^2 } $, 
lag-0 covariance evaluates to\footnote{
	The generalised result for covariance at lag $\lag$ evaluates to 
	$\cov{ \Lrate{\postN}{\layer{B}}{\filt}\left(\lag\right) }
	     { \Lrate{\postNtwo}{\layer{B}}{\filt} }
   = \frac{ \exp{ -\frac{\timesym}{\taupsp} } }{ \taupsp \Tmax } $
}
\begin{align}\label{eq:covExpPSPcoeff}
   \PSPlayer
 &= 
   \frac{1}{2 \Tmax \taupsp}.
\end{align}

\paragraph{Network incorporating propagation delay and a general postsynaptic response}
\label{sec:covDelayPSP}

We relaxed the assumption of homogeneous propagation delay for spikes from 
presynaptic neurons by incorporating a distance-dependent delay in the neuron 
model, \eq{\ref{eq:Linsker_rateDelayPSP}}. An expression for covariance between 
layer $\layer{B}$ neurons can be obtained by again using the frequency domain 
expression for covariance in \eq{\ref{eq:cov_FFT}} with respect to the spectral 
variates for the neuron model incorporating delay, \eq{\ref{eq:rate_FFT_DelayPSP}}, 
\begin{align}\label{eq:covDelayPSPcoeff}
   \scov{\Lrate{\postN}{\layer{B}}{\del,\filt}}
	    {\Lrate{\postNtwo}{\layer{B}}{\del,\filt}}
 &= 
   \frac{1}{\Tmax^2} 
   \Sumk \Rb{\inputlayer}{\layer{B}} 
         \conj{\atFreq{\PSP} \sum\limits_{\preN}
                      \atFreq{\LRATE{\preN}{\inputlayer}} 
                      \exp{-2\pi \ii \delay{\preN}{\postN}{}{} {\freqT}} }
         \Rb{\inputlayer}{\layer{B}} \atFreq{\PSP} 
		 \sum\limits_{\preNtwo} \atFreq{\LRATE{\preNtwo}{\inputlayer}} 
                                \exp{-2\pi \ii \delay{\preNtwo}{\postN}{}{} 
									   {\freqT}}.
\end{align}
Given that the delay variables, $\delay{\preN}{\postN}{}{}$ and 
$\delay{\preNtwo}{\postN}{}{}$, are statistically independent of the layer 
$\inputlayer$ rate values, $\LRATE{\preN}{\inputlayer}$ and 
$\LRATE{\preNtwo}{\inputlayer}$, respectively, the inner sum terms can be 
separated to give 
\begin{align}
   \scov{\Lrate{\postN}{\layer{B}}{\del,\filt}}
	    {\Lrate{\postNtwo}{\layer{B}}{\del,\filt}}
 &= 
   \frac{1}{\Tmax^2} 
   \Sumk \RbSq{\inputlayer}{\layer{B}}
	     \conj{\atFreq{\PSP}} \atFreq{\PSP}
         \conj{\expect{\exp{-2\pi \ii \delay{\preN}{\postN}{}{} {\freqT}}}}
         \expect{\exp{-2\pi \ii \delay{\preNtwo}{\postN}{}{} {\freqT}}}
         \conj{\sum\limits_{\preN} \atFreq{\LRATE{\preN}{\inputlayer}} }
	     \sum\limits_{\preNtwo} \atFreq{\LRATE{\preNtwo}{\inputlayer}}     \notag \\ 
 &= 
   \frac{1}{\Tmax^2} 
   \Sumk \conj{\atFreq{\PSP}} \atFreq{\PSP}
         \conj{\expect{\exp{-2\pi \ii \delay{\preN}{\postN}{}{} {\freqT}}}}
         \expect{\exp{-2\pi \ii \delay{\preNtwo}{\postN}{}{} {\freqT}}}
         \conj{\atFreq{\LRATE{\postN}{\layer{B}}}} 
		 \atFreq{\LRATE{\postNtwo}{\layer{B}}} .
\end{align}

Using the expected value of delay denoted by 
$  \atFreqAndRadius{\delaymean}{\inputlayer}{\layer{B}}
 = \expect{\exp{-2\pi \ii \delay{}{}{}{} {\freqT}}}$ 
and given in \eq{\ref{eq:avgDelayAB_reducedDim}}, covariance 
between layer $\layer{B}$ neurons in the presence of delay and a \gls{psp} can 
be expressed as 
\begin{align}\label{eq:covDelPSP}
   \scov{ \Lrate{\postN}{\layer{B}}{\del,\filt} }
        { \Lrate{\postNtwo}{\layer{B}}{\del,\filt} }
 &= 
   \frac{1}{\Tmax^2} 
   \Sumk \conj{\atFreq{\PSP}} \atFreq{\PSP}
         \conj{\atFreqAndRadius{\delaymean}{\inputlayer}{\layer{B}}}
         \atFreqAndRadius{\delaymean}{\inputlayer}{\layer{B}} 
         \conj{\atFreq{\LRATE{\postN}{\layer{B}}}} 
		 \atFreq{\LRATE{\postNtwo}{\layer{B}}}		\notag \\
 &= 
   \frac{1}{\Tmax^2} 
   \Sumk \conj{\atFreq{\PSP}} \atFreq{\PSP}
         \conj{\atFreqAndRadius{\delaymean}{\inputlayer}{\layer{B}}} 
		 \atFreqAndRadius{\delaymean}{\inputlayer}{\layer{B}}
         \conj{\atFreq{\LRATE{\postN}{\layer{B}}}} 
		 \atFreq{\LRATE{\postNtwo}{\layer{B}}}       \notag \\
 &= 
   \scov{\Lrate{\postN}{\layer{B}}{}}
		{\Lrate{\postNtwo}{\layer{B}}{}}
   \frac{\Sumk \conj{\atFreq{\PSP}} \atFreq{\PSP} 
	           \conj{\atFreqAndRadius{\delaymean}{\inputlayer}{\layer{B}}} 
			   \atFreqAndRadius{\delaymean}{\inputlayer}{\layer{B}}} 
		{\Tmax}										 \notag \\
 &= 
   \Dellayer{\inputlayer}{\layer{B}} \: 
   \scov{\Lrate{\postN}{\layer{B}}{}}
		{\Lrate{\postNtwo}{\layer{B}}{}}, 
\end{align}
where $\Dellayer{\inputlayer}{\layer{B}}$ expresses the attenuation of  
covariance in neural activity between postsynaptic neurons in layer 
$\layer{B}$ resulting from propagation delay between layers $\inputlayer$ 
and $\layer{B}$, and is defined by 
\begin{align}\label{eq:covDelCoeff}
   \Dellayer{\inputlayer}{\layer{B}} 
  &= 
	\frac{ \sum\limits_{\freq=0}^{\Tmax}
	           \conj{\atFreqAndRadius{\delaymean}{\inputlayer}{\layer{B}}} 
			    \atFreqAndRadius{\delaymean}{\inputlayer}{\layer{B}}
			    \conj{\atFreq{\PSP}} \atFreq{\PSP} }
		{ \Tmax }
	\leq 1. 
\end{align}

For a \gls{psp} modelled as a one-sided exponential decay parameterized by 
$\taupsp$, this evaluates to 
\begin{align}\label{eq:covExpPSPDelCoeff}
   \Dellayer{\inputlayer}{\layer{B}} 
 &= 
	\frac{ \taupsp - \tau_D }{ 2\Tmax ( \taupsp^2 - \tau_D^2 ) }, 
\end{align}
where $\tau_D$ is determined from the propagation delay between layers, 
$\taul{}{}$, and the radial propagation delay, $\taur{}{}$, according to the 
approximation in \eq{\ref{eq:approxDelay}}\footnote{
	Note that for lagged covariance this more generally evaluates to 
	$\scov{ \Lrate{\postN}{\layer{B}}{\del,\filt}\left( \lag \right) }
	      { \Lrate{\postNtwo}{\layer{B}}{\del,\filt} }
   = \Dellayer{ \inputlayer }{ \layer{B} } \left( \lag \right)
     \scov{ \Lrate{\postN}{\layer{B}}{} }{ \Lrate{\postNtwo}{\layer{B}}{} }$, 
   where 
	$\Dellayer{ \inputlayer }{ \layer{B} } \left( \lag \right)
   = \frac{ \exp{ -\frac{ \lag }{ \taupsp } } \taupsp 
          - \exp{ -\frac{ \lag }{ \tau_D  } } \tau_D  }
		  { 2 \left( \taupsp^2 - \tau_D^2 \right) } $
	   for a lag of $\lag$.}. 
Note that, if propagation delay is assumed to be $0$, this reduces to the network
incorporating a \gls{psp}-only case detailed in \eq{\ref{eq:covPSPcoeff}}. 

\begin{figure}[!htb]
\centering
	\includegraphics[width=0.6\textwidth]{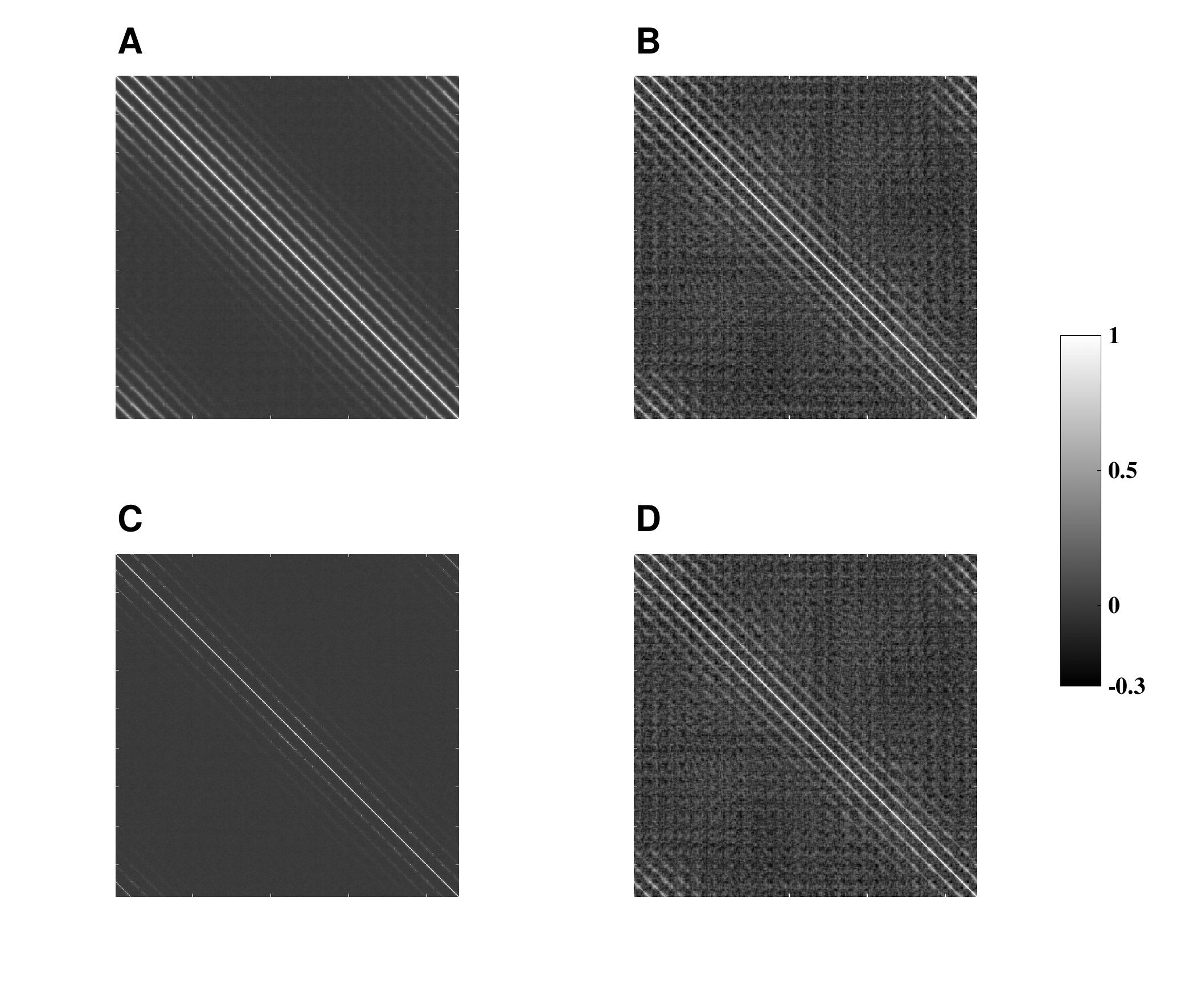}
	\caption{Normalised covariance between layer $\layer{B}$ neurons with and 
		 	 without incorporating distance-dependent propagation delay. The
			 additional correlation on the off diagonal edges reflects the
			 periodic boundaries used in simulations to avoid edge artifacts. 
			 Each axis is neuron index in the layer.  
			 \figpanel{A} Correlations between layer $\layer{B}$ neurons for
			 \citepos{Lin86a} network, \eq{\ref{appEq:Linsker_cov}}. 
			 \figpanel{B} Correlations between layer $\layer{B}$ neurons for
			 a network with postsynaptic potential modelled by an exponential 
			 decay with a $10$ms time constant, \eq{\ref{eq:covPSP}}. 
			 \figpanel{C} Correlations between layer $\layer{B}$ neurons for
			 a network with distance-dependent propagation delay and a trivial 
			 delta function model for the postsynaptic potential,  
			 \eq{\ref{eq:covDelPSP}}. 
			 \figpanel{D} Correlations between layer $\layer{B}$ neurons for
			 a network with distance-dependent propagation delay and a 
			 postsynaptic potential modelled by an exponential  
			 decay with a $10$ms time constant, \eq{\ref{eq:covDelPSP}}. 
		\label{fig:corrAB_DelPSP}}
\end{figure}

\fig{\ref{fig:corrAB_DelPSP}} shows correlation between layer $\layer{B}$
neurons for \citepos{Lin86a} network, for a network incorporating a
non-trivial postsynaptic potential, and for a network incorporating
both distance-dependent propagation delay and a non-trivial post-synaptic
potential. \fig{\ref{fig:corrAB_DelPSP}}B shows that, for a network with 
no propagation delay, an exponentially decaying postsynaptic potential 
spreads the impact of each spike over time, which does not impact expected 
correlation when there is no propagation delay. This is intuitive for white 
input; however, the correlation matrix is noisier, reflecting the increased 
variance in sample correlation owing to the loss of signal power in the input 
after low pass filtering \citep{DavGraEgaJoh13}.

\fig{\ref{fig:corrAB_DelPSP}}C shows the impact of propagation delay when 
a trivial delta function postsynaptic potential is used. Propagation delay spreads 
out the arrival of spikes to the neuron but, since the spikes are instantaneous, the
majority of correlation between layer $\layer{B}$ neurons is destroyed. 
Finally, \fig{\ref{fig:corrAB_DelPSP}}D shows that, while propagation delay spreads out the arrival
times of spikes to layer $\layer{B}$ postsynaptic neurons, the postsynaptic
potential spreads the impact of each spike out over time, mitigating the impact
of propagation delay. Note that the correlation matrix is also noisy because the
sample variance of correlation increases from the low-pass filtering. 

   \subsubsection{Learning equation with propagation delay}\label{sec:DelPSP_learningEqn} 
     
Thus far, we have derived the covariance between the neural activities of layer 
$\layer{B}$ neurons. However, when these neurons propagate spikes to a postsynaptic 
layer $\layer{C}$ neuron, there will be a delay as the signal propagates from layer 
$\layer{B}$ to layer \layer{C}, where arrival at the postsynaptic neuron will cause 
the signal to be further filtered by \gls{psp} of the postsynaptic cell. To 
evaluate the impact of this, it is necessary to consider the origin of the learning 
rule in \eq{\ref{eq:learningRuleBC_Linsker}}, outlined in 
\eq{\ref{eq:learningRuleBC_Linsker}}, in the context of the presynaptic output 
being delayed before arriving at the synapse where learning occurs, and the 
postsynaptic neural activity being filtered again at the layer $\layer{C}$ site. 
Using the model for layer $\layer{C}$ neural activity given in 
\eq{\ref{eq:rateC_DelayPSP}} and assuming that the plasticity process of synapse 
weight change is initiated near the cell body of the post-synaptic neuron, the 
learning equation becomes 
\begin{align}\label{eq:learningRule_DelPSP_nearCellBody}
    \Lwchange{\layer{B}}{\layer{C}}{\postN}{\postpostN} 
 &= 
    \etalayer{} \left(
		\kone{\layer{B}}{\layer{C}} 
	  + \frac{1}{\N{\layer{B}}{\layer{C}}}
		\sum \Lweight{\layer{B}}{\layer{C}}{\postNtwo}{\postpostN}
			 \left( \lcov{ \atDelayTime{\Lrate{\postNtwo}{\layer{B}}{\del,\filt}}
									   {\delay{\postNtwo}{\postpostN}{\layer{B}}{\layer{C}}} }
						 { \atDelayTime{\Lrate{\postN}{\layer{B}}{\del,\filt}}
									   {\delay{\postN}{\postpostN}{\layer{B}}{\layer{C}}} }
				  + \ktwo{\layer{B}}{\layer{C}} \right)
	\right) ,
\end{align}
where $\lcov{}{}$ is the normalized form of covariance between layer $\layer{B}$ 
neurons. 

To evaluate this covariance in the presence of propagation delay and 
arbitrary \gls{psp}, we can use the framework established in \sec{\ref{sec:covDelayPSP}}. 
Application of the result in \eq{\ref{eq:covDelPSP}} to calculate covariance 
between layer \layer{B} output and layer \layer{C} input, and using 
\atFreqAndRadius{\delaymean}{\layer{\inputlayer}}{\layer{B}} from 
\eq{\ref{eq:avgDelayAB}}, we obtain 
\begin{align}\label{eq:covC_DelayPSP}
 & \scov{ \atDelayTime{\Lrate{\postNtwo}{\layer{B}}{\del,\filt}}
					  {\delay{\postNtwo}{\postpostN}{\layer{B}}{\layer{C}}} }
		{ \atDelayTime{\Lrate{\postN}{\layer{B}}{\del,\filt}}
					  {\delay{\postN}{\postpostN}{\layer{B}}{\layer{C}}} } \notag \\
 &= 
   \frac{1}{\Tmax^2} 
   \Sumk \Rb{\inputlayer}{\layer{B}} 
         \conj{ \atFreq{\PSP} 
               \exp{-2\pi \ii \delay{\postNtwo}{\postpostN}{\layer{B}}{\layer{C}} 
				      {\freqT} } 
			   \sum\limits_{\preNtwo}
                   \atFreq{\LRATE{\preNtwo}{\inputlayer}} 
                   \exp{-2\pi \ii \delay{\preNtwo}{\postNtwo}{\inputlayer}{\layer{B}} 
					      {\freqT} }
		       }																		
         \Rb{\inputlayer}{\layer{B}} \atFreq{\PSP} 
		 \exp{-2\pi \ii \delay{\postN}{\postpostN}{\layer{B}}{\layer{C}} 
			    {\freqT} }
		 \sum\limits_{\preN} \atFreq{\LRATE{\preN}{\inputlayer}} 
               \exp{-2\pi \ii \delay{\preN}{\postN}{\inputlayer}{\layer{B}} 
				      {\freqT} }                                      \notag \\
 &= 
   \frac{1}{\Tmax^2} 
   \Sumk \conj{\atFreq{\PSP}} \atFreq{\PSP}
         \conj{\atFreqAndRadius{\delaymean}{\inputlayer}{\layer{B}}}
		 \atFreqAndRadius{\delaymean}{\inputlayer}{\layer{B}}
		 \exp{-2\pi \ii {\freqT} 
			    \left( \delay{\postN}{\postpostN}{\layer{B}}{\layer{C}} 
				     - \delay{\postNtwo}{\postpostN}{\layer{B}}{\layer{C}} 
				\right)}
         \conj{\atFreq{\LRATE{\postN}{\layer{B}}}} 
		 \atFreq{\LRATE{\postNtwo}{\layer{B}}}		\notag \\
 &= 
   \scov{\Lrate{\postN}{\layer{B}}{}}
		{\Lrate{\postNtwo}{\layer{B}}{}}
   \frac{1}{\Tmax} 
   \Sumk \conj{\atFreq{\PSP}} \atFreq{\PSP}
         \conj{\atFreqAndRadius{\delaymean}{\inputlayer}{\layer{B}}}
		 \atFreqAndRadius{\delaymean}{\inputlayer}{\layer{B}}
		 \exp{-2\pi \ii {\freqT} 
			    \left( \delay{\postN}{\postpostN}{\layer{B}}{\layer{C}} 
				     - \delay{\postNtwo}{\postpostN}{\layer{B}}{\layer{C}} 
				\right)}							\notag \\
 &= 
   \DelClayer{\postN}{\postNtwo}
   \scov{\Lrate{\postN}{\layer{B}}{}}
		{\Lrate{\postNtwo}{\layer{B}}{}},
\end{align}
where $\DelClayer{\postN}{\postNtwo}$ captures attenuation of covariance
in activity of layer $\layer{B}$ presynaptic neurons $\postN$ and $\postNtwo$
on arrival at the
postsynaptic neuron in layer $\layer{C}$ by propagation delay between both 
layers $\inputlayer$ and $\layer{B}$ and layers $\layer{B}$ and $\layer{C}$. 
Propagation delay is compounded from the input layer $\inputlayer$ all the  
way to layer $\layer{C}$ because covariance is induced by common synapses from 
layer $\inputlayer$ to layer $\layer{B}$ neurons. 

\eq{\ref{eq:covC_DelayPSP}} describes the expected covariance between 
inputs arriving from two layer $\layer{B}$ neurons, $\postN$ and $\postNtwo$, at 
the synapse of a given layer $\layer{C}$ cell, $\postpostN$, taking
into consideration distance-dependent axonal propagation delays. Therefore, 
the result depends on the position of the layer $\layer{B}$ presynaptic 
neurons relative to the layer $\layer{C}$ postsynaptic neuron $\postpostN$. More 
precisely, the result depends on the difference in spike arrival times between 
neuron $\postN$ and neuron $\postNtwo$ at the neuron $\postpostN$ in layer 
$\layer{C}$, given by 
$ \delay{\postN}{\postpostN}{\layer{B}}{\layer{C}} 
- \delay{\postNtwo}{\postpostN}{\layer{B}}{\layer{C}}$, which itself is a function 
of the radial distance between $\postN$ and $\postNtwo$, the interlaminar distance 
between layers $\layer{B}$ and $\layer{C}$, and the spike propagation velocity; i.e., 
$ \delay{\postN}{\postpostN}{\layer{B}}{\layer{C}} 
- \delay{\postNtwo}{\postpostN}{\layer{B}}{\layer{C}} 
= \frac{( (\distance{\postN}{\postpostN}{\inputlayer}{\layer{B}})^2 
	    + \distlayerSqu{\layer{B}}{\layer{C}} )^{\frac{1}{2}} 
	 -  ( (\distance{\postNtwo}{\postpostN}{\inputlayer}{\layer{B}})^2 
	    + \distlayerSqu{\layer{B}}{\layer{C}} )^{\frac{1}{2}} }
       {\velocity} $. 
This has the effect of further attenuating covariance between the presynatic
layer $\layer{B}$ neurons because, if the difference in spike arrival times is not
within the smoothing period of the \gls{psp}, the covariance is destroyed.


To analytically examine the impact of propagation delay on learning, we use the 
decaying sinusoidal approximation of the real and imaginary components of mean 
delay in the frequency spectrum, which results in a frequency spectrum of
$ \conj{\atFreqAndRadius{\delaymean}{\inputlayer}{\layer{B}}}
		\atFreqAndRadius{\delaymean}{\inputlayer}{\layer{B}} $ 
that is well approximated by $\frac{1}{ 1 + (2\pi\freq\tau_D)^2 }$ 
(see \eq{\ref{eq:approxDelay}}).  
We also introduce a simple one-sided exponential \gls{psp}
such that $\psp = \frac{1}{\taupsp} \exp{-\frac{\timesym}{\taupsp}}$ and 
$  \conj{\atFreq{\PSP}} \atFreq{\PSP} 
=  \frac{1}{1 + \left( 2\pi {\freqT} \taupsp \right)^2 } $. 
Therefore, we can write the expression for attenuation of covariance between 
inputs from $\layer{B}$ neurons, $\postN$ and $\postNtwo$, arriving at the 
postsynaptic neuron, 
\eq{\ref{eq:covC_DelayPSP}}, as 
\begin{align}\label{eq:covC_DelayExpPSP}
   \DelClayer{\postN}{\postNtwo} 
 &= 
	\sum\limits_{\freq=0}^{\Tmax} 
	   \frac{1}{ \left( 1 + \left( 2\pi {\freqT} \tau_D  \right)^2 \right) } 
	   \frac{1}{ \left( 1 + \left( 2\pi {\freqT} \taupsp \right)^2 \right) } 
	   \exp{-2\pi \ii {\freqT} 
	  		  \left( \delay{\postN}{\postpostN}{\layer{B}}{\layer{C}} 
				   - \delay{\postNtwo}{\postpostN}{\layer{B}}{\layer{C}} 
			  \right)}	    \notag \\
 &= 
	\frac{ \exp{ -\Delta_{\postN\postNtwo}  
	\frac{ \taupsp + \tau_D }{ \taupsp \tau_D } } 
		 \left( \exp{ \frac{ \Delta_{\postN\postNtwo} }
						   { \tau_D  } }\taupsp 
			  - \exp{ \frac{ \Delta_{\postN\postNtwo} }
						   { \taupsp } }\tau_D
		 \right) }
	{2\left( \taupsp - \tau_D \right) 
	  \left( \taupsp + \tau_D \right) } 
\end{align}
for lag-0 covariance\footnote{
	For lagged covariance this result generalises to  
	$ \scov{ \atDelayTime{\Lrate{\postNtwo}{\layer{B}}{\del,\filt}}
						 {( \delay{\postNtwo}{\postpostN}{\layer{B}}{\layer{C}} 
						  - \lag ) } }
		   { \atDelayTime{\Lrate{\postN}{\layer{B}}{\del,\filt}}
					     {\delay{\postN}{\postpostN}{\layer{B}}{\layer{C}}} }
	= \frac{ \exp{ -(\Delta_{\postN\postNtwo} + \lag) 
	                \frac{ \taupsp + \tau_D }{ \taupsp \tau_D } } 
			 \left( \exp{ \frac{ (\Delta_{\postN\postNtwo} + \lag) }
							   { \tau_D  } }\taupsp 
				  - \exp{ \frac{ (\Delta_{\postN\postNtwo} + \lag) }
							   { \taupsp } }\tau_D
		     \right) }
		   {2\left( \taupsp - \tau_D \right) 
		     \left( \taupsp + \tau_D \right) } $, for lag $\lag$.
} and for 
$ \Delta_{\postN\postNtwo} 
      = \abs{ \delay{\postN}{\postpostN}{\layer{B}}{\layer{C}} 
            - \delay{\postNtwo}{\postpostN}{\layer{B}}{\layer{C}} }$.
For the case of zero propagation delay, this reduces to the covariance between
neurons in the presence of propagation delay only, \eq{\ref{eq:covExpPSPcoeff}}. 

The rate of change of synaptic weights between layers \layer{B} and \layer{C} can  
then be given as 
\begin{align}\label{eq:learningEqn_DelPSP}
    \Lwchange{\layer{B}}{\layer{C}}{\postN}{\postpostN} 
 \triangleq 
	\etalayer{} \ensemble{ \left( \Delta_{\postN\postNtwo} 
	                              \Lweight{}{}{\postN}{\postpostN} \right) }
 &= 
    \etalayer{} \left( 
		\kone{ \layer{B} }{ \layer{C} } 
	  + \frac{ 1 }{ \N{\layer{B}}{\layer{C}} }
		\sum \Lweight{\layer{B}}{\layer{C}}{\postNtwo}{\postpostN}
			 \left( 
				 \DelClayer{\postN}{\postNtwo}
				 \lcov{ \Lrate{\postNtwo}{\layer{B}}{} }{ \Lrate{\postN}{\layer{B}}{} }
			   + \ktwo{\layer{B}}{\layer{C}} \right) 
	\right) .
\end{align}
The receptive fields in a three-layer \cite{Lin86a} model are well known to evolve 
into spatial opponent cells. Since attenuation of covariance from propagation 
delay is circularly symmetric, we do not expect the shape of the resulting 
receptive fields to change. However, we examine properties such as the size of the 
on-centre in the resulting spatial opponent cell and time to convergence for 
synaptic weights.


\subsection{On-center size}\label{sec:RFsize}
   	
For cortical neurons, the term `receptive field' is assumed to refer to an 
inter-layer receptive field, defined by a cortical neuron's input synaptic weight 
structure. We wish to analytically derive an expression for the size of the on-center 
of a neuron's receptive field to determine the impact of propagation delay on the 
evolution of layer $\layer{C}$ spatial-opponent neurons. To this end, we consider 
the neuron when it has undergone sufficient learning (see the learning equation, 
\eq{\ref{eq:learningRuleBC_Linsker}}) such that the weights of all synapses 
connecting to the layer $\layer{C}$ neuron have diverged to the upper or lower bound. 
This assumption is valid because the learning equation is unstable and, hence, all 
weights, or all but one, will reach a stable limiting value (see \citet{Lin86a}, p.~7510). 

\subsubsection{Linsker's network}
To determine the size of a neuron's on-center, it is necessary to determine 
its average synaptic weight. We assume that synaptic weights are independent of 
covariance between neural output early in the learning process. The mean weight 
converges during this early stage \citep{KemGerHem99} and, thus, from 
\eq{\ref{eq:learningRuleBC_Linsker}}, we can write 
\begin{align}\label{eq:meanweight}
    \meanLwchange{\layer{B}}{\layer{C}} 
 &= 
    \etalayer{} \left( 
		\kone{\layer{B}}{\layer{C}} + \meanLweight{\layer{B}}{\layer{C}} 
		\left( \ktwo{\layer{B}}{\layer{C}} + \meanCov{\layer{B}}{\layer{B}}{} \right),
		\quad 
		\LwgtMin{\layer{B}}{\layer{C}} \leq \meanLweight{\layer{B}}{\layer{C}} 
									   \leq \LwgtMax{\layer{B}}{\layer{C}} 
	\right) , 
\end{align}
where $\meanLweight{\layer{B}}{\layer{C}}$ denotes the mean weight of synapses 
between layer $\layer{B}$ and layer $\layer{C}$ and $\meanCov{\layer{B}}{\layer{B}}{}$ 
is the mean covariance between layer $\layer{B}$ neural outputs. As noted by 
\citet{Lin86a}, for $\ktwo{\layer{B}}{\layer{C}}>0$, the weights are unstable and 
grow to the bounds. \citet{Lin86a} showed that $\ktwo{\layer{B}}{\layer{C}}$ can 
be derived from first-order firing rate statistics and is positive if the 
postsynaptic neuron is firing at a rate that is greater than a given constant 
benchmark rate, and vice-versa. This means that, for the mean weight to be stable, 
the mean rate of neural firing must be less than this benchmark value. 

The fixed point for the mean 
weight follows immediately, 
\begin{align}\label{eq:linsker_fixedPt}
   \FP{\meanLweight{\layer{B}}{\layer{C}}} 
 &= 
   \frac{-\kone{\layer{B}}{\layer{C}}}
        { \ktwo{\layer{B}}{\layer{C}} + \meanCov{\layer{B}}{\layer{B}}{} }.
\end{align}

To determine the fixed point for the mean weight, we need to calculate the mean 
covariance between layer \layer{B} neurons. As per \citet{Lin86a}, we normalize the 
temporal covariance between layer \layer{B} neurons by dividing by input variance 
and number of input connections (\eqs{\ref{eq:cov_temp2ensemble}}{\ref{eq:f0_normCov}}). 
Consequently, for the case of a delta function \gls{psp} and instantaneous 
propagation delay, the covariance of a neuron with itself is 1. 
\ifx\isEmbedded\undefined
	
The average covariance between layer $\layer{B}$ neurons in the network 
assumed by \citet{Lin86a} can be found by integrating the expression for 
covariance over the laminar, 
\begin{align}
   \meanCov{\layer{B}}{\layer{B}}{} 
 &= 
   \frac{1}{ \left( \pi \radiusvar{\layer{B}}{\layer{C}} \right)^2 }
   \int_{-\infty}^{\infty} 
     \int_{-\infty}^{\infty} 
       \exp{- \frac{ \left| \poscont{}{} - \postwocont{}{} \right|^2 }
			       {2 \radiusvar{\inputlayer}{\layer{B}} } } 
	   \exp{-\frac{\left( \poscont{}{}^2 + \postwocont{}{}^2 \right)}
		          {\radiusvar{\layer{B}}{\layer{C}} } }
	  d\postwocont{}{}	d\poscont{}{}	                                      \notag \\ 
 &= 
   \frac{1}{ \left( \pi \radiusvar{\layer{B}}{\layer{C}} \right)^2 }
   \exp{-\frac{ \poscont{}{}^2 }{\alpha}}
   \int_{-\infty}^{\infty} 
     \int_{-\infty}^{\infty} 
	   \exp{-\frac{ \postwocont{}{}^2 }{\alpha} }
	   \exp{ \frac{ \postwocont{}{} \poscont{}{} }{ \radiusvar{\inputlayer}{\layer{B}} }}
	  d\postwocont{}{}	d\poscont{}{}
\end{align}
where $\alpha=\frac{2\radiusvar{\inputlayer}{\layer{B}} \radiusvar{\layer{B}}{\layer{C}}}
                   {2\radiusvar{\inputlayer}{\layer{B}}+\radiusvar{\layer{B}}{\layer{C}}  } $ 
for convenience. Furthermore, the radial symmetry of the covariance and probability 
connection functions renders the result directionally invariant. Therefore, we 
calculate the result in one dimension and square it to generalize it to two 
dimensions, 
\begin{align}
   \meanCov{\layer{B}}{\layer{B}}{} 
 &= 
   \frac{1}{ \left( \pi \radiusvar{\layer{B}}{\layer{C}} \right)^2 }
   \left( 
   \int_{-\infty}^{\infty} 
     \exp{-\frac{ \posxcont^2 }{\alpha}}
     \exp{ \frac{ \posxcont^2 \alpha }{ 4\radiusquad{\inputlayer}{\layer{B}} }}
     \int_{-\infty}^{\infty} 
	   \exp{-\left( \frac{\posxcont}{\alpha} 
			      - \frac{\sqrt{\alpha}\posxtwocont}
			 	         {2 \radiusvar{\inputlayer}{\layer{B}} } 
		 	 \right)^2}
	  d\posxtwocont	d\posxcont
   \right)^2						     \notag \\
 &= 
   \frac{\pi \alpha }{ \left( \pi \radiusvar{\layer{B}}{\layer{C}} \right)^2 }
   \left( 
	   \int_{-\infty}^{\infty} 
		  \exp{-\frac{ \posxcont^2 }{\alpha}}
		  \exp{ \frac{ \posxcont^2 \alpha }{ 4\radiusquad{\inputlayer}{\layer{B}} }}
		  \exp{ -\chi^2}       
		  d\chi d\posxcont						
   \right)^2						     \notag \\
 &= 
   \frac{\alpha}{ \pi \radiusquad{\layer{B}}{\layer{C}} }
   \left( 
	   \int_{-\infty}^{\infty} 
		  \exp{-\posxcont^2 \frac{4 \radiusquad{\inputlayer}{\layer{B}} - \alpha^2 }
		                         {4 \radiusvar{\inputlayer}{\layer{B}}   \alpha}}
		  d\posxcont	                          
   \right)^2            \notag \\
 &= 
   \frac{\alpha}{ \pi \radiusquad{\layer{B}}{\layer{C}} }
   \left( 
	   \int_{-\infty}^{\infty} 
		  \exp{-\posxcont^2 
		        \frac{2 \radiusvar{\inputlayer}{\layer{B}} + \radiusvar{\layer{B}}{\layer{C}}}
		             {2 \radiusvar{\inputlayer}{\layer{B}} \radiusvar{\layer{B}}{\layer{C}}
					  + \radiusquad{\layer{B}}{\layer{C}} }}
		  d\posxcont	         
   \right)^2                             \notag \\
 &= 
   \frac{\alpha}{ \pi \radiusquad{\layer{B}}{\layer{C}} }
   \left( 
	   \frac{ \pi \left( 2\radiusvar{\inputlayer}{\layer{B}} \radiusvar{\layer{B}}{\layer{C}}
					   +  \radiusquad{\layer{B}}{\layer{C}} \right) }
			{2\radiusvar{\inputlayer}{\layer{B}} + 2\radiusquad{\layer{B}}{\layer{C}} } 
   \right)^2							 \notag \\
 &= 
   \frac{ 2 \radiusquad{\inputlayer}{\layer{B}} 
	    +   \radiusvar{\inputlayer}{\layer{B}} \radiusvar{\layer{B}}{\layer{C}} }
        { 2 \radiusquad{\inputlayer}{\layer{B}} 
		+ 3 \radiusvar{\inputlayer}{\layer{B}} \radiusvar{\layer{B}}{\layer{C}} 
		+   \radiusquad{\layer{B}}{\layer{C}} }			\notag \\
 &= 
   \frac{ 2 +   \frac{ \radiusvar{\layer{B}}{\layer{C}} }{ \radiusvar{\inputlayer}{\layer{B}} } }
        { 2 + 3 \frac{ \radiusvar{\layer{B}}{\layer{C}} }{ \radiusvar{\inputlayer}{\layer{B}} } 
		    +   \frac{ \radiusquad{\layer{B}}{\layer{C}} }
			         { \radiusquad{\inputlayer}{\layer{B}} } }  \notag \\
 &= 
   \frac{ 2 + \frac{ \radiusvar{\layer{B}}{\layer{C}} }{ \radiusvar{\inputlayer}{\layer{B}} } }
        { \left( 2 + \frac{ \radiusvar{\layer{B}}{\layer{C}} }
				          { \radiusvar{\inputlayer}{\layer{B}} } \right) 
          \left( 1 + \frac{ \radiusvar{\layer{B}}{\layer{C}} }
				          { \radiusvar{\inputlayer}{\layer{B}} } \right) }   \notag \\
 &=
   \frac{ 1 } 
        { 1 + \frac{ \radiusvar{\layer{B}}{\layer{C}} }
			       { \radiusvar{\inputlayer}{\layer{B}} } }. 
\end{align}

\else
	The average covariance between layer \layer{B} neurons can now be found by 
	integrating covariance over the laminar (see \app{\ref{app:avgCov}} for details), 
	\begin{align}\label{eq:meanCov_layerB}
	   \meanCov{\layer{B}}{\layer{B}}{} 
	 &= 
	   \frac{1}{ \left( \pi \radiusvar{\layer{B}}{\layer{C}} \right)^2 }
	   \int_{-\infty}^{\infty} 
		 \int_{-\infty}^{\infty} 
		   \exp{- \frac{ \left| \poscont{}{}^2 - \postwocont{}{}^2 \right| }
					   {2 \radiusvar{\inputlayer}{\layer{B}} } } 
		   \exp{-\frac{\poscont{}{}^2}{\radiusvar{\layer{B}}{\layer{C}} } }
		   \exp{-\frac{\postwocont{}{}^2}{\radiusvar{\layer{B}}{\layer{C}} } }
		  d\postwocont{}{}	d\poscont{}{}	                      \notag \\ 
	 &= 
	    \frac{ 1 } 
			 { 1 + \frac{ \radiusvar{\layer{B}}{\layer{C}} }
					    { \radiusvar{\inputlayer}{\layer{B}} } }. 
	\end{align}
\fi
From \eq{\ref{eq:meanCov_layerB}}, it can be seen that mean covariance between 
layer $\layer{B}$ neurons is maximized by either maximizing the connection 
radius between layers $\inputlayer$ and $\layer{B}$ or minimizing the connection 
radius between layers $\layer{B}$ and $\layer{C}$. 

Now it is necessary to calculate the average weight of a layer $\layer{C}$ synapse 
with receptive field size of $\RF{}$. Knowing that synapses within the on-center 
have reached the upper limiting value of $\LwgtMax{\layer{B}}{\layer{C}}$ while 
synapses outside of it have reached the lower limiting value of 
$\LwgtMax{\layer{B}}{\layer{C}}$, we can determine 
the average synaptic weight by integrating individual synapse weights scaled by 
the probability of each synapse being connected to the layer \layer{C} neuron. 
Implementing the integral in polar coordinates this time, we obtain 
\begin{align}\label{eq:linsker_RFintegral}
    \FP{\meanLweight{\layer{B}}{\layer{C}}} 
 &= 
    \frac{ \LwgtMax{\layer{B}}{\layer{C}} }{\pi \radiusvar{\layer{B}}{\layer{C}} }
    \int_{0}^{\RF{}} \int_{0}^{2\pi} 
        \exp{- \frac{ \posintr^2 } { \radiusvar{\layer{B}}{\layer{C}} }}
	   d\posintr d\postwoth
  - \frac{ \LwgtMin{\layer{B}}{\layer{C}} }{\pi \radiusvar{\layer{B}}{\layer{C}} }
    \int_{\RF{}}^{\infty} \int_{0}^{2\pi} 
        \exp{- \frac{ \posintr^2 } { \radiusvar{\layer{B}}{\layer{C}} }}
 	   d\posintr d\postwoth                                           \notag \\
 &= 
    \LwgtMax{\layer{B}}{\layer{C}} 
    \left( 1 - \exp{ -\frac{ \RFvar{} }{ \radiusvar{\layer{B}}{\layer{C}} } } \right) 
  + \LwgtMin{\layer{B}}{\layer{C}} 
    \left( \exp{ -\frac{ \RFvar{} }{ \radiusvar{\layer{B}}{\layer{C}} } } \right).
\end{align}

For a lower weight bound of 0 and an upper weight bound of 1, this gives
\begin{align}\label{linsker_meanWgt}
    \FP{\meanLweight{\layer{B}}{\layer{C}}} 
 &= 
    1 - \exp{ -\frac{ \RFvar{} }{ \radiusvar{\layer{B}}{\layer{C}} } } . 
\end{align}

Finally, to determine the size of the on-center as a function of covariance, 
equate the two equations for mean synaptic weight, 
\eqs{\ref{eq:linsker_fixedPt}}{\ref{eq:linsker_RFintegral}}, and rearrange to
obtain  
\begin{align}
    \RF{}
 &= 
    \radius{\layer{B}}{\layer{C}}
	\sqrt{ \log \left( 
					\frac{ \ktwo{\layer{B}}{\layer{C}} 
					     + \meanCov{\layer{B}}{\layer{B}}{} }
						 { \LwgtMax{\layer{B}}{\layer{C}} 
						   \left( \ktwo{\layer{B}}{\layer{C}} 
								+ \meanCov{\layer{B}}{\layer{B}}{} \right) 
						 + \kone{\layer{B}}{\layer{C}} } 
			    \right) }		 
	\label{eq:linsker_RF} \\
  &= 
    \radius{\layer{B}}{\layer{C}}
	\sqrt{ \log \left( \frac{1}{ \LwgtMax{\layer{B}}{\layer{C}} 
			                   - \FP{\meanLweight{\layer{B}}{\layer{C}}} } 
			    \right) } \, , 
	\label{eq:linsker_RF_FP}
\end{align}
where we have used the assumption that 
$\LwgtMax{\layer{B}}{\layer{C}} - \LwgtMin{\layer{B}}{\layer{C}} = 1$ 
\citep{Lin86a} and applied the equation for the fixed point to go from 
\eq{\ref{eq:linsker_RF}} to \eq{\ref{eq:linsker_RF_FP}}. 
Note that, since the numerator of \eq{\ref{eq:linsker_RF}} must be negative for 
the mean weight to be stable (see \eq{\ref{eq:meanweight}}), the denominator 
must also be negative for the size of the on-center to be a real number.

\subsubsection{Network incorporating a general post-synaptic response}

To calculate receptive field size for a network with an arbitrary \gls{psp} and 
identical propagation delays between all neurons, \eq{\ref{eq:linsker_RF}} 
can be trivially extended to the case for which an arbitrary \gls{psp} is 
incorporated into the neuron model since the impact of the \gls{psp} on covariance 
is to scale it by a constant factor, \eq{\ref{eq:covPSP}}. Consequently, the 
expected size of a layer $\layer{C}$ neuron's receptive field becomes
\begin{align}\label{eq:linskerPSP_RF}
    \RF{\psp}
 &= 
    \radius{\layer{B}}{\layer{C}}
	\sqrt{ \log \left( 
					\frac{ \ktwo{\layer{B}}{\layer{C}} 
					     + \PSPlayer \meanCov{\layer{B}}{\layer{B}}{} }
						 { \LwgtMax{\layer{B}}{\layer{C}} \left( \ktwo{\layer{B}}{\layer{C}} 
								         + \PSPlayer \meanCov{\layer{B}}{\layer{B}}{} 
						            \right) 
						 + \kone{\layer{B}}{\layer{C}} } 
		        \right) }. 
\end{align}

\subsubsection{Network incorporating propagation delay and a general post-synaptic response}

We now determine expressions for the receptive field size for an arbitrary \gls{psp} 
function and propagation delay that is proportional to the three-dimensional 
distance between neurons. Using these expressions, a neuron's receptive field size 
will be calculated in this more realistic context. 

Introduce the term, $\delpsp = \velocity \distlayer{\layer{B}}{\layer{C}} \taupsp$, 
which captures the elements that minimize the spread in arrival time of input 
spikes to the layer $\layer{C}$ post-synaptic neuron, As inter-layer distance, 
$\distlayer{\layer{B}}{\layer{C}}$, and velocity, $\velocity$, increase, the 
impact of radial propagation delay decreases because there is reduced spread in
arrival times of spikes that were initiated simultaneously in layer $\layer{B}$. 
As the \gls{psp} time constant, $\taupsp$, increases, the impact of radial 
propagation delay is reduced because the input signals are low-pass filtered. 

Using the expressions for covariance derived in \sec{\ref{sec:covDelPSP}}, the
average covariance into a layer $\layer{C}$ cell when propagation delay and
\gls{psp} are incorporated are found to be (see \sec{\ref{subapp:avgCovDelPSP}}  
for details)
\begin{align}\label{eq:avgCovLayerC}
   \meanCov{\layer{B}}{\layer{B}}{\del,\psp}
 &=
   \frac{1}{ (\taupsp - \tau_D)^2 (\taupsp + \tau_D)^2 } 
   \left( 
	   \frac{4 \left( 2 + \frac{ \radiusvar{\layer{B}}{\layer{C}} } 
							 { \radiusvar{\inputlayer}{\layer{B}} }
			   \right) \taupsp^2
			}
			{\left(  \frac{ \radiusvar{\layer{B}}{\layer{C}} }{ \velocity \distlayer{\layer{B}}{\layer{C}} \taupsp } 
				 - 2 \right)
			 \left(  \frac{ \radiusvar{\layer{B}}{\layer{C}} }{ \velocity \distlayer{\layer{B}}{\layer{C}} \taupsp }
				  -  2 
				  -2 \frac{ \radiusvar{\layer{B}}{\layer{C}} } 
						  { \radiusvar{\inputlayer}{\layer{B}} } 
			 \right) 
			 \left(  \frac{ \radiusvar{\layer{B}}{\layer{C}} }{ \velocity \distlayer{\layer{B}}{\layer{C}} \taupsp }
				  +  2 
				  +  \frac{ \radiusvar{\layer{B}}{\layer{C}} }
						  { \radiusvar{\inputlayer}{\layer{B}} }
			 \right) 
			}  
   \right. \notag \\
 &+ 
   \left.
   \frac{4 \left( 2 + \frac{ \radiusvar{\layer{B}}{\layer{C}} } 
				           { \radiusvar{\inputlayer}{\layer{B}} }
		   \right) \tau_D^2 
        }
        {\left(  \frac{ \radiusvar{\layer{B}}{\layer{C}} }
		              { \velocity \distlayer{\layer{B}}{\layer{C}} \tau_D }  
			 - 2 \right)
		 \left(  \frac{ \radiusvar{\layer{B}}{\layer{C}} }{ \velocity \distlayer{\layer{B}}{\layer{C}} \tau_D }
		      -  2 
		      -2 \frac{ \radiusvar{\layer{B}}{\layer{C}} } 
			          { \radiusvar{\inputlayer}{\layer{B}} } 
		 \right) 
	     \left(  \frac{ \radiusvar{\layer{B}}{\layer{C}} }{ \velocity \distlayer{\layer{B}}{\layer{C}} \tau_D }
		      +  2 
			  +  \frac{ \radiusvar{\layer{B}}{\layer{C}} }
			          { \radiusvar{\inputlayer}{\layer{B}} }
		 \right)
		} 
   \right) .
\end{align}
Note that this result is undefined for $\taupsp = \tau_D$ and, for zero
propagation delay and a long inter-laminar distance, this reduces to the same
attenuation as for a network with an exponential \gls{psp}, 
$\PSPlayer \meanCov{\layer{B}}{\layer{B}}{} $. 

The impact of delay from propagation between layers is to attenuate covariance, 
so that 
$\meanCov{\layer{B}}{\layer{B}}{\del,\psp} \leq \meanCov{\layer{B}}{\layer{B}}{}$. 
Attenuation resulting from delay between layers $\inputlayer$ and $\layer{B}$ 
is captured in $\Dellayer{\inputlayer}{\layer{B}}$, which is itself a function 
of radial propagation delay, $\taur{\inputlayer}{\layer{B}}$, and inter-laminar 
propagation delay, $\taul{\inputlayer}{\layer{B}}$. For delay between layers 
$\layer{B}$ and $\layer{C}$, this is captured by the relationship between 
$\radius{\layer{B}}{\layer{C}}$ and $\velocity \distlayer{\layer{B}}{\layer{C}} \tau_D$ 
(velocity, distance between the layers and the time constant of propagation delay) 
such that the larger the connectivity radius, the larger the denominator and the 
more covariance is attenuated. The radial propagation delay spreads out the arrival 
time in every consecutive layer pair, compounding the impact of radial delay. 

Importantly, for large inter-laminar delay, mean covariance in the presence of
propagation delay approaches mean covariance in \citepos{Lin86a} network, 
\eq{\ref{eq:meanCov_layerB}}, since the impact of radial propagation delay 
becomes negligible, such that 
\begin{equation}\label{eq:avgCovLayerC_largeDelta}
    \meanCov{\layer{B}}{\layer{B}}{\del,\psp} \;\;\; 
	\overset{ \distlayer{\layer{B}}{\layer{C}} \rightarrow \; \infty}{=}  \;\;\; 
	\frac{ 1 } 
		 { 1 + \frac{ \radiusvar{\layer{B}}{\layer{C}} }
					{ \radiusvar{\inputlayer}{\layer{B}} } }. 
\end{equation}

The mean synapse weight converges in the early stages of learning, having a much 
faster time constant than that of the individual synapses \citep{KemGerHem99}. 
The learning equation for the mean weight synapse in the presence of propagation 
delay can be expressed as  
\begin{align}\label{eq:learningMeanWgt_DelPSP}
     \meanLwchange{\layer{B}}{\layer{C}} 
 &= 
	\etalayer{} \left( 
		\kone{\layer{B}}{\layer{C}}  + \meanLweight{\layer{B}}{\layer{C}} 
		\left( \ktwo{\layer{B}}{\layer{C}} 
			 + \meanCov{\layer{B}}{\layer{B}}{\del,\psp}
		\right) 
	\right),
	\quad 
    \LwgtMin{\layer{B}}{\layer{C}} \leq \meanLweight{\layer{B}}{\layer{C}} 
	                               \leq \LwgtMax{\layer{B}}{\layer{C}}, 
\end{align}
so that
\begin{align}\label{eq:linsker_fixedPt_DelPSP}
   \FP{\meanLweight{\layer{B}}{\layer{C}}} 
 &= 
   \frac{-\kone{\layer{B}}{\layer{C}}}
        { \ktwo{\layer{B}}{\layer{C}} 
		+ \meanCov{\layer{B}}{\layer{B}}{\del,\psp} }.
\end{align}
Consequently, the fixed point of the mean weight is modified by the impact of 
propagation delay. Since the denominator is negative and 
$\ktwo{\layer{B}}{\layer{C}} < 0$, reducing the size of the mean covariance 
decreases the size of the mean weight to which the network converges. 

\begin{figure}[!tbp]
  \centering
  \includegraphics[width=0.6\textwidth]{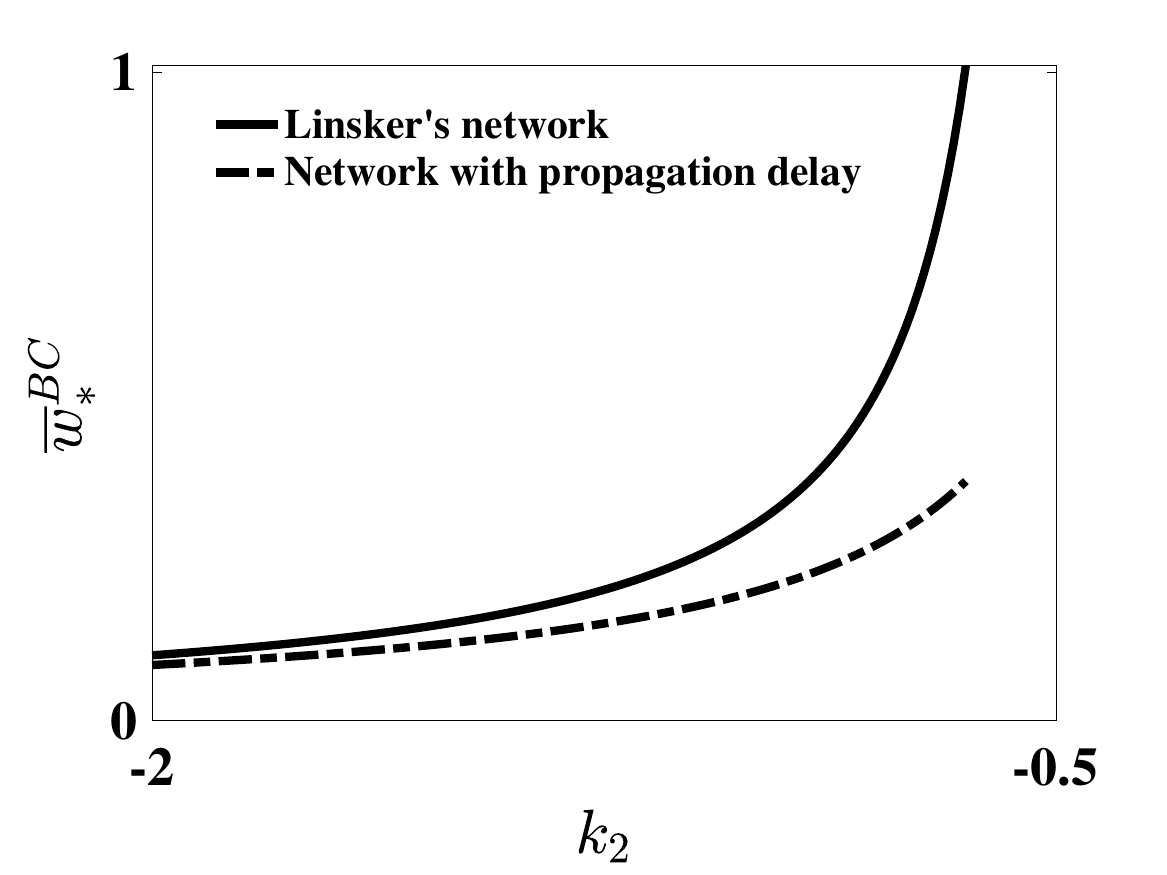}
	 \caption{Fixed point mean weight for \citepos{Lin86a} network 
		(\eq{\ref{eq:linsker_fixedPt}}, solid line) compared to the fixed point 
		mean weight for the network incorporating propagation delay and an 
		arbitrary \gls{psp} (\eq{\ref{eq:linsker_fixedPt_DelPSP}}, dot-dashed line). 
		The figure was generated with 
		$\kone{\layer{B}}{\layer{C}} = \SI{-0.5}{\per\second}$ and 
		$\overline{Q^{BB}} = \SI{0.5}{\per\second}$, 
		\eq{\ref{eq:meanCov_layerB}}. 
		Mean covariance for the network with delay was 
		$\overline{Q^{BB}}_{\Delta,\epsilon} = \SI{0.24}{\per\second}$, 
		\eq{\ref{eq:avgCovLayerC}}, calculated from 
		$\taupsp = 1 \si{\milli\meter^2}$, $\tau_D = 0.1 \si{\milli\meter^2}$, 
		$\radiusvar{\inputlayer}{\layer{B}} = 100$ grid spaces, 
		$\radiusvar{\layer{B}}{\layer{C}} = 200$ grid spaces, 
		$\distlayer{\layer{B}}{\layer{C}} = 400$ grid spaces, and  
		$\velocity = 10 \si{\meter/\second}$.
		\label{fig:weight_FP}}
\end{figure}

\fig{\ref{fig:weight_FP}} shows a comparison of the fixed point mean synapse weight 
for \citepos{Lin86a}, \eq{\ref{eq:meanCov_layerB}}, and the network incorporating 
propagation delay and arbitrary \psp, \eq{\ref{eq:linsker_fixedPt_DelPSP}}. 
Large changes in the homeostatic constant, $\ktwo{\layer{B}}{\layer{C}}$, result
in relatively small changes in the fixed point mean weight for the network with
propagation delay. In contrast, the same change in homeostatic constant causes
the mean weight in \citepos{Lin86a} network to change from a value near the
lower weight bound to a value to the upper weight bound. 

The expected receptive field size after incorporating both propagation delay and 
an arbitrary \gls{psp} function is 
\begin{align}
    \RF{\delsym,\psp}
 &= 
    \radius{\layer{B}}{\layer{C}}
	\sqrt{ \log \left( 
					\frac{ \ktwo{\layer{B}}{\layer{C}} 
						 + \meanCov{\layer{B}}{\layer{B}}{\del,\psp} }
						 { \LwgtMax{\layer{B}}{\layer{C}} 
						   \left( \ktwo{\layer{B}}{\layer{C}} 
								+ \meanCov{\layer{B}}{\layer{B}}{\del,\psp}  
						   \right) 
						 + \kone{\layer{B}}{\layer{C}} } 
			    \right) }
	\label{eq:delayPSP_RF} \\
  &= 
    \radius{\layer{B}}{\layer{C}}
	\sqrt{ \log \left( \frac{1}{ \LwgtMax{\layer{B}}{\layer{C}} 
			                   - \FP{\meanLweight{\layer{B}}{\layer{C}}} } \right) } \, . 
	\label{eq:delayPSP_RF_FP}
\end{align}
This equation shows that receptive field size depends on the mean covariance, 
$\meanCov{\layer{B}}{\layer{B}}{\del,\psp}$, which was derived to be a function
of the radial versus inter-laminar propagation delays and the \gls{psp} decay
time constant. The fixed point mean weight is more stable for the network with
propagation delay and, therefore, the range in receptive field sizes will also be
smaller. Consequently, propagation delay has the effect of limiting the impact
of correlation from distal neurons, which stabilises the fixed point mean weight
and receptive field sizes.


\glsresetall 

\section{Discussion}
   
\citet{Lin86a} proposed a simple rate-based, three-layer, feed-forward network 
to show how spatial opponent cells emerge after a period of learning based purely 
on spontaneous activity and in the absence of external environmental input. 
\citet{Lin86a} demonstrated that Gaussian distributed synaptic connectivity 
between the layers was sufficient to introduce spatially-dependent correlation 
in neural outputs, which in turn prompted a self-organizing network. 
\citet{MacMil90} provided a mathematical framework from which to study 
\citepos{Lin86a} network, while \citet{WimGerHem98} extended the analysis to 
included lateral connectivity in the third layer. While these works were formative 
in establishing a possible mechanism underlying synaptic learning of simple cells 
prior to birth, they assumed that propagation delay between all neurons was 
identical and, consequently, had negligible impact. 

In this study the assumption of identical propagation delays in \citepos{Lin86a} 
network, introducing delays that account for both radial distance (i.e., separation 
within the lamina) and inter-laminar distance (i.e., separation between the layers) 
is relaxed. The results demonstrate that the impact of propagation delay is to 
attenuate high frequency information in the neural activity. This is intuitive since 
propagation delay spreads out the arrival time of input to the neurons: when the 
inputs are summed together as they arrive at the postsynaptic neuron, the spread in 
arrival times filters out the high frequency components of the signal. As the impact 
of radial propagation delay increases by enlarging the synaptic connectivity radius, 
reducing neuron density, reducing propagation velocity, or reducing the inter-laminar 
distance, the spread in signal arrival times from the previous layer becomes more 
dispersed and the higher frequencies are increasingly attenuated. Consequently, 
propagation delay acts as a low-pass filter, where the cut-off frequency is determined 
by the relationship of the radial propagation delay to the inter-laminar propagation 
delay. Our results demonstrate that the ratio of inter-laminar propagation delay to 
within layer, radial propagation delay, is the crucial factor in determining the 
extent of the low-pass filtering. To facilitate the application of our results to 
analytical models of learning, we derived a simplified model of the impact of 
propagation delay using an effective delay time constant, determined from the 
inter-laminar propagation delay and the radial propagation delay. 

A general \gls{psp} with a finite time course was introduced, which was interpreted 
as spreading the probability of a spike over a finite period and thus low-pass filtering the input. The values of the \gls{psp} time constants were chosen to reflect 
the effective membrane time constant, which is shorter than the passive membrane time 
constant due to the neuron becoming more leaky as more synaptic channels open with 
increased activity \citep{BurMefGra03}. Where the \gls{psp} has a lower cut-off 
frequency than the effective cut-off frequency imposed by propagation delay, 
differences in the synaptic connectivity radius and inter-laminar distance had
negligible effect on the frequency spectra of the postsynaptic neuron's activity. 
Consequently, differences in neural activity resulting from propagation delay 
became minimal in these circumstances. Thus, the \gls{psp} can play an important 
role in standardising the frequency spectra of neural activity in the presence of 
differing propagation delay properties amongst neurons. Conversely, where propagation 
delay has a longer effective time constant than that of the \gls{psp} and, therefore, 
a lower cut-off frequency, propagation delay can significantly attenuate frequencies 
that are present after convolution with the \gls{psp}. Therefore, the effective time 
constant of each enables the processing of similar scales of temporal information. 
These results can apply to any topographical multi-layered network, and is seen 
in the retina, where time constants at the periphery have been found to be up to 
twice as long as those in the fovea, where neurons are much more densely packed
\citep{SinHooBauOkaWonRie17}. 

\citet{Lin86a} showed that, under Gaussian synaptic connectivity distributions,
covariance between firing rates of two neurons in the second layer is a Gaussian 
function of their radial separation distance in the lamina. In this study we showed 
that, when distance-dependent propagation delay between the first and second layers 
is considered and there is a more biophysically realistic \gls{psp}, covariance 
between the outputs of neurons in the second layer can be expressed as a scaled 
function of covariance between neurons in \citepos{Lin86a} network. That is, 
when propagation delay and \gls{psp} are accounted for, covariance between two 
neurons in the second layer remains a Gaussian function of their separation 
distance in the lamina, but is attenuated by a constant determined by the mean 
propagation delay between the first and second layers and the \gls{psp} function. 
The \gls{psp} was found to be important because, in its absence, if two presynaptic 
neurons at different distances from the postsynaptic neuron were instantaneously 
correlated, propagation delay reduced the correlation, since the activity from the 
two neurons arrived at the postsynaptic neurons at different times. Thus, the 
\gls{psp} plays a crucial role in enabling correlation at non-zero lags to be 
incorporated in the output of a postsynaptic neurons.

The change in frequency content of the summed input to a postsynaptic neuron as
radial propagation delay becomes more prominent may explain some experimental
results in the visual system. In the visual system, it has been found that 
contrast sensitivity is frequency dependent in the periphery, with sensitivity 
decreasing for higher frequencies \citep{VenLewUnsLun17}. This accords with our 
results, since neurons that are more spread out receive input that is attenuated 
at higher frequencies and thus higher amplitude input is required to obtain a 
similar response. Additionally, \citet{ThiStilBra96} demonstrated that cut-off 
frequency is a function of eccentricity, decreasing as distance from 
the fovea increases. This likewise accords with our results: as the impact of radial
delay increases with distance from the fovea, delay acts like a low-pass filter,
where the cut-off frequency decreases as neurons become more spread across the
laminar. 

A further contribution of this paper was to consider the size of the on-center
of the spatial opponent neuron that emerges in the third layer. This was determined 
by calculating the fixed point for the mean synaptic weight and establishing how 
large the on component was required to be in order to establish the fixed point mean 
weight. The learning equation for the mean weight is stable only when the competitive 
plasticity component is negative and larger in magnitude than the non-competitive 
component. The competitive component comprises a negative constant and the positive 
mean covariance between the presynaptic inputs. If the mean covariance approaches 
the magnitude of the negative constant in magnitude, the fixed-point weight becomes 
increasingly large until it reaches a point of instability. Furthermore, as the 
fixed-point weight approaches the upper weight bound, the size of the neuron's 
on-center increases rapidly. 

When propagation delay is incorporated into the network, the fixed point weight, 
and hence the on-center size of the spatial opponent neuron, is much more stable
to changes in homeostatic parameters and factors that determine covariance. A small 
change in the homeostatic constant parameters ($k_1 , k_2$) can trigger a large change 
in the fixed point weight in \citepos{Lin86a} network and, hence, a large change in 
the receptive field structure of the cell. The fixed-point weight for the network 
with propagation delay is comparatively stable. The region of instability for 
\citepos{Lin86a} is largely avoided since covariance is attenuated, and so the fixed 
point weight does not approach the upper bound. 

The results of this study show that a small connectivity radius in addition to densely 
packed cells, and consequently negligible radial propagation delay, both act to maintain 
large covariances between presynaptic inputs. The consequence of this is a larger
mean synaptic weight and comparatively larger on-center spatial extent. Therefore, 
densely packed cells with small connectivity radii have mostly excitatory input and
very little inhibitory input. Conversely, cells that are more spread out and
more likely to connect to neurons further away have a smaller mean synaptic
weight, which results in a smaller on-center. Therefore, these neurons have more
inhibitory input. Interestingly, \citet{SunUenWagGarTanChe07} found that, in human 
\gls{v1}, cells responding to 
high spatial frequency preferred low speed input, while cells responding to low 
spatial frequency preferred high speed input, similar to what is found in cats 
\citep{ShoHubSchGriBon97}. Our results provide a potential explanation for this 
finding, in that simple cells preferring low spatial frequency input have a larger 
on-centre, requiring a higher mean covariance of presynaptic inputs, which in 
turn requires high speed input so that presynaptic spikes are arrival times are 
not too spread out.

Finally, although we assume here that axonal propagation dominates spike delay, our
results nevertheless generalise to the case in which dendrite propagation dominates 
delay, as the analytical results remain unchanged, but the additional delay between 
layers two and three is not required in the learning equation, simplifying the result. 

   \label{sec:delayDiscussion}

\section*{Acknowledgements}

	The authors acknowledge support under the Australian Research Council's Discovery 
	Projects funding scheme (Project DP140102947).
	CED acknowledges support of a University of Melbourne Research Fellowship.

\ifx\isEmbedded\undefined

\else



	\section*{Appendix}
	\begin{appendices}

	\section{Expected number of shared inputs}\label{app:sharedInputs}

	\section{Variance of neural activity}\label{app:var}

	\section{Covariance of neural activity}\label{app:cov}
		
We wish to derive expressions for the covariance of layer $\layer{B}$ neurons. 
Sample covariance between two postsynaptic neuron rates in layer $\layer{B}$, say 
$\Lrate{\postN}{\layer{B}}{}$ and $\Lrate{\postNtwo}{\layer{B}}{}$, for neurons 
$\postN$ and $\postNtwo$, respectively, is calculated as 
\begin{align}
   \scov{\Lrate{\postN}{\layer{B}}{}}{\Lrate{\postNtwo}{\layer{B}}{}}
 &= 
   \expect{\Lrate{\postN}{\layer{B}}{} \Lrate{\postNtwo}{\layer{B}}{}}
 - \expect{\Lrate{\postN}{\layer{B}}{}} \expect{\Lrate{\postNtwo}{\layer{B}}{}}.
 \end{align}
For unitary weights from layer $\inputlayer$ to layer $\layer{B}$,  
\eq{\ref{eq:Linsker_rate}} can be employed to give 
\begin{align}
   \scov{\Lrate{\postN}{\layer{B}}{}}{\Lrate{\postNtwo}{\layer{B}}{}}
 &= 
   \expect{\left(\Ra{\layer{B}} + \Rb{\inputlayer}{\layer{B}}
		   \sum\limits_{\preN} \Lrate{\preN}{\layer{A}}{} \right)
               \left(\Ra{\layer{B}} + \Rb{\inputlayer}{\layer{B}}
			   \sum\limits_{\preNtwo} \Lrate{\preNtwo}{\layer{A}}{}} \right)
 - \expect{\Ra{\layer{B}} + \Rb{\inputlayer}{\layer{B}}
   \sum\limits_{\preN} \Lrate{\preN}{\layer{A}}{}} 
   \expect{\Ra{\layer{B}} + \Rb{\inputlayer}{\layer{B}}
   \sum\limits_{\preNtwo} \Lrate{\preNtwo}{\layer{A}}{}} \notag \\
 &= 
   (\Ra{\layer{B}})^2 + 2\Ra{\layer{B}} \Rb{\inputlayer}{\layer{B}} \meanrate{\layer{B}} 
 + \RbSq{\inputlayer}{\layer{B}}
   \expect{\sum\limits_{\preN}\sum\limits_{\preNtwo}
	           \Lrate{\preN}{\layer{A}}{}\Lrate{\preNtwo}{\layer{A}}{}}
 - \left((\Ra{\layer{B}})^2 
	   + 2\Ra{\layer{B}} \Rb{\inputlayer}{\layer{B}} \meanrate{\layer{B}} 
       + (\N{\inputlayer}{\layer{B}} \Rb{\inputlayer}{\layer{B}} \meanrate{\layer{B}} )^2 
   \right)										                              \notag \\
 &=
   \RbSq{\inputlayer}{\layer{B}}
   \left( 
	   \expect{\sum\limits_{\preN} \sum\limits_{\preNtwo}
				   \Lrate{\preN}{\layer{A}}{}\Lrate{\preNtwo}{\layer{A}}{}}
	 - (\N{\inputlayer}{\layer{B}} \meanrate{\layer{B}})^2
   \right). 
\end{align}

Using the expression for the expected number of shared connections between layer 
$\layer{B}$ neurons in \eq{\ref{eq:sharedInputs}}, the covariance is given by 
\begin{align}\label{eq:Linsker_cov}
   \scov{\Lrate{\postN}{\layer{B}}{}}{\Lrate{\postNtwo}{\layer{B}}{}}
 &= 
   \frac{(\Rb{\inputlayer}{\layer{B}} \N{\inputlayer}{\layer{B}})^2 \layerrateSq{\inputlayer}}
		{2\pi\radiusvar{\inputlayer}{\layer{B}}}
   \exp{-\frac{\distance{}{}{}{}^2}{2\radiusvar{\inputlayer}{\layer{B}}}} . 
\end{align}

	\section{Covariance of neural activity in the frequency domain}\label{app:covFreq}

		Since instantaneous covariance is equivalent to the cross-correlation of 
		zero-meaned variates at lag 0, we can express covariance between two 
		layer $\layer{B}$ neurons as 
		\begin{align}\label{appEq:cov_FFT}
		   \cov{\Lrate{\postN}{\layer{B}}{}}{\Lrate{\postNtwo}{\layer{B}}{}}
		 &= 
		   \frac{1}{\Tmax^2} \Sumk \conj{\atFreq{\LRATE{\postN}{\layer{B}}}}
								\atFreq{\LRATE{\postNtwo}{\layer{B}}}
								\exp{2\pi \ii 0 \frac{\freq}{\Tmax}}		       \notag \\
		 &= 
		   \frac{1}{\Tmax^2} \Sumk \conj{\atFreq{\LRATE{\postN}{\layer{B}}}}
								\atFreq{\LRATE{\postNtwo}{\layer{B}}}.
		\end{align}

	\section{Derivation of delay in the frequency domain}\label{app:delayDerivation}
		
We wish to calculate the expected value of delay in the frequency domain. This 
requires an expression of delay as a function of distance in the frequency domain 
and the probability of obtaining each distance. The product of these two functions 
can then be integrated over the laminar to give the expected value. Since we are 
finding the expected value in the frequency domain, we can expect values to be 
complex. Consequently, contour integration will be used. 

To calculate the expected value of delay in the frequency domain, use 
$\expect{\distance{}{}{}{}}=\int \distfunc(\distance{}{}{}{})p(\distance{}{}{}{}) dr$ 
for some  function, $\distfunc(\distance{}{}{}{})$, of distance, $\distance{}{}{}{}$, 
with probability density, $p(\distance{}{}{}{})$. Using the known distribution of 
distance, given in \eq{\ref{eq:dist_Rayl}}, the linear transformation from distance 
to delay, given in \eq{\ref{eq:dist_3D}}, and the expression of a temporal delay 
in the frequency domain, given by 
$\fourier{\atDelayTime{x}{\del}}=\atFreq{X}\exp{-2\pi \ii\del\freqT}$, 
the function can be identified as 
$\distfunc(\distance{}{}{}{}) = 
    \exp{-2\pi \ii\nicefrac{\freq}{T}
         \frac{ \left( \distlayerSqu{\inputlayer}{\layer{B}} 
				     + \distance{}{}{}{} \right)^{\nicefrac{1}{2}} \gridspace{} }
			  { \velocity } }$
Consequently, the expected value of delay in the frequency domain can be determined 
by evaluating 
\begin{align}\label{eq:meanDelayEqn}
   \expect{\exp{-2\pi \ii \del \freqT}}
&= 
   \int\limits_0^{\infty}
       \exp{ -2\pi \ii\freqT
            \frac{ \left( \distlayerSqu{\inputlayer}{\layer{B}} 
		                + \distance{}{}{}{}^2 \right)^{\nicefrac{1}{2}} 
				   \gridspace{} }
			     { \velocity } }
       \frac{ 2\distance{}{}{}{} }{ \radiusvar{\inputlayer}{\layer{B}} }
       \exp{ -\frac{\distance{}{}{}{}^2}
			       {\radiusvar{\inputlayer}{\layer{B}}}}
	   d\distance{}{}{}{},
\end{align}
where we have used 
$\distance{\preN}{\postN}{\inputlayer}{\layer{B}} = \distance{}{}{}{}$. 
Introduce a change of variable, 
$\distCOV = \left(\distlayerSqu{\inputlayer}{\layer{B}} 
		        + \distance{}{}{}{}^2
		    \right)^{\nicefrac{1}{2}}$, so that 
$d\distance{}{}{}{} = 
    d\distCOV\left(\distCOV^2 - 
		      	   \distlayerSqu{\inputlayer}{\layer{B}} 
		     \right)^{-\nicefrac{1}{2}}\distCOV$, 
and
\begin{align}\label{eq:exp_delay_int}
   \expect{ \exp{-2\pi \ii \del \freqT} }
 &= 
   \frac{2}{ \radiusvar{\inputlayer}{\layer{B}} }
   \int\limits_{\distlayer{\inputlayer}{\layer{B}}}^{\infty} \distCOV
       \exp{-2\pi \ii\freqT \frac{ \distCOV \gridspace{} }{ \velocity }}
       \exp{-\frac{\distCOV^2 - \distlayerSqu{\inputlayer}{\layer{B}}}
			      {\radiusvar{\inputlayer}{\layer{B}}}}
	   d\distCOV			  \notag \\
 &= 
   \frac{2}{\radiusvar{\inputlayer}{\layer{B}}}
   \exp{\frac{\distlayerSqu{\inputlayer}{\layer{B}}}
	         {\radiusvar{\inputlayer}{\layer{B}}}} 
   \int\limits_{\distlayer{\inputlayer}{\layer{B}}}^{\infty} \distCOV
       \exp{-\frac{\distCOV^2}
			      {\radiusvar{\inputlayer}{\layer{B}}}}
       \exp{-2\pi i\freqT \frac{ \distCOV \gridspace{} }{ \velocity }}
	   d\distCOV.			 \\
\end{align}

This integral has the form 
$\int\limits_{x_0}^{\infty} x e^{-ax^2} e^{-\ii bx} dx$, which we will use for the 
interim for readability. Using integration by parts,
\begin{align}
   \int\limits_{x_0}^{\infty} f(x) g'(x) dx 
 &= 
   f(x) g(x) \bigg]_{x_0}^{\infty} - \int\limits_{x_0}^{\infty} g(x)f'(x) dx,
\end{align}
and setting $g(x)=-\frac{1}{2a}e^{-ax^2}$, $f(x)=e^{-\ii bx}$, so that 
$g'(x)=xe^{-ax^2}$, and $f'(x)=-\ii be^{-\ii bx}$, 
and
\begin{align}\label{eq:complex_exp_integral}
     \int\limits_{x_0}^{\infty} x e^{-ax^2} e^{-\ii bx} dx 
 &= 
   - \frac{1}{2a} e^{-\ii bx} e^{-ax^2} \bigg]_{x_0}^{\infty} 
   -  
     \frac{1}{2a} \int\limits_0^{\infty} \ii be^{-\ii bx} e^{-ax^2} dx  \notag \\
 &= 
   + \frac{1}{2a} e^{-ax_0^2 - \ii bx_0}
   + \frac{ib}{2a}
   - \int\limits_{x_0}^{\infty} e^{-\left(ax^2 + \ii bx\right)} dx.     
\end{align}
Since the integral in \eq{\ref{eq:complex_exp_integral}} is over a complex domain, 
it can be solved 
using contour integration. Cauchy's theorem states that, for an 
analytic function that is differentiable everywhere, a closed line integral 
of the function evaluates to zero. Consequently, in general, 
\begin{align}
  0 &= \oint e^{-ax^2} dx                                      \\
 &= 
   \int\limits_{x_0}^{\infty} e^{-a\left(x + \ii z\right)^2} dx 
							   \bigg]_{z=0}
 + \ii \int\limits_{0}^p      e^{-a\left(x + \ii z\right)^2} dz
                               \bigg]_{x=\infty} 
 - \int\limits_{x_0}^{\infty} e^{-a\left(x+\ii z\right)^2} dx
							   \bigg]_{z=p}
 - \ii \int\limits_{0}^p      e^{-a\left(x + \ii z\right)^2} dz
                               \bigg]_{x=0},                    \\
\end{align}
where the contour is from 
$x_0    + \ii 0 \rightarrow \infty + \ii 0 $, then 
$\infty + \ii 0 \rightarrow \infty + \ii p $ , from 
$\infty + \ii p \rightarrow x_0    + \ii p $, and finally 
$x_0    + \ii p \rightarrow x_0    + \ii 0 $. 
The second term on the right-hand side evaluates to $0$ since it contains 
$e^{-\infty}$, so that 
\begin{align}
  0 &= \oint e^{-ax^2} dy   \\
 &= 
   \int\limits_{x_0}^{\infty} e^{-ax^2} dy 
 - \int\limits_{x_0}^{\infty} e^{-a\left(x+\ii p\right)^2} dy
 - \ii \int_0^p e^{-a\left(x_0 + \ii z\right)^2} dz. \\
\end{align}
Using the fact that 
$\int_{-\infty}^{\infty}e^{-ax^2} = \sqrt{\frac{\pi}{a}}$, 
in conjunction with $e^{-ax^2}$ being a real function, we know that 
$\int_0^{\infty}e^{-ax^2} = \sqrt{\frac{\pi}{4a}}$. 
Therefore, 
\begin{align}\label{eq:contour_integral}
   \int\limits_{x_0}^{\infty} e^{-a\left(x + \ii p\right)^2} dx
 &= 
   \int\limits_{x_0}^{\infty} e^{-ax^2} dy 
 - \ii\int_0^p e^{-a\left(x_0 + \ii z\right)^2} dz                   \notag \\
 &=
   \sqrt{\frac{\pi}{4a}} \erfc{\sqrt{a}x}
 - \ii \sqrt{\frac{\pi}{4a}}
   \left( \erf{\sqrt{a}p\left( x_0 + ip \right)} 
		- \erf{\sqrt{a}x_0}\right)  \notag \\
 &=
   \sqrt{\frac{\pi}{4a}}
   \left(\erfc{\sqrt{a}x} 
	     - \ii\erf{\sqrt{a}\left( x_0 + \ii p \right)} 
		 + \ii\erf{\sqrt{a}x_0}\right)  \notag \\
\end{align}
where we have used the definition of 
$\erfi{z}=\frac{2}{\sqrt{\pi}}\int\limits_0^z e^{t^2} dt$, followed by 
$\erfi{z}=-\ii\erf{\ii z}$.

For the specific integral of $e^{-\left(ax^2+\ii bx\right)}$, set $p=\frac{b}{2a}$,
\begin{align}
    e^{-a \left(x + ip \right)^2} 
 &= e^{-a \left(x + \ii\frac{b}{2a} \right)^2} \\
 &= e^{-ax^2 - \ii bx} e^{\frac{b^2}{4a}}, 
\end{align}
so that, from \eq{\ref{eq:contour_integral}}, 
\begin{align}
  e^{-\left(ax^2 + ibx\right)} e^{\frac{b^2}{4a}} 
 &= 
   \sqrt{\frac{\pi}{4a}}
   \left(\erfc{\sqrt{a}x} 
	     - \ii\erf{\sqrt{a} \left( x_0 + \ii\frac{b}{2a} \right)}
		 + \ii\erf{\sqrt{a}x_0}
   \right), 
\end{align}
 or
\begin{align}
   e^{-ax^2} e^{-\ii bx}
 &= 
   e^{\frac{-b^2}{4a}} \sqrt{\frac{\pi}{4a}} 
   \left(\erfc{\sqrt{a}x} 
	     - \ii\erf{\sqrt{a}x_0 + \ii\frac{b}{2\sqrt{a}}} 
		 + \ii\erf{\sqrt{a}x_0}
   \right).  
\end{align}

Substituting this result back into \eq{\ref{eq:complex_exp_integral}}, we get
\begin{align}\label{eq:complex_exp_result}
   \int\limits_{x_0}^{\infty} x e^{-ax^2} e^{- \ii bx} dx 
 &= 
   \frac{1}{2a} - \frac{ \ii b\sqrt{\pi}}{4a^{\nicefrac{3}{2}}}
   e^{\frac{-b^2}{4a}} 
   \left(\erfc{\sqrt{a}x_0} 
	     -  \ii\erf{\sqrt{a}x_0 +  \ii\frac{b}{2\sqrt{a}}} 
		 +  \ii\erf{\sqrt{a}x_0}
   \right)                                                   \notag \\
 &= 
    \frac{1}{2a} e^{-ax_0^2 - \ii bx_0}
  - \frac{b\sqrt{\pi}}{4a^{\nicefrac{3}{2}}}
     e^{\frac{-b^2}{4a}} 
    \left( \ii \erfc{\sqrt{a}x_0} 
 	      + \erf{\sqrt{a}x_0 +  \ii\frac{b}{2\sqrt{a}}} 
		  - \erf{\sqrt{a}x_0}
    \right).  
\end{align}

Applying this result to the original equation in \eq{\ref{eq:exp_delay_int}}
using $a = \frac{1}{\radiusvar{\inputlayer}{\layer{B}}}$, 
	  $b = 2\pi \freqT \frac{ \gridspace{} }{ \velocity }$, and 
	  $x_0 = \distlayer{\inputlayer}{\layer{B}}$, the expected value 
of delay between layers $\inputlayer$ and layer $\layer{B}$ is 
\begin{align}\label{eq:DelPSPCoeff}
   \atFreqAndRadius{\delaymean}{\inputlayer}{\layer{B}}
 =& \, 
   \expect{ \exp{-2\pi \ii \del \freqT} }    \notag \\
 =& \, 
   \exp{-2\pi \ii \freqT \frac{ \distlayer{\inputlayer}{\layer{B}} \gridspace{} }
	                          { \velocity } }					   \notag \\
  & \qquad 
  - \pi^{\nicefrac{3}{2}} \freqT \frac{ \radius{\inputlayer}{\layer{B}} \gridspace{} }
									  { \velocity }  
    \exp{ \frac{\distlayerSqu{\inputlayer}{\layer{B}}}
	           {\radiusvar{\inputlayer}{\layer{B}}} } 
    \exp{ -\left( \pi \freqT \frac{ \radius{\inputlayer}{\layer{B}} \gridspace{} }
			                      { \velocity} \right)^2 }
    \left( \ii\erfc{\frac{\distlayer{\inputlayer}{\layer{B}}}
	                     {\radius{\inputlayer}{\layer{B}}}}  
		 + \erf{ \frac{\distlayer{\inputlayer}{\layer{B}}}
	                  {\radius{\inputlayer}{\layer{B}}}
			   + \ii\pi \freqT \frac{ \radius{\inputlayer}{\layer{B}} \gridspace{} } 
			                        { \velocity } }					  
		 - \erf{ \frac{\distlayer{\inputlayer}{\layer{B}}}
	                  {\radius{\inputlayer}{\layer{B}}}} \right).
\end{align}

For $\distlayer{\inputlayer}{\layer{B}} >> \radius{\inputlayer}{\layer{B}}$, this 
reduces to 
$ \expect{ \exp{-2\pi \ii \freqT \frac{ \distlayer{\inputlayer}{\layer{B}} \gridspace{} }
	                          { \velocity } } }$
which has an absolute value of 1.

	\section{Derivation of average covariance}\label{app:avgCov}
	   \subsection{Linsker's network}
	      \label{subapp:avgCovLinsker}

	   \subsection{Network with propagation delay and arbitary postsynaptic potential function}
	      \label{subapp:avgCovDelPSP}
		  
For a network that incorporates the impact of delay and an arbitrary \gls{psp} 
function, the average covariance of the pre-synaptic inputs to the post-synaptic
neurons in layer $\layer{C}$ must be calculated using the expression for covariance 
between two filtered and delayed layer $\layer{B}$ neuron outputs, since the 
signals arrive at the synapse where learning is assumed to occur. This expression 
was determined in \eq{\ref{eq:covC_DelayExpPSP}}, where it was assumed that synapse 
weight changes are initiated near the cell body of the post-synaptic neuron. To 
calculate the average normalized covariance, \eq{\ref{eq:cov_temp2ensemble}}, we 
use Cartesian coordinates, recognizing that the expression for covariance is 
circularly symmetric. Therefore, we determine the result for a single dimension 
and square it. We first find the average covariance for a neuron at position 
$\poscont{}{}$ with all other neurons in the laminar, and then find the average 
across all neurons. Propagation delay for spikes between the pre-synaptic neuron 
at $\poscont{}{}$ and the post-synaptic neuron in layer $\layer{C}$ can be 
expressed as 
$ \delay{\poscont}{}{\layer{B}}{\layer{C}} 
  = 
  \frac{ \left( \distlayerSqu{\layer{B}}{\layer{C}}
			  + \norm{\poscont}^2 \right)^{\nicefrac{1}{2}} \gridspace{} }
       { \velocity} \,$ , 
which can be approximated by 
$ \delay{\poscont}{}{\layer{B}}{\layer{C}} 
  \approx 
  \frac{ \norm{\poscont}^2 }
       {2\distlayer{\layer{B}}{\layer{C}} \velocity} \, $  
for inter-laminar distances, $\distlayer{\layer{B}}{\layer{C}}$, significantly 
larger than the pre-synaptic neuron's radial distance to the post-synaptic neuron, 
$\norm{\poscont{}{}}$. Mean covariance can then be expressed as 
\begin{align}
   \meanCov{\layer{B}}{\layer{B}}{\del,\psp}
 \approx & 
   \frac{1}{ \pi^2 \radiusquad{\layer{B}}{\layer{C}} } 
   \left( 
	   \int_{-\infty}^{\infty} d\posxcont  \,
	   \int_{-\infty}^{\infty} d\posxtwocont  \,
		   \exp{- \frac{ \posxcont^2 + \posxtwocont^2}
					   {2\radiusvar{\inputlayer}{\layer{B}}} }
		   \exp{  \frac{ \posxcont \posxtwocont}
					   { \radiusvar{\inputlayer}{\layer{B}}} }
		   \exp{- \frac{ \posxcont^2 + \posxtwocont^2}
					   { \radiusvar{\layer{B}}{\layer{C}} } } 
        \DelClayer{\posxcont}{\posxtwocont}
   \right)^2  \notag \\
 =&
   \frac{1}{ \pi \radiusquad{\layer{B}}{\layer{C}} 
	         (\taupsp^2 - \tau_D^2)^2 } 
   \left( 
	   \int_{-\infty}^{\infty} d\posxcont  \,
	   \int_{-\infty}^{\infty} d\posxtwocont  \,
		   \exp{- \frac{ \posxcont^2 + \posxtwocont^2}
					   {2\radiusvar{\inputlayer}{\layer{B}}} }
		   \exp{  \frac{ \posxcont \posxtwocont}
					   { \radiusvar{\inputlayer}{\layer{B}}} }
		   \exp{- \frac{ \posxcont^2 + \posxtwocont^2}
					   { \radiusvar{\layer{B}}{\layer{C}} } }  
   \right.  \notag \\
 &\times 
   \left.
		   \left( \exp{- \frac{ \abs{\posxcont^2 - \posxtwocont^2} }
					   {2\distlayer{\layer{B}}{\layer{C}} \velocity \taupsp} }
				  \taupsp
		        - \exp{- \frac{ \abs{\posxcont^2 - \posxtwocont^2} }
					   {2\distlayer{\layer{B}}{\layer{C}} \velocity \tau_D} }
				  \tau_D
		   \right)
   \right)^2  \notag \\
 =&
   \frac{1}{ \pi \radiusquad{\layer{B}}{\layer{C}} 
	         (\taupsp^2 - \tau_D^2)^2 } 
   \left[ 
	   \int_{-\infty}^{\infty} d\posxcont  \,
	   \int_{-\infty}^{\posxcont} d\posxtwocont  \,
		   \exp{- \frac{ \posxcont^2 + \posxtwocont^2}
					   {2\radiusvar{\inputlayer}{\layer{B}}} }
		   \exp{  \frac{ \posxcont \posxtwocont}
					   { \radiusvar{\inputlayer}{\layer{B}}} }
		   \exp{- \frac{ \posxcont^2 + \posxtwocont^2 }
					   { \radiusvar{\layer{B}}{\layer{C}} } }  
   \right.  \notag \\
 &\times 
   \left.
		   \left( \exp{- \frac{ \posxcont^2 - \posxtwocont^2 } 
					          { 2\distlayer{\layer{B}}{\layer{C}} \velocity \taupsp} }
				  \taupsp
		        - \exp{- \frac{ \posxcont^2 - \posxtwocont^2 }
					  { 2\distlayer{\layer{B}}{\layer{C}} \velocity \tau_D} }
				  \tau_D
		   \right)
   \right.  \notag \\
 &+
   \left. 
	   \int_{-\infty}^{\infty} d\posxcont  \,
	   \int_{ \posxcont}^{\infty} d\posxtwocont  \,
		   \exp{- \frac{ \posxcont^2 + \posxtwocont^2}
					   {2\radiusvar{\inputlayer}{\layer{B}}} }
		   \exp{  \frac{ \posxcont \posxtwocont}
					   { \radiusvar{\inputlayer}{\layer{B}}} }
		   \exp{- \frac{ \posxcont^2 + \posxtwocont^2 }
					   { \radiusvar{\layer{B}}{\layer{C}} } }  
   \right.  \notag \\
 &\times 
   \left.
		   \left( \exp{- \frac{ \posxtwocont^2 - \posxcont^2 } 
					          { 2\distlayer{\layer{B}}{\layer{C}} \velocity \taupsp} }
				  \taupsp
		        - \exp{- \frac{ \posxtwocont^2 - \posxcont^2 }
					  { 2\distlayer{\layer{B}}{\layer{C}} \velocity \tau_D} }
				  \tau_D
		   \right)
   \right]^2  \notag \\
 =& 
   \frac{1}{ \pi \radiusquad{\layer{B}}{\layer{C}} 
	         (\taupsp^2 - \tau_D^2)^2 } 
   \left[ 
   \left( 
	   \int_{-\infty}^{\infty} d\posxcont  \,
		   \exp{- \frac{ \posxcont^2 }{ \beta^2_{\psp} } }
	   \int_{-\infty}^{\posxcont} d\posxcont  \,
		   \exp{- \frac{ \posxtwocont^2}{ \tilde{\beta}^2_{\psp} } }
		   \exp{  \frac{ \posxcont \posxtwocont}
					   { \radiusvar{\inputlayer}{\layer{B}}} }
		   \taupsp 
   \right. \right. \notag \\
  -&
   \left.
	   \int_{-\infty}^{\infty} d\posxcont  \,
		   \exp{- \frac{ \posxcont^2 }{ \beta^2_{D} } }
	   \int_{-\infty}^{\posxcont} d\posxcont  \,
		   \exp{- \frac{ \posxtwocont^2}{ \tilde{\beta}^2_{D} } }
		   \exp{  \frac{ \posxcont \posxtwocont}
					   { \radiusvar{\inputlayer}{\layer{B}}} }
		   \tau_D 
   \right) \notag \\
  & \, +
   \left( 
	   \int_{-\infty}^{\infty} d\posxcont  \,
		   \exp{- \frac{ \posxcont^2 }{ \tilde{\beta}^2_{\psp} } }
	   \int_{\posxcont}^{\infty} d\posxcont  \,
		   \exp{- \frac{ \posxtwocont^2}{ \beta^2_{\psp} } }
		   \exp{  \frac{ \posxcont \posxtwocont}
					   { \radiusvar{\inputlayer}{\layer{B}}} }
		   \taupsp 
   \right. \notag \\
  -&
   \left. \left.
	   \int_{-\infty}^{\infty} d\posxcont  \,
		   \exp{- \frac{ \posxcont^2 }{ \tilde{\beta}^2_{D} } }
	   \int_{\posxcont}^{\infty} d\posxcont  \,
		   \exp{- \frac{ \posxtwocont^2}{ \beta^2_{D} } }
		   \exp{  \frac{ \posxcont \posxtwocont}
					   { \radiusvar{\inputlayer}{\layer{B}}} }
		   \tau_D 
   \right) 
   \right]^2  \notag \\
\end{align}
where 
$\beta_{\psp}^2
	= \frac{2 \velocity \distlayer{\layer{B}}{\layer{C}} \taupsp 
		      \radiusvar{\inputlayer}{\layer{B}} 
	          \radiusvar{\layer{B}}{\layer{C}}}
           {  \velocity \distlayer{\layer{B}}{\layer{C}} \taupsp 
			  \radiusvar{\layer{B}}{\layer{C}} 
			+2\velocity \distlayer{\layer{B}}{\layer{C}} \taupsp 
			  \radiusvar{\inputlayer}{\layer{B}} 
		    + \radiusvar{\inputlayer}{\layer{B}} 
			  \radiusvar{\layer{B}}{\layer{C}}   } $, 
$\tilde{\beta}^2_{\psp} 
    = \frac{2 \velocity \distlayer{\layer{B}}{\layer{C}} \taupsp 
              \radiusvar{\inputlayer}{\layer{B}} 
              \radiusvar{\layer{B}}{\layer{C}}}
           {  \velocity \distlayer{\layer{B}}{\layer{C}} \taupsp 
		      \radiusvar{\layer{B}}{\layer{C}} 
			+2\velocity \distlayer{\layer{B}}{\layer{C}} \taupsp 
			  \radiusvar{\inputlayer}{\layer{B}} 
			- \radiusvar{\inputlayer}{\layer{B}} 
			  \radiusvar{\layer{B}}{\layer{C}}   } $, 
\newline 
$\beta_{D}^2
	= \frac{2 \velocity \distlayer{\layer{B}}{\layer{C}} \tau_D
		      \radiusvar{\inputlayer}{\layer{B}} 
	          \radiusvar{\layer{B}}{\layer{C}}}
           {  \velocity \distlayer{\layer{B}}{\layer{C}} \tau_D 
			  \radiusvar{\layer{B}}{\layer{C}} 
			+2\velocity \distlayer{\layer{B}}{\layer{C}} \tau_D 
			  \radiusvar{\inputlayer}{\layer{B}} 
		    + \radiusvar{\inputlayer}{\layer{B}} 
			  \radiusvar{\layer{B}}{\layer{C}}   } $, and 
$\tilde{\beta}^2_{D} 
    = \frac{2 \velocity \distlayer{\layer{B}}{\layer{C}} \tau_D 
              \radiusvar{\inputlayer}{\layer{B}} 
              \radiusvar{\layer{B}}{\layer{C}}}
           {  \velocity \distlayer{\layer{B}}{\layer{C}} \tau_D 
		      \radiusvar{\layer{B}}{\layer{C}} 
			+2\velocity \distlayer{\layer{B}}{\layer{C}} \tau_D 
			  \radiusvar{\inputlayer}{\layer{B}} 
			- \radiusvar{\inputlayer}{\layer{B}} 
			  \radiusvar{\layer{B}}{\layer{C}}   } $, 
all have units of \si{\meter^2} and capture the extent of the radial spread of
covariance between the layer $\layer{B}$ neurons with the \gls{psp} time
constant, $\taupsp$ and the propagation delay time constant, $\tau_D$. The 
$\beta^2$ terms capture covariance of each $\posxcont$ neuron in the layer 
with $\posxtwocont$ neurons that are radially closer to the postsynaptic 
neuron (which, without loss of generality is assumed to be located at $(0.0)$). 
Conversely, the $\tilde{\beta}^2$ terms capture covariance of each $\posxcont$
neuron with other neurons in the layer that are radially further from the
postsynaptic neuron. 
Complete the square in the exponent for $\posxtwocont$ to obtain
\begin{align}
   \meanCov{\layer{B}}{\layer{B}}{\del,\psp}\left( \posxcont, \posycont \right)
 =& 
   \frac{1}{ \pi \radiusquad{\layer{B}}{\layer{C}} 
	         (\taupsp^2 - \tau_D^2)^2 } 
   \left[ 
	   \int_{-\infty}^{\infty} d\posxcont  \,
		   \exp{- \posxcont^2 
				  \left( \frac{1}{\beta_{\psp}^2}
					   - \frac{\tilde{\beta}_{\psp}^2}{4\radiusquad{\inputlayer}{\layer{B}}} 
				  \right) }
	   \int_{-\infty}^{\posxcont \left( \frac{1}{\tilde{\beta}_{\psp}}
									  - \frac{\tilde{\beta}_{\psp}}
									         {2\radiusvar{\inputlayer}{\layer{B}}} 
				                 \right) } d\chi  \,
			\beta_{\psp} \exp{-\chi^2} \taupsp
	\right. \notag \\
 & \, -
	   \int_{-\infty}^{\infty} d\posxcont  \,
		   \exp{- \posxcont^2 
				  \left( \frac{1}{\beta_{D}^2}
					   - \frac{\tilde{\beta}_{D}^2}{4\radiusquad{\inputlayer}{\layer{B}}} 
				  \right) }
	   \int_{-\infty}^{\posxcont \left( \frac{1}{\tilde{\beta}_{D}}
									  - \frac{\tilde{\beta}_{D}}{2\radiusvar{\inputlayer}{\layer{B}}} 
				                 \right) } d\chi  \,
			\beta_{D} \exp{-\chi^2} \tau_D 
	\notag \\
  & \, + 
	   \int_{-\infty}^{\infty} d\posxcont  \,
		   \exp{- \posxcont^2 
				  \left( \frac{1}{\tilde{\beta}_{\psp}^2}
					   - \frac{\beta_{\psp}^2}{4\radiusquad{\inputlayer}{\layer{B}}} 
				  \right) }
	   \int_{\posxcont \left( \frac{1}{\beta_{\psp}}
							- \frac{\beta_{\psp}}{2\radiusvar{\inputlayer}{\layer{B}}} 
				       \right) }^{\infty} d\tilde{\chi}  \,
			\tilde{\beta}_{\psp} \exp{-\tilde{\chi}^2} \taupsp 
	\notag \\
  & \, -
	\left. 
	   \int_{-\infty}^{\infty} d\posxcont  \,
		   \exp{- \posxcont^2 
				  \left( \frac{1}{\tilde{\beta}_{D}^2}
					   - \frac{\beta_{D}^2}{4\radiusquad{\inputlayer}{\layer{B}}} 
				  \right) }
	   \int_{\posxcont \left( \frac{1}{\beta_{D}}
							- \frac{\beta_{D}}{2\radiusvar{\inputlayer}{\layer{B}}} 
				       \right) }^{\infty} d\tilde{\chi}  \,
			\tilde{\beta}_{D} \exp{-\tilde{\chi}^2} \tau_D 
	\right]^2  \notag \\
  =&
   \frac{1}{ \pi \radiusquad{\layer{B}}{\layer{C}} 
	         (\taupsp^2 - \tau_D^2)^2 } 
 	\left[ 
	   \int_{-\infty}^{\infty} d\posxcont \, \taupsp \,
		   \exp{- \posxcont^2 
				  \left( \frac{1}{\beta_{\psp}^2}
					   - \frac{\tilde{\beta}_{\psp}^2}{4\radiusquad{\inputlayer}{\layer{B}}} 
				  \right) }
	   \left( 1 + \erf{\posxcont \left( \frac{1}{\tilde{\beta}_{\psp}}
									  - \frac{\tilde{\beta}_{\psp}}
									         {\radiusvar{\inputlayer}{\layer{B}}} 
								 \right) } 
	   \right) \beta_{\psp} 
	\right.  \notag \\
 & \, - 
	   \int_{-\infty}^{\infty} d\posxcont \, \tau_D \,
		   \exp{- \posxcont^2 
				  \left( \frac{1}{\beta_{D}^2}
					   - \frac{\tilde{\beta}_{Dn}^2}{4\radiusquad{\inputlayer}{\layer{B}}} 
				  \right) }
	   \left( 1 + \erf{\posxcont \left( \frac{1}{\tilde{\beta}_{D}}
									  - \frac{\tilde{\beta}_{D}}
									         {2\radiusvar{\inputlayer}{\layer{B}}} 
								 \right) } 
	   \right) \beta_{D}
	\notag \\
  & \; +  
	   \int_{-\infty}^{\infty} d\posxcont \, \taupsp \, 
		   \exp{- \posxcont^2 
				  \left( \frac{1}{\tilde{\beta}_{\psp}^2}
					   - \frac{\beta_{\psp}^2}{4\radiusquad{\inputlayer}{\layer{B}}} 
				  \right) }
	   \erfc{\posxcont \left( \frac{1}{\beta_{\psp}}
							- \frac{\beta_{\psp}}
								   {2\radiusvar{\inputlayer}{\layer{B}}} 
								 \right) } 
	   \tilde{\beta}_{\psp}
	\notag \\
  & \; -  
	\left. 
	   \int_{-\infty}^{\infty} d\posxcont \, \tau_D \,
		   \exp{- \posxcont^2 
				  \left( \frac{1}{\tilde{\beta}_{D}^2}
					   - \frac{\beta_{D}^2}{4\radiusquad{\inputlayer}{\layer{B}}} 
				  \right) }
	   \erfc{\posxcont \left( \frac{1}{\beta_{D}}
							- \frac{\beta_{D}}
								   {2\radiusvar{\inputlayer}{\layer{B}}} 
								 \right) } 
	   \tilde{\beta}_{D}
	\right]^2  \notag \\
  =&
   \frac{1}{ \pi \radiusquad{\layer{B}}{\layer{C}} 
	         (\taupsp^2 - \tau_D^2)^2 } 
 	\left[ 
	   \int_{-\infty}^{\infty} d\posxcont \, \taupsp \,
		   \exp{- \posxcont^2 
				  \left( \frac{1}{\beta_{\psp}^2}
					   - \frac{\tilde{\beta}_{\psp}^2}{4\radiusquad{\inputlayer}{\layer{B}}} 
				  \right) }
	   \left( 1 + \erf{\posxcont \left( \frac{1}{\tilde{\beta}_{\psp}}
									  - \frac{\tilde{\beta}_{\psp}}
									         {2\radiusvar{\inputlayer}{\layer{B}}} 
								 \right) } 
	   \right) \beta_{\psp} 
	\right.  \notag \\
 & \, - 
	   \int_{-\infty}^{\infty} d\posxcont \, \tau_D \,
		   \exp{- \posxcont^2 
				  \left( \frac{1}{\beta_{D}^2}
					   - \frac{\tilde{\beta}_{Dn}^2}{4\radiusquad{\inputlayer}{\layer{B}}} 
				  \right) }
	   \left( 1 + \erf{\posxcont \left( \frac{1}{\tilde{\beta}_{D}}
									  - \frac{\tilde{\beta}_{D}}
									         {2\radiusvar{\inputlayer}{\layer{B}}} 
								 \right) } 
	   \right) \beta_{D}
	\notag \\
  & \; +  
	   \int_{-\infty}^{\infty} d\posxcont \, \taupsp \, 
		   \exp{- \posxcont^2 
				  \left( \frac{1}{\tilde{\beta}_{\psp}^2}
					   - \frac{\beta_{\psp}^2}{4\radiusquad{\inputlayer}{\layer{B}}} 
				  \right) }
	   \left( 1 - \erf{\posxcont \left( \frac{1}{\beta_{\psp}}
									  - \frac{\beta_{\psp}}
									         {2\radiusvar{\inputlayer}{\layer{B}}} 
								 \right) } 
	   \right) \tilde{\beta}_{\psp}
	\notag \\
  & \; -  
	\left. 
	   \int_{-\infty}^{\infty} d\posxcont \, \tau_D \,
		   \exp{- \posxcont^2 
				  \left( \frac{1}{\tilde{\beta}_{D}^2}
					   - \frac{\beta_{D}^2}{4\radiusquad{\inputlayer}{\layer{B}}} 
				  \right) }
	   \left( 1 - \erf{\posxcont \left( \frac{1}{\beta_{D}}
									  - \frac{\beta_{D}}
									         {2\radiusvar{\inputlayer}{\layer{B}}} 
								 \right) } 
	   \right) \tilde{\beta}_{D}
	\right]^2  
\end{align}
The error function components can be shown to integrate to $0$, since 
$\int\limits_{\infty}^{\infty} \exp{-ay^2} \erf{ay}dy = 0$, giving 
\begin{align}
   \meanCov{\layer{B}}{\layer{B}}{\del,\psp}\left( \posxcont, \posycont \right)
 =&
   \frac{1}{ \radiusquad{\layer{B}}{\layer{C}} 
	         (\taupsp^2 - \tau_D^2)^2 } 
	\left[ 
	   \taupsp \, \beta_{\psp} 
	   \left( \frac{1}{\beta_{\psp}^2}
	 	    - \frac{\tilde{\beta}_{\psp}^2}{4\radiusquad{\inputlayer}{\layer{B}}} 
	   \right)^{ -\frac{1}{2} } 
   - 
	   \tau_D \, \beta_{D}
	   \left( \frac{1}{\beta_{D}^2}
		    - \frac{\tilde{\beta}_{D}^2}{4\radiusquad{\inputlayer}{\layer{B}}} 
	   \right)^{ -\frac{1}{2} } 
	\right.  \notag \\
  & \; +  
	   \taupsp \, \tilde{\beta}_{\psp}  
	   \left( \frac{1}{\tilde{\beta}_{\psp}^2}
		    - \frac{\beta_{\psp}^2}{4\radiusquad{\inputlayer}{\layer{B}}} 
	   \right)^{ -\frac{1}{2} } 
  -  
    \left.
	\tau_D \, \tilde{\beta}_{D} 
	   \left( \frac{1}{\tilde{\beta}_{D}^2}
		    - \frac{\beta_{D}^2}{4\radiusquad{\inputlayer}{\layer{B}}} 
	   \right)^{ -\frac{1}{2} } 
	\right]^2  \notag \\  
 =& 
	\frac{1}{ \radiusquad{\layer{B}}{\layer{C}} 
	         (\taupsp^2 - \tau_D^2)^2 } 
	\left(
	    \frac{ \taupsp \left( \beta_{\psp}^2 + \tilde{\beta}_{\psp}^2 \right) }
			 { \left( 1 - \frac{ \beta_{\psp}^2 \tilde{\beta}_{\psp}^2 }
						{ 4 \radiusquad{\inputlayer}{\layer{B}} }  
			   \right)^{\frac{1}{2} } }
  -
	    \frac{ \tau_D \left( \beta_{D}^2 + \tilde{\beta}_{D}^2 \right) }
			 { \left( 1 - \frac{ \beta_{D}^2 \tilde{\beta}_{D}^2 }
						       { 4 \radiusquad{\inputlayer}{\layer{B}} }
			   \right)^{\frac{1}{2}} }
	\right)^2  \notag \\
 =& 
	\frac{1}{ \radiusquad{\layer{B}}{\layer{C}} 
	         (\taupsp^2 - \tau_D^2)^2 } 
	    \frac{ \taupsp^2  \left( \beta_{\psp}^2 + \tilde{\beta}_{\psp}^2 \right)^2 }
			 { \left( 1 - \frac{ \beta_{\psp}^2 \tilde{\beta}_{\psp}^2 }
						{ 4 \radiusquad{\inputlayer}{\layer{B}} }  
			   \right) } 
  +
	    \frac{ \tau_D^2   \left( \beta_{D}^2 + \tilde{\beta}_{D}^2 \right)^2 }
			 { \left( 1 - \frac{ \beta_{D}^2 \tilde{\beta}_{D}^2 }
						       { 4 \radiusquad{\inputlayer}{\layer{B}} }
			   \right) }
  -
	    \frac{2\tau_D \taupsp \left( \beta_{\psp}^2 + \tilde{\beta}_{\psp}^2 \right)
							  \left( \beta_{D}^2 + \tilde{\beta}_{D}^2 \right) }
			 { \left( 1 - \frac{ \beta_{\psp}^2 \tilde{\beta}_{\psp}^2 }
						       { 4 \radiusquad{\inputlayer}{\layer{B}} }
			   \right)^{ \frac{1}{2} } 
               \left( 1 - \frac{ \beta_{D}^2 \tilde{\beta}_{D}^2 }
						       { 4 \radiusquad{\inputlayer}{\layer{B}} }
			   \right)^{ \frac{1}{2} }
			   }
\end{align}
Substituting back in the definitions for $\beta_{\taupsp}$ and 
$\tilde{\beta}_{\taupsp}$, $\beta_{D}$, and $\tilde{\beta}_{D}$ and 
simplifying gives
\begin{align}\label{app:avgCovLayer}
   \meanCov{\layer{B}}{\layer{B}}{\del,\psp}
 =&
   \frac{1}{ (\taupsp - \tau_D)^2 (\taupsp + \tau_D)^2 } 
   \left( 
	   \frac{4 \left( 2 + \frac{ \radiusvar{\layer{B}}{\layer{C}} } 
							 { \radiusvar{\inputlayer}{\layer{B}} }
			   \right) \taupsp^2
			}
			{\left(  \frac{ \radiusvar{\layer{B}}{\layer{C}} }{ \velocity \distlayer{\layer{B}}{\layer{C}} \taupsp } 
				 - 2 \right)
			 \left(  \frac{ \radiusvar{\layer{B}}{\layer{C}} }{ \velocity \distlayer{\layer{B}}{\layer{C}} \taupsp }
				  -  2 
				  -2 \frac{ \radiusvar{\layer{B}}{\layer{C}} } 
						  { \radiusvar{\inputlayer}{\layer{B}} } 
			 \right) 
			 \left(  \frac{ \radiusvar{\layer{B}}{\layer{C}} }{ \velocity \distlayer{\layer{B}}{\layer{C}} \taupsp }
				  +  2 
				  +  \frac{ \radiusvar{\layer{B}}{\layer{C}} }
						  { \radiusvar{\inputlayer}{\layer{B}} }
			 \right) 
			}  
   \right. \notag \\
 &+ 
   \left.
   \frac{4 \left( 2 + \frac{ \radiusvar{\layer{B}}{\layer{C}} } 
				           { \radiusvar{\inputlayer}{\layer{B}} }
		   \right) \tau_D^2 
        }
        {\left(  \frac{ \radiusvar{\layer{B}}{\layer{C}} }
		              { \velocity \distlayer{\layer{B}}{\layer{C}} \tau_D }  
			 - 2 \right)
		 \left(  \frac{ \radiusvar{\layer{B}}{\layer{C}} }{ \velocity \distlayer{\layer{B}}{\layer{C}} \tau_D }
		      -  2 
		      -2 \frac{ \radiusvar{\layer{B}}{\layer{C}} } 
			          { \radiusvar{\inputlayer}{\layer{B}} } 
		 \right) 
	     \left(  \frac{ \radiusvar{\layer{B}}{\layer{C}} }{ \velocity \distlayer{\layer{B}}{\layer{C}} \tau_D }
		      +  2 
			  +  \frac{ \radiusvar{\layer{B}}{\layer{C}} }
			          { \radiusvar{\inputlayer}{\layer{B}} }
		 \right)
		} 
   \right) .
\end{align}
The final expression for mean covariance between pre-synaptic inputs on arrival 
at the post-synaptic neuron is a function of the following components: the size 
of the radial propagation delay to inter-laminar propagation delay between 
layers $\inputlayer$ and $\layer{B}$, captured in $\tau_D$, the time constant of
the exponential \gls{psp}, $\taupsp$, the ratio of the connectivity radii 
between each layer, 
$\frac{\radiusvar{\layer{B}}{\layer{C}}}{\radiusvar{\inputlayer}{\layer{B}}}$, 
the ratio of the connectivity radius between layers $\layer{B}$ and
$\layer{C}$ to the inter-laminar propagation delay, 
$\velocity \distlayer{\layer{B}}{\layer{C}} \tau_D$,  
which captures the impact of propagation delay between layers $\layer{B}$ and 
$\layer{C}$, and the ratio of the connectivity radius between layers $\layer{B}$ 
and $\layer{C}$ to the \gls{psp} , 
$\velocity \distlayer{\layer{B}}{\layer{C}} \taupsp$, which captures the temporal 
spread of incoming spike to layer $\layer{C}$ resulting from the \gls{psp}, 
relative to the propagation delay between layers $\layer{B}$ and $\layer{C}$.

	\end{appendices}

\fi

\bibliographystyle{plainnat}
\bibliography{stdp_refs}


\end{document}